\documentclass[journal,final,twoside]{IEEEtran}
\usepackage[normalem]{ulem}

\usepackage{multirow}
\pdfoutput=1 

\usepackage{graphicx} 
\DeclareRobustCommand*{\IEEEauthorrefmark}[1]{\raisebox{0pt}[0pt][0pt]
{\textsuperscript{\footnotesize\ensuremath{\ifcase#1\or 1\or 2\or 3\or%
    4\or 5\or 6\or 7\or 8
\else\textsuperscript{\expandafter\romannumeral#1}\fi}}}}
\usepackage{flushend}
\usepackage[table,svgnames]{xcolor} % JYK 28.06.2013, 11:17

\newcommand{\Case}[1]{\emph{Case~#1}}
\usepackage[cal=boondoxo]{mathalfa}
\usepackage[
colorlinks=true,
allcolors=DarkBlue,
urlbordercolor={1 0 0}, % hyperlink borders will be red
pdfborderstyle={/S/U/W 0.5},% border style will be underline of width 0.5pt
]{hyperref}
\hypersetup{pdflinkmargin=-0.5pt}

\hypersetup{%
  pdfsubject={This paper investigates the application of fast convolution (FC) filtering schemes for flexible and effective waveform generation and processing in 5th generation (5G) systems.},
  pdfkeywords={5G, physical layer, 5G New Radio, 5G-NR, multicarrier, waveforms, filtered-OFDM, fast-convolution},
  pdfauthor={Juha Yli-Kaakinen, Toni Levanen, Sami Valkonen, Kari Pajukoski, Juho Pirskanen,\\
Markku Renfors, and Mikko Valkama}%
}

\def\kay{\ensuremath{\mathcal k}}

\usepackage{siunitx}
\usepackage[utf8]{inputenc} % JYK 27.11.2016, 16:48
\usepackage[labelformat=simple]{subcaption}
\DeclareCaptionFormat{rs}{\footnotesize#1#2#3} 
\usepackage{booktabs}
\usepackage{amsmath}
\usepackage{amssymb}
\usepackage{newtxmath}

\providecommand{\iu}{\ensuremath{{\mathop{\mspace{1mu}\mathrm{j}\mspace{0.5mu}}\nolimits}}}
\providecommand*{\eu}{\ensuremath{\mathrm{e}}}
\DeclareMathOperator{\diag}{diag}% 

\newcommand{\norm}[1]{\left\lVert#1\right\rVert}

\makeatletter
\newcommand*{\transpose}{%
  {\mathpalette\@transpose{}}%
}
\newcommand*{\@transpose}[2]{%
  \raisebox{\depth}{$\m@th#1\intercal$}%
}
\makeatother
\usepackage[nolist,nohyperlinks]{acronym} 

\makeatletter
\DeclareMathSizes{\@xpt}{\@xpt}{7}{5}
\makeatother

\usepackage{xkvltxp}
\def\figwidth{0.4905\textwidth}
\def\FIGURE_WIDTH{0.945\columnwidth}
\begin{document}

\title{Efficient Fast-Convolution Based Waveform Processing for 5G Physical Layer}
\author{Juha Yli-Kaakinen, Toni Levanen, Sami Valkonen, Kari Pajukoski, Juho Pirskanen,\\
Markku~Renfors,~\IEEEmembership{Fellow,~IEEE}, and Mikko~Valkama,~\IEEEmembership{Senior~Member,~IEEE}

\thanks{This work was partially supported by the Finnish Funding Agency for Technology and Innovation (Tekes) and Nokia Bell Labs, under the projects ``Phoenix+'', ``5G Radio Systems Research'', and ``Wireless for Verticals (WIVE)'', and in part by the Academy of Finland under the project no. 284694 and no. 284724. Early stage results of this paper have been submitted to EUCNC 2017, Oulu, Finland \cite{C:Yli-Kaakinen17:EUCNCsubm}}

\thanks{Juha Yli-Kaakinen, Toni Levanen, Sami Valkonen, Markku Renfors, and Mikko Valkama are with the Department of Electronics and Communications Engineering, Tampere University of Technology, FI-33101 Tampere, Finland \, (e-mail: $\lbrace$juha.yli-kaakinen; toni.levanen; sami.valkonen; markku.renfors; mikko.e.valkama$\rbrace$@tut.fi)}

\thanks{Kari Pajukoski and Juho Pirskanen are with the Nokia Bell Labs, Finland \, \, (e-mail: $\lbrace$kari.pajukoski; juho.pirskanen$\rbrace$@nokia-bell-labs.com)}

\thanks{Digital Object Identifier \href{http://dx.doi.org/10.1109/JSAC.2017.2687358}{10.1109/JSAC.2017.2687358}}
}  
\hypersetup{pdflinkmargin=0.5pt}

\IEEEpubid{%
  {\footnotesize
    \begin{minipage}{\textwidth}\ \\[12pt]
      \centering
      \copyright 2017 IEEE. Personal use of this material is permitted. Permission from IEEE must be obtained for all other users, including reprinting/republishing this material for advertising or promotional purposes, creating new collective works for resale or redistribution to servers or lists, or reuse of any copyrighted components of this work in other works.
    \end{minipage}
  }
}

\markboth{IEEE Journal on Selected Areas in Communications, Vol.~35, No.~6,  2017}%
{Yli-Kaakinen \MakeLowercase{\text\it{et al.}}: Efficient Fast-Convolution Based Waveform Processing for 5G Physical Layer}

\maketitle
\begin{abstract}
This paper investigates the application of \ac{fc} filtering schemes for flexible and effective waveform generation and processing in \ac{5g} systems. \ac{fc} based filtering is presented as a generic multimode waveform processing engine while, following the progress of \ac{5g} new radio (NR) standardization in \ac{3gpp}, the main focus is on efficient generation and processing of subband-filtered \ac{cp-ofdm} signals. First, a matrix model for analyzing \ac{fc} filter processing responses is presented and used for designing optimized multiplexing of filtered groups of \ac{cp-ofdm} \acp{prb} in a spectrally well-localized manner, i.e., with narrow guardbands. Subband filtering is able to suppress interference leakage between adjacent subbands, thus supporting independent waveform parametrization and different numerologies for different groups of \acp{prb}, as well as asynchronous multiuser operation in uplink. These are central ingredients in the \ac{5g} waveform developments, particularly at sub-\SI{6}{GHz} bands. The \ac{fc} filter optimization criterion is passband error vector magnitude minimization subject to a given subband band-limitation constraint. Optimized designs with different guardband widths, \ac{prb} group sizes, and essential design parameters are compared in terms of interference levels and implementation complexity. Finally, extensive coded \ac{5g} radio link simulation results are presented to compare the proposed approach with other subband-filtered \ac{cp-ofdm} schemes and time-domain windowing methods, considering cases with different numerologies or asynchronous transmissions in adjacent subbands. Also the feasibility of using independent transmitter and receiver processing for \ac{cp-ofdm} spectrum control is demonstrated. 
\end{abstract}  


% \fi
% \IEEEpeerreviewmaketitle
\section{Introduction} %----------------------------------------------------------------------
\label{sec:introduction} 
\IEEEPARstart{O}{rthogonal} frequency-division multiplexing (OFDM)\acused{ofdm} is extensively utilized in modern radio access systems. This is due to the high flexibility and efficiency in allocating spectral resources to different users, simple and robust way of channel equalization, as well as simplicity of combining multiantenna schemes with the core physical layer processing \cite{B:Toskala09}. However, due to limited spectrum localization, \ac{ofdm} has major limitations in challenging new spectrum use scenarios, like asynchronous multiple access, as well as mixed numerology cases aiming to use adjustable \ac{scs}, symbol length, and \acf{cp} length, depending on the service requirements \cite{J:20145GNOW,J:Banelli14:HeirOfOFDM}.

\IEEEpubidadjcol

The debate over different waveform candidates for the \ac{5g-nr} physical layer has been exhaustive during the last few years. Researchers have revised their knowledge over different waveforms thoroughly, evaluating, e.g., filter bank based orthogonal and non-orthogonal multicarrier waveforms \cite{J:20145GNOW,J:Banelli14:HeirOfOFDM}.  Regarding \ac{ofdm} based advanced waveform candidates, subband-filtered \ac{cp-ofdm} schemes are receiving great attention in the \ac{5g} waveform development, due to their ability to address the mentioned issues while maintaining high level of commonality with legacy \ac{ofdm} systems. Generally, these schemes apply filtering at subband level, over a physical \acl{rbg} including single or multiple \acp{prb}. In the existing studies, different window based time-domain filtering schemes have been considered in \cite{C:2016_Coexistence_UFOFDM_CPOFDM} (referred to as \ac{uf-ofdm}) and in \cite{C:2015_Zhang_f-OFDM_for_5G} (referred to as \ac{f-ofdm}). Effective uniform polyphase filter bank structures have been considered in \cite{J:Li2014:RB-F-OFDM} (referred to as RB-F-OFDM). In our recent study, flexible and effective frequency-domain filtering scheme, based on \acf{fc}, was proposed for filtered \ac{ofdm} in \cite{C:Renfors2015:fc-f-ofdm} whereas the possibility to design parametrizations supporting adjustable \ac{cp} lengths {with fixed overall \ac{cp}-\ac{ofdm} symbol duration} was demonstrated in \cite{C:Renfors16:adjustableCP}. \ac{fc} based realization of \ac{uf-ofdm} has also been proposed in \cite{2016RAN1UF-OFDM}. We see these as alternative implementations of the same idea. While noting that subband-filtered zero-prefix \ac{ofdm} variants have also been considered, \ac{5g-nr} development focuses on \ac{cp-ofdm} and we refer to these schemes jointly as \ac{F-ofdm} throughout this paper. As an alternative approach for enhancing the \ac{cp-ofdm} spectrum, \ac{wola} based \ac{cp-ofdm} \cite{2016RAN1WOLA} is regarded as another strong \ac{5g} waveform candidate.

Less studies have been devoted to situations where the transmitter and receiver utilize different waveform processing techniques. The recent development in the specification of \ac{5g-nr} in \ac{3gpp} \ac{tsg}-\ac{ran} WG1 stated that  the baseline assumption of the waveform for below \SI{40}{GHz} communications is \ac{cp-ofdm} and that the \ac{tx} processing has to be transparent to the \ac{rx} \cite{2016RAN1IndependentDesignofTxandRx, 2016RAN1WayForwardWaveforms}. This means that the possible spectrum enhancement function performed in  \ac{tx} is not signaled to  \ac{rx} and it is generally an unknown, implementation dependent function. This implies that  \ac{tx} and  \ac{rx} waveform processing are typically not matched, and that they need to be evaluated separately. It is expected that \ac{F-ofdm} or \ac{wola} is applied on the \ac{tx} side when improved spectrum localization is required, and the \ac{rx} processing is selected without knowledge of the detailed transmission scheme. 

In this paper, we develop and provide a generic and universal optimization-based framework for {\ac{fc-f-ofdm}} waveform processing for \ac{5g} physical layer and evaluate its performance in   scenarios following the test cases defined by \ac{3gpp} \cite[Annex A]{3GPPTR38802}. More specifically, the main contributions of the present paper can be listed as follows:
\begin{itemize}
\item \ac{fc} based frequency-domain windowing is shown to be an effective way to realize subband-filtered \ac{ofdm} schemes, with significantly lower computational complexity and highly increased flexibility compared to time-domain filtering approaches.   
\item Analytical models are developed for the essential responses of \ac{fc} based synthesis and analysis processing and for the resulting passband \ac{evm} and \ac{sblr} performance.
\item These models are used for effectively optimizing subband filtering with very narrow guardbands (1\,--\,7 subcarriers) between groups of \acp{prb}. The optimization minimizes directly the distortion introduced by partial suppression of the sidelobes of the subcarriers close to the subband edges. The number of non-trivial frequency-domain windowing weights is minimized, resulting in substantially reduced processing complexity and memory requirements.
\item Comprehensive coded \ac{5g} radio link simulations are carried out for evaluating the performance of \ac{fc-f-ofdm}, based on tentative \ac{3gpp} \ac{5g-nr} numerology and test cases. Also comparisons with \ac{wola} and other existing \ac{F-ofdm} proposals are included. 
\item In general, these results demonstrate the good performance, flexibility, and efficiency of the proposed scheme with different sizes of \ac{prb} groups  facilitating better radio link performance with lower complexity compared to other existing \ac{F-ofdm} schemes.
\item Fast-convolution filter bank (FC-FB)\acused{fc-fb} is presented (see Section II-C) as a generic waveform processing engine for evolving cellular mobile communications systems, supporting also, e.g., traditional single-carrier waveforms with frequency-domain equalization (considered in \ac{5g} for above-\SI{40}{GHz} frequency bands) in an effective and flexible manner.
\end{itemize}

In the proposed approach, frequency-domain windowing is defined by a very low number of weight coefficients, which facilitates efficient implementation and direct filter optimization for specific \ac{evm}, \ac{sblr} and \ac{oob} emission requirements. The widths and center frequencies can be adjusted individually for each subband, which is not possible in \ac{F-ofdm} schemes based on traditional uniform filter bank approaches, like \cite{J:Li2014:RB-F-OFDM}. On the other hand, individual time-domain filtering, as proposed in \cite{C:2016_Coexistence_UFOFDM_CPOFDM,C:2015_Zhang_f-OFDM_for_5G}, becomes overly complicated to implement. Even in case of a single subband (e.g., channelization filtering for the whole carrier) \ac{fc} filtering provides competitive performance vs.  complexity tradeoffs. Time-domain windowing methods, like \ac{wola}, exhibit excellent passband \ac{evm} performance but their spectrum localization capability, in terms of \ac{sblr} and \ac{oob} emissions, is rather limited compared to all \ac{F-ofdm} schemes, unless the symbol duration is greatly extended to accommodate long window transition intervals. 

The remainder of this paper is organized as follows. In Section II, the multirate \ac{fc} idea and \ac{fc-fb} concept are introduced, along with a basic matrix model for analysis and optimization purposes. This model was developed originally in \cite{J:Yli-Kaakinen17:TCASsubm} for \ac{fbmc} cases and is here specifically tailored for efficient \ac{F-ofdm} processing and optimization. Also the use of \ac{fc-fb} as a generic waveform processing engine is discussed. Section III develops analytical matrix models for evaluating the \ac{evm} and \ac{sblr} in the {\ac{fc-f-ofdm}} systems. Then the {\ac{fc-f-ofdm}} physical layer processing design is formulated as an optimization problem. In this problem the goal is to minimize the maximum of the subcarrier-level passband \ac{evm} subject to the given subband band-limitation constraint. Numerical results for the optimized scheme with tentative \ac{3gpp} \ac{5g-nr} numerology and with alternative design parameters are provided as well. In Section IV, extensive coded \ac{5g} radio link simulations are reported for evaluating the performance of \ac{fc} based \ac{F-ofdm} and DFT-spread-OFDM in various different test cases. Finally, the conclusions are drawn in Section V.

\begin{figure*}[t!]     
  \centering
  \includegraphics[trim=0 0 0 0,clip,width=\textwidth]%
  {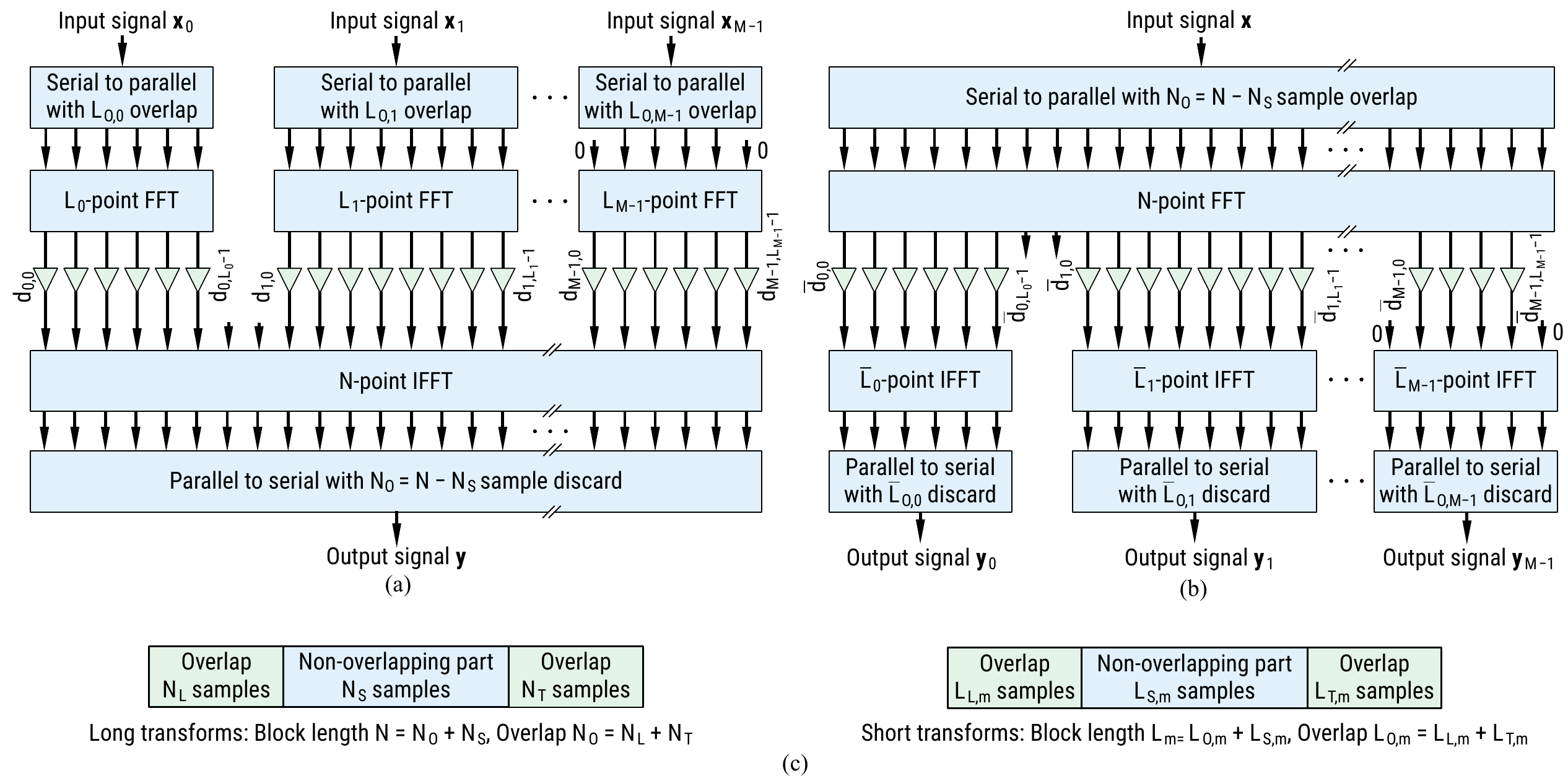}
  \caption{(a) \ac{fc} based flexible \acl{sfb} structure. (b) Corresponding \ac{fc} \acl{afb} structure. (c) Notations used for the number of samples in different parts of the overlap-save blocks.}
  \label{fig:Structure}   
\end{figure*}  

\section{Multirate Fast-Convolution and Filter Banks}
%%%%%%%%%%%%%%%%%%%%%%%%%%%%%%%%%%%%%%%%%%%%%%%%%%%%%
The main idea of \ac{fc} is that a high-order filter can be implemented effectively through multiplication in frequency domain, after taking \acp{dft} of the input sequence and the filter impulse response. Then the time-domain output signal is obtained by \ac{idft}. In practice, efficient implementation techniques, like \ac{fft} and \ac{ifft}, are used for the transforms, and overlap-save processing is applied for long sequences \cite{B:Rabiner75}.

The application of \ac{fc} to multirate filters has been presented in \cite{J:Borgerding06}, and \ac{fc} implementations of channelization filters have been considered in \cite{J:Boucheret99,C:Zhang00:Fast-FD-filter,C:Pucker03}. The authors have introduced the idea of \ac{fc}-implementation of nearly perfect-reconstruction filter bank systems and detailed analysis and \ac{fc-fb} optimization methods are developed in \cite{J:Renfors14:FC}. In \cite{C:Shao2015ISCAS} \ac{fc} approach has been applied for filter bank multicarrier waveforms and in \cite{C:Renfors2015:ICC} for flexible \ac{sc} waveforms.  These papers demonstrate the flexibility and efficiency of \ac{fc-fb} in communications signal processing, in general. In this article, the focus is fully on subband-filtered \ac{cp-ofdm}.
 
\subsection{Fast-Convolution Filter Bank Schemes} 
%%%%%%%%%%%%%%%%%%%%%%%%%%%%%%%%%%%%%%%%%%%%%%%%%
Fig.~\ref{fig:Structure}(a) shows the structure of \ac{fc}-based flexible \acf{sfb}, for a case where the $M$ incoming low-rate, narrowband signals $\mathbf{x}_m$ for $m=0,1,\dots,M-1$ with adjustable frequency responses and adjustable sampling rates are to be combined into single wideband signal $\mathbf{y}$. The dual structure show in Fig.~\ref{fig:Structure}(b) can be used on the receiver side as an \ac{afb} for splitting the incoming high-rate, wideband signal into several narrowband signals \cite{J:Yli-Kaakinen16:JSPS}. The cascade of \ac{sfb} and \ac{afb} is often called transmultiplexer. 

In the \ac{sfb} case, each of the $M$ incoming signals is first segmented into overlapping blocks of length $L_m$. Then, each input block is transformed to frequency-domain using \ac{dft} of length $L_m$.  The frequency-domain bin values of each  converted subband signal are multiplied by the weight values corresponding to the \ac{dft} of the finite-length linear filter impulse response, $d_{\ell,m}=\sum_{n=0}^{L_m-1}h_m[n]\eu^{-\iu 2\pi(\ell+L_m/2) n/L_m}$ for $\ell=0,1,\dots,L_m-1$ and for $m=0,1,\dots,M-1$.\footnote{For convenience of notation, we use the ``FFT-shifted'' indexing scheme in this context, i.e., index $0$ corresponds to the lower edge of the subband.}  Here, $\ell$ is the \ac{dft} bin index within the subband and $m$ is the subband index. Finally, the weighted signals are combined and converted back to time-domain using \ac{idft} of length $N$ and the resulting time-domain output blocks are concatenated using the overlap-save principle \cite{B:Rabiner75,J:Daher10:OLS}.

The multirate \ac{fc}-processing of Fig.~\ref{fig:Structure}(a) increases the sampling rates of the subband signals by the factors of
\begin{equation}
  \label{eqn:Rk}
  I_m=N/L_m=N_\text{S}/L_{\text{S},m},
\end{equation}  
where $L_{\text{S},m}$ and $N_{\text{S}}$ are the numbers of non-overlapping input and output samples, respectively.  The number of overlapping samples $L_{\text{O},m}=L_m-L_{\text{S},m}$ is divided into leading and tailing overlapping parts as follows:
\begin{align}
  L_{\text{L},m}=\lceil (L_m-L_{\text{S},m})/2 \rceil
  \quad\text{and}\quad 
  L_{\text{T},m}=\lfloor (L_m-L_{\text{S},m})/2 \rfloor.
\end{align}
Given the \ac{idft} length $N$, the sampling rate conversion factor is determined by the \ac{dft} length $L_m$, and it can be configured for each subband individually. Generally, there is no need to restrict the sampling rate conversion factor to take integer values. Naturally, $L_m$ determines the maximum number of non-zero frequency bins, i.e., the bandwidth of the subband. 

We assume an \ac{fc-fb} parametrization where the overlapping block structures at low-rate and high-rate sides match exactly, such that overlapping and non-overlapping parts can be expressed as an integer number of samples, corresponding to the same time duration on both sides.  This is reached if $N$, $L$, $N_\text{S}$, and $L_{\text{S},m}$ take integer values in (\ref{eqn:Rk}) for all subbands.  Generally, $N=p\Gamma$ and $N_\text{S}=q\Gamma$, where $p$ and $q$ are two relatively prime integers and $\Gamma=\gcd(N,N_\text{S})$, where $\gcd(\cdot)$ is the greatest common divisor. Then for the narrowest possible subband case satisfying the integer-length criterion, $L_m=p$ and $L_{\text{S},m}=q$. Generally, $L_m$ has to be a multiple of $N/\Gamma$, that is, the configurability of the subband sampling rates depends greatly on the choice of $N$ and $N_\text{S}$.

In the \ac{afb} case, it is assumed that the forward transform length is larger than the inverse transform lengths and, therefore, the above process reduces the sampling rate of the subband signal by factors of
\begin{equation}
  \label{eqn:Dk}
  D_m = N/\bar{L}_m = N_\text{S}/\bar{L}_{\text{S},m}.
\end{equation}
Here, the \ac{idft} lengths on the \ac{afb} side are denoted by $\bar{L}_m$'s. For simplicity, it is assumed that the long transform length $N$ for \ac{sfb} and \ac{afb} is the same.

In \cite{J:Renfors14:FC} and \cite{J:Yli-Kaakinen16:JSPS}, the performance of the \ac{fc-fb} was analyzed using a periodically time-variant model and tools for frequency response analysis and \ac{fc-fb} optimization were developed.  In the following sections, we first summarize the generic matrix model for \ac{fc-fb} analysis, and then develop \ac{fc-fb} analysis and optimization tools for the \ac{F-ofdm} based \ac{5g-nr} physical layer.

\subsection{Matrix Model for FC-FB Analysis and Optimization} 
%%%%%%%%%%%%%%%%%%%%%%%%%%%%%%%%%%%%%%%%%%%%%%%%%
In the \ac{fc} \ac{sfb} case, the block processing of $m$th subband signal $\mathbf{x}_m$ for the generation of high-rate subband waveform $\mathbf{w}_m$ can be represented as 
\begin{subequations}
  \label{eq:BDM}
  \begin{equation} 
    \mathbf{w}_m = \mathbf{F}_m\mathbf{x}_m,
  \end{equation}
  where $\mathbf{F}_m$ is the block diagonal transform matrix of the form
  \begin{align}
    \label{eq:tx_matrix1}
    \mathbf{F}_m &= \diag(\mathbf{F}_{m,0}, \mathbf{F}_{m,1}, \dots, \mathbf{F}_{m,R_m-1})
  \end{align}
\end{subequations}
with $R_m$ blocks. Here, the dimensions and locations of the $\mathbf{F}_{m,r}$'s are determined by the overlapping factor of the overlap-save processing, defined as 
\begin{equation}
  \lambda = 1-L_{\text{S},m}/L_m = 1-N_\text{S}/N.
\end{equation}

The multirate version of the \ac{fc} \ac{sfb} can be represented using block processing by decomposing the $\mathbf{F}_{m,r}$'s as the following $N_\text{S}\times L_m$ matrix
\begin{equation}
  \label{eq:synthesis}
  \mathbf{F}_{m,r} = 
  \mathbf{S}_{N} \mathbf{W}^{-1}_{N}
  \mathbf{M}_{m,r} \mathbf{D}_m 
  \mathbf{P}^{(L_m/2)}_{L_m} \mathbf{W}_{L_m}.
\end{equation}
Here, $\mathbf{W}_{L_m}$ and $\mathbf{W}_N^{-1}$ are the $L_m\times L_m$ \ac{dft} matrix (with $[\mathbf{W}_{L_m}]_{p,q}=\eu^{-\iu{2}\pi(p-1)(q-1)/L_m}$) and $N\times N$ inverse \ac{dft} matrix, respectively. The \ac{dft} shift matrix $\mathbf{P}^{(L_m/2)}_{L_m}$ is circulant permutation matrix obtained by cyclically left shifting the $L_m\times L_m$ identity matrix by $L_m/2$ positions. $\mathbf{D}_m$ is the $L_m\times L_m$ diagonal matrix with the frequency-domain window weights of the subband $m$ on its diagonal.  The $N\times L_m$ frequency-domain mapping matrix $\mathbf{M}_{m,r}$ maps $L_m$ frequency-domain bins of the input signal to frequency-domain bins $(c_m-\lceil L_m/2\rceil+\ell)_N$ for $\ell=0,1,\dots,L_m-1$ of output signal. Here $c_m$ is the center bin of the subband $m$ and $(\cdot)_N$ denotes the modulo-$N$ operation. In addition, this matrix rotates the phases of the block by
\begin{equation} 
  \label{eq:phase_rot}
  \theta_m(r) = \exp(\iu 2\pi{r}\theta_{m})\quad\text{with}\quad\theta_m = c_mL_{\text{S},m}/L_m
\end{equation} 
in order to maintain the phase continuity between the consecutive overlapping processing blocks \cite{J:Renfors14:FC}. The $N_\text{S}\times N$ selection matrix $\mathbf{S}_N$ selects the desired $N_\text{S}$ output samples from the inverse transformed signal corresponding to overlap-save processing. The \ac{fc} \ac{sfb} processing in the \ac{fc-f-ofdm} context is illustrated later in Fig.~\ref{fig:FC-F-OFDM_TXblock}.

\begin{figure}[t] 
  \centering
  \includegraphics[clip,trim=0 0 0 0,width=1.0\columnwidth]{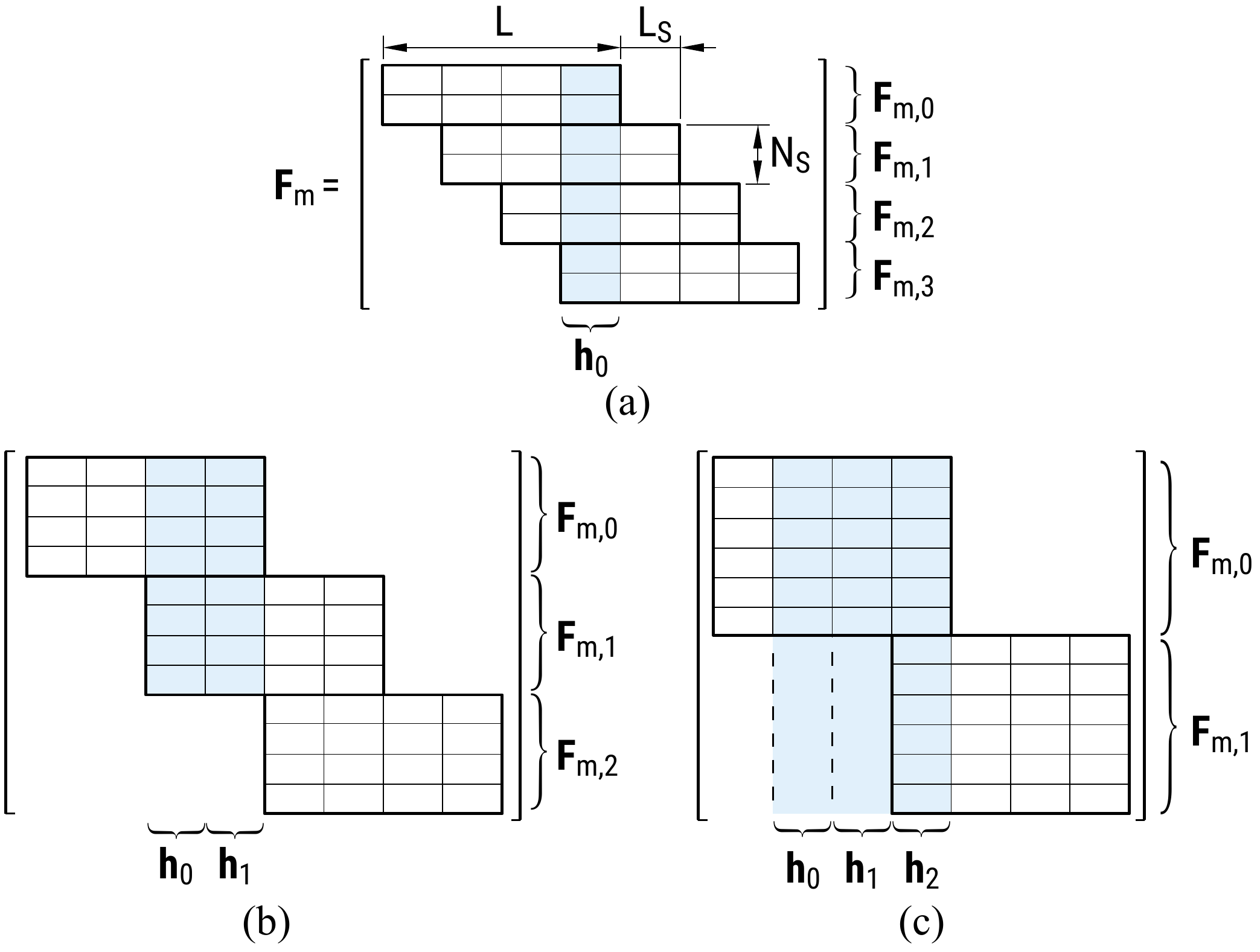} 
  \caption{Structure of block-diagonal synthesis matrix $\mathbf{F}_m$ for $L_m=4$, $N=8$, and $L_{\text{S},m}=1,2,3$ ($\lambda=3/4,2/4,1/4$). The colored columns illustrate the elements of the synthesis matrix that form the shift-variant impulse responses of the synthesis filter bank. (a) For $L_{\text{S},m}=1$, the system has only one impulse response of eight samples. (b) For $L_{\text{S},m}=2$, the system has two eight-samples long responses. (c) For $L_{\text{S},m}=3$, two among the three responses are six samples long and the remaining one is 12 samples long.}
  \label{fig:analysisBlock}      
\end{figure}  

In the \ac{afb} case, the corresponding analysis sub-block matrix of size $\bar{L}_{\text{S},m}\times N$ can be decomposed as 
\begin{equation}
  \label{eq:analysis}
  \mathbf{G}_{m,r} = 
  \mathbf{S}_{\bar{L}_m} \mathbf{W}^{-1}_{\bar{L}_m}
  \mathbf{P}^{(N/2)}_{N} \mathbf{D}_m
  \mathbf{M}^\transpose_{m,r} \mathbf{W}_N,
\end{equation}
where $\mathbf{P}^{(N/2)}_{N}$ is the $N\times{N}$ inverse Fourier-shift matrix and the $\bar{L}_{\text{S},m}\times \bar{L}_m$ selection matrix $\mathbf{S}_{\bar{L}_m}$ selects the desired $\bar{L}_{\text{S},m}$ output samples from the inverse transformed output signal. 

In general, the above \ac{fc} based synthesis and analysis filter banks are \ac{lpsv} systems with period of $L_{\text{S},m}$, that is, the systems have $L_{\text{S},m}$ different impulse responses. In the \ac{sfb} case, the impulse responses are given by the $L_{\text{S},m}$ shift-variant columns of the $\mathbf{F}_m$ as illustrated Fig.~\ref{fig:analysisBlock}.

\begin{figure}[t]
  \centering
  \makeatletter% 
  \if@twocolumn%
    \includegraphics[clip,trim=0 0 0 0,width=1.0\columnwidth]{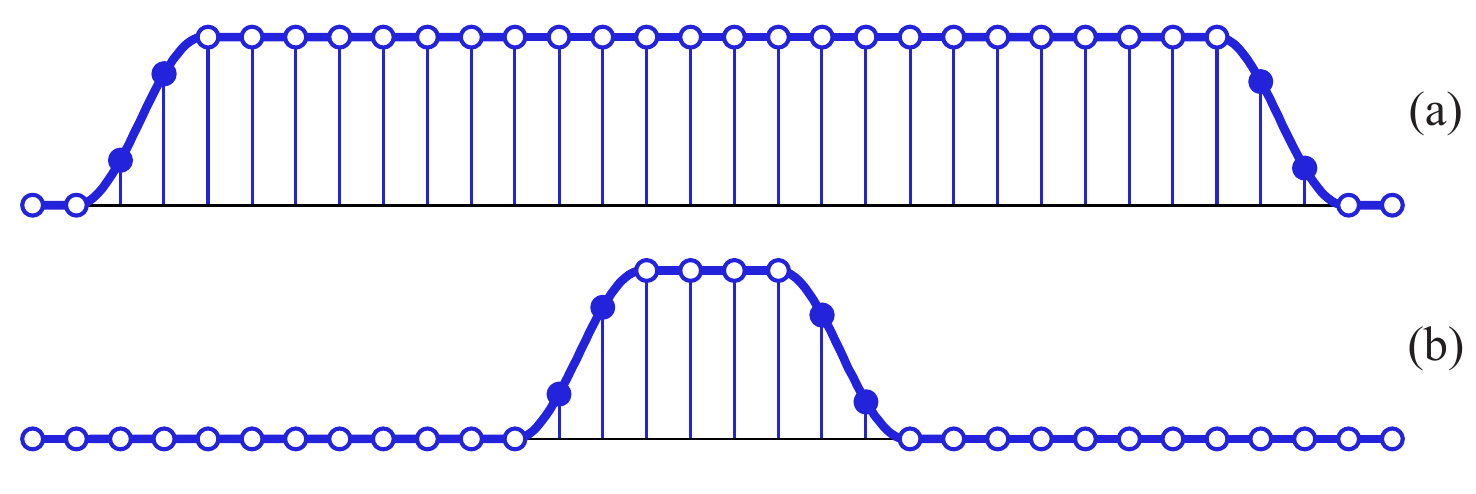} 
  \else% \@twocolumnfalse
     \includegraphics[clip,trim=0 0 0 0,width=1.0\textwidth]{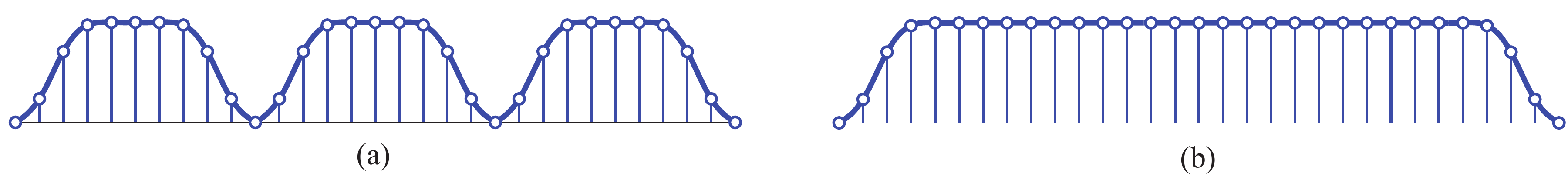} 
  \fi
  \makeatother
  \caption{Examples of FFT-domain weight masks.  (a) Single subband with wider bandwidth. (b) Single subband with narrower bandwidth. The ($L_\text{TBW}=2$) non-trivial transition-band weights are denoted by filled circles.}
  \label{fig:masks}     
\end{figure}

In our approach, \ac{fc} design is done in frequency-domain by defining/optimizing the weight coefficients. Generally, the frequency-domain weights consist of two symmetric transition bands with $L_{\text{TBW},m}$ non-trivial weights, where $L_{\text{TBW},m}$ also defines the \ac{tbw} for subband $m$. All passband weights are set to one, and all stopband weights are set to zero. The number of stopband weights (and the corresponding transform length) can be selected to reach a feasible subband oversampling factor. Now the diagonal weighting matrix in \eqref{eq:synthesis} and \eqref{eq:analysis} is expressible as 
\begin{equation}
	\label{eq:param_vect}
    \mathbf{D}_m = \diag
    \left(
    \begin{bmatrix}
     	\mathbf{0}_{(\lceil[L_m-L_{\text{ACT},m}]/2\rceil-L_{\text{TBW},m})\times1}\\ 
     	d_{0,m} \\
        \cdots & \\
        d_{L_\text{TBW}-1,m} \\
        \mathbf{1}_{L_{\text{ACT},m}\times1} \\
        d_{L_\text{TBW}-1,m} \\
        \cdots \\
    	d_{0,m} \\
        \mathbf{0}_{(\lfloor[L_m-L_{\text{ACT},m}]/2\rfloor-L_{\text{TBW},m})\times1}\\ 
   \end{bmatrix}\right),
\end{equation}
where $L_{\text{ACT},m}$ is the number of active subcarriers on subband $m$, and $\mathbf{0}_{q\times 1}$ is the column vector of $q$ zeros whereas  $\mathbf{1}_{p\times 1}$ is the vector of $p$ ones. Fig.~\ref{fig:masks} shows the weights for two example cases with different bandwidths. 

\subsection{FC-FB as a Generic Waveform Processing Engine}
%%%%%%%%%%%%%%%%%%%%%%%%%%%%%%%%%%%%%%%%%%%%%%%%%%%%%%%%%%
\ac{fc-fb} was originally applied for channelization filtering \cite{J:Boucheret99,C:Zhang00:Fast-FD-filter,C:Pucker03}, and it can be used for that purpose in a very flexible manner, allowing simultaneous transmission/reception of different waveforms with arbitrary bandwidths and channel rasters. After establishing the idea that \ac{fc-fb} can be used for realizing nearly perfect-reconstruction filter bank systems \cite{J:Renfors14:FC}, it was utilized for the generation and detection of \ac{fbmc/oqam} and \ac{fmt} type multicarrier waveforms \cite{C:Shao2015ISCAS}. It was applied also for traditional Nyquist pulse shaping based \ac{sc} waveforms, with adjustable and possibly very small roll-off \cite{C:Renfors2015:ICC}. Basically the same processing structure, but without overlap processing, can be used also for implementing circular multicarrier waveforms like \ac{gfdm}~\cite{J:20145GNOW}, and block-CP variants of \ac{fbmc/oqam} and \ac{fmt}~\cite{J:Lin2014}.  In the cellular mobile communication context the flexibility of \ac{fc-fb} can be exploited, for example, in multi-standard broadband base-station transmitters and receivers for simultaneous channelization of GSM, WCDMA, \ac{lte} carriers, as well as subbands of \ac{prb} groups in \ac{5g-nr}.

Fig.~\ref{fig:masks} illustrates how different \ac{fc} subband filters can be constructed using a fixed transition band weight mask. The subband frequency responses are constructed by adding one-valued weights between symmetric transition bands and zero-valued weights outside the transition bands. In this way the bandwidth, sampling rate conversion factor, and roll-off can adjusted in a flexible manner. While the weights can be optimized separately for each case, the performance reduction is usually rather small when using a well-optimized fixed weight mask. This implies that the whole \ac{fc-fb} can be stored to device memory only with $L_{\text{TBW}}$ non-trivial transition band weights which are used with all different subband widths. In general, this is a remarkable implementation benefit.

The parametrization for a flexible \ac{fc-fb} processing engine depends mainly on the needed spectral resolution: The narrowest transition bandwidth should be in the order of 2\,--\,7 \ac{fft} bin spacings, depending on the passband \ac{evm} and subband band limitation requirements. After fixing the \ac{fft} bin spacing, the long transform length follows directly from the targeted overall bandwidth. 

Also synchronization and channel equalization functions can be integrated with the \ac{fft}-domain processing of \ac{fc-fb}. Timing offsets in different subbands can be compensated by introducing a proper linearly frequency-dependent phase term in the weights \cite{C:Renfors2013}. Also a way to compensate fractional frequency offsets (with respect to the \ac{fft} bin spacing) has been presented in \cite{C:Yli-Kaakinen14:PMRwithTETRA}. Channel equalization can be realized for different waveforms in a unified manner by combining the channel equalization weights with the subband weight masks \cite{C:Renfors14:EmbeddedEq}. However, in case of \ac{fc-f-ofdm}, it is straightforward and computationally more effective to do the channel equalization in the traditional way in the \ac{rx} \ac{ofdm} processing module. 

\section{FC-based F-OFDM for 5G Physical Layer}
%%%%%%%%%%%%%%%%%%%%%%%%%%%%%%%%

\subsection{\ac{fc}-based \ac{F-ofdm}}
%%%%%%%%%%%%%%%%%%%%%%%%%%%%%%%%%%%%%%
In \ac{fc}-based \ac{F-ofdm} (\ac{fc-f-ofdm}), we apply \ac{fc-fb} based filtering at subband level, which means one or multiple contiguous \acp{prb}, while utilizing normal \ac{cp}-\ac{ofdm} waveform for the \acp{prb}~\cite{C:Renfors2015:fc-f-ofdm,C:Renfors16:adjustableCP}. One clear application is in cellular uplink scenarios, in which the different \acp{ue} utilize different sets of \acp{prb} for their simultaneous transmissions. In such cases, an individual \ac{ue} can adopt the subband filtering separately for each contiguous set of \acp{prb} allocated to it. This, together with similar \ac{fc}-based subband filtering at the base-station receiver, allow for reduced uplink timing synchronization requirements, or even completely asynchronous uplink, as well as enable adopting different numerologies (e.g., different subcarrier spacings, \ac{cp}-lengths, and/or frame structures) \cite{C:Renfors16:adjustableCP}, for different \acp{ue} and services simultaneously inside the carrier. Furthermore, if the allocated \ac{ul} \acp{prb} are at the channel edge, subband filtering at the \ac{ue} transmitter contributes to reducing the \ac{oob} emissions. Additionally, in dynamic time-division duplexing (TDD) networks with frequency reuse~1, better band-limitation of the uplink subband signals also helps in reducing the intercell interference between neighboring subbands.

Concerning the cellular downlink scenarios, synchronization is basically not an issue, but the filtered \ac{ofdm} idea would still make it possible to parametrize individually the subsignals of different groups of \acp{prb}, inside the carrier, and thus facilitate flexible multiplexing of \acp{ue} and services also in the downlink. Interestingly, in downlink CoMP type scenarios, this would also allow for tuning the downlink transmit signal timing in an individual transmission point, separately for different \acp{ue} at different subbands.

\begin{figure*}[t]  
\centering
  \includegraphics[clip,trim=0 0 0 0,width=0.95\textwidth]{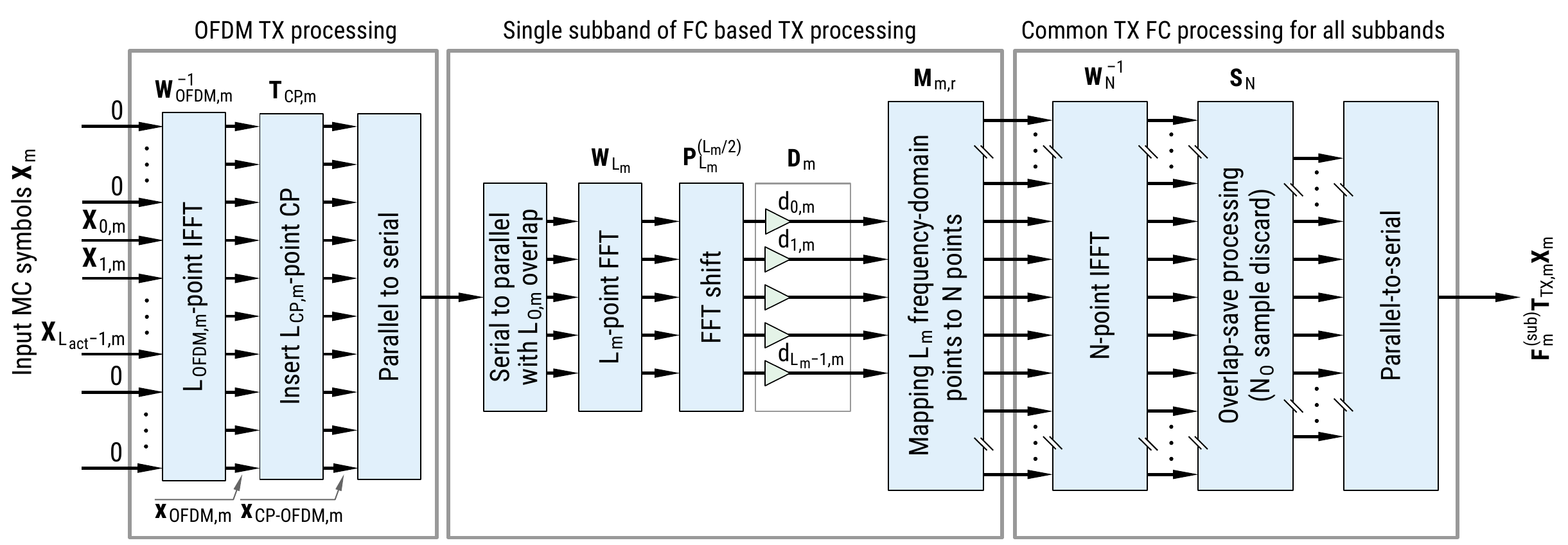}
  \caption{{\ac{fc-f-ofdm}} transmitter. The processing structure for one filtered group of \acp{prb} is shown (cf. Fig.~\ref{fig:Structure}(a) for the overall synthesis processing). The long $N$-point \ac{ifft} is common for all the subbands.} 
  \label{fig:FC-F-OFDM_TXblock}    
\end{figure*}

With {\ac{fc-f-ofdm}} processing, it is easy to adjust the filtering bandwidth for the subbands individually. This is very useful in \ac{prb}-filtered \ac{ofdm} because there is no need to realize filter transition bands and guardbands between equally parametrized, synchronous \acp{prb}. In the extreme case, the group of filtered \acp{prb} could cover the full carrier bandwidth, and \ac{fc} processing would implement tight channelization filtering for the whole carrier. Fig.~\ref{fig:FC-F-OFDM_TXblock} shows a generic block diagram of a \ac{fc-f-ofdm} transmitter. The \ac{fc}'s long transform length $N$ is assumed to be fixed in the following discussions.
  
\begin{figure}[t]  
\centering
  \includegraphics[clip,trim=0 0 0 0,width=0.95\columnwidth]{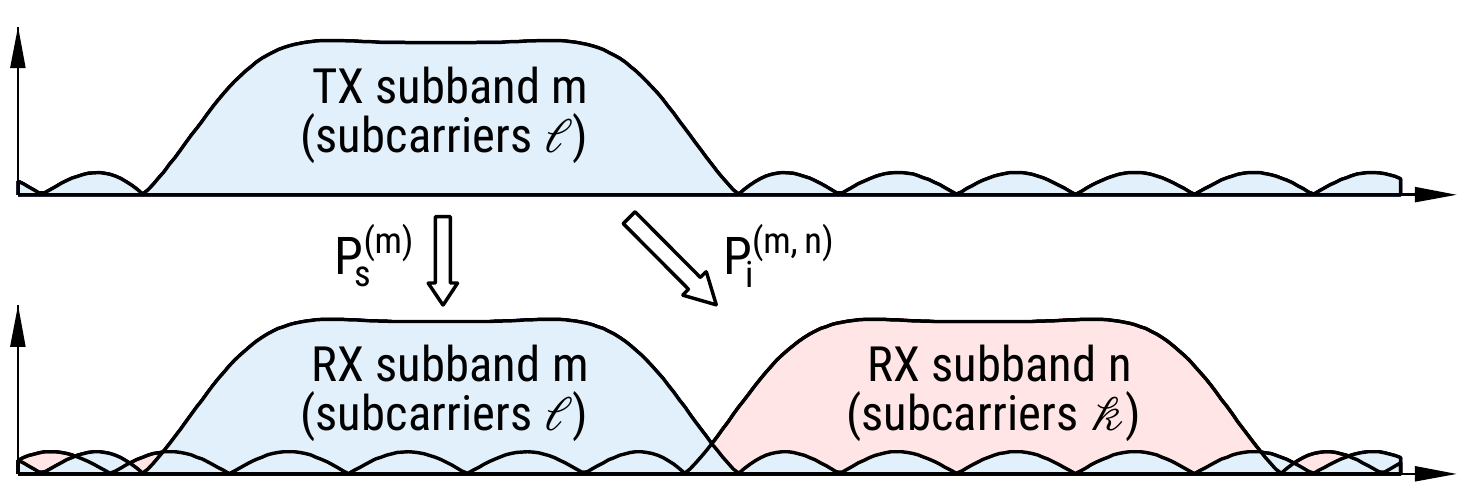}
  \caption{Illustration of the subband leakage ratio evaluation.}
  \label{fig:SBLRdiagram}    
\end{figure} 

\subsection{Transmultiplexer Optimization for {\ac{fc-f-ofdm}}}
%%%%%%%%%%%%%%%%%%%%%%%%%%%%%%%%%%%%%%%%%%%%%%%%%%%%%%%%%%%%%%%
The \ac{cp}-\ac{ofdm} processing of the $m$th \ac{ofdm} subband on the transmitter side can be expressed as 
\begin{subequations}
  \begin{equation}
    \mathbf{T}_{\text{TX},m} = \mathbf{T}_{\text{CP},m}\mathbf{W}^{-1}_{\text{OFDM},m},
  \end{equation}
  where $\mathbf{W}^{-1}_{\text{OFDM},m}$ is the $L_{\text{OFDM},m}\times L_{\text{OFDM},m}$ \ac{idft} matrix and the $(L_{\text{OFDM},m}+L_{\text{CP},m})\times L_{\text{OFDM},m}$ \ac{cp} insertion matrix is given by
  \begin{equation} 
    \mathbf{T}_{\text{CP},m} = 
    \Big[ 
    \begin{matrix}
      \begin{bmatrix}
        \mathbf{0}_{L_{\text{CP},m}\times(L_{\text{OFDM},m}-L_{\text{CP},m})} & \mathbf{I}_{L_{\text{CP},m}}
      \end{bmatrix}^\transpose
      & \mathbf{I}_{L_{\text{OFDM},m}}
    \end{matrix}
    % }
    \Big]^\transpose.
  \end{equation}
\end{subequations}
Here, $\mathbf{0}_{q\times p}$ and $\mathbf{I}_{r}$ are $q\times p$ zero matrix and $r\times r$ identity matrix, respectively.  The effective response from the $\ell$th \ac{ofdm} subcarrier on subband $m$ to the \ac{fc-fb} output can be expressed as
\begin{subequations}
  \begin{equation}
    \label{eq:TX}
    \mathbf{f}^\text{(OFDM)}_{\ell,m}=\mathbf{F}^\text{(sub)}_m \mathbf{T}_{\text{TX},m}
    \begin{bmatrix}
      \mathbf{0}_{1\times{\ell-1}} & 1 & \mathbf{0}_{1\times{L_{\text{OFDM},m}-\ell}}
    \end{bmatrix}^\transpose,
  \end{equation}
  where the $P_m\times (L_{\text{OFDM},m}+L_{\text{CP},m})$ sub-matrix $\mathbf{F}_m^\text{(sub)}$ with $P_m=N_{\text{S}}B_{\text{F},m}$ defining the number of useful samples in the output and $B_{\text{F},m}=\lceil (L_{\text{OFDM},m}+L_{\text{CP},m})/L_{\text{S},m}\rceil$ defining the number of \ac{fc} processing blocks, is obtained  by selecting the desired rows and columns from  $\mathbf{F}_m$ as follows: 
  \begin{equation} 
    [\mathbf{F}_m^\text{(sub)}]_{p,q} = [\mathbf{F}_m]_{p,q+S_{\text{F},m}} 
  \end{equation}
\end{subequations}
with $S_{\text{F},m} = L_m-L_{\text{S},m}$ for $p=1,2,\dots,P_m$ and for $q=1,2,\dots,L_{\text{OFDM},m}+L_{\text{CP},m}$. 

On the receiver side the \ac{ofdm} processing can be expressed as 
\begin{subequations}
  \begin{equation}
    \mathbf{T}_{\text{RX},m} = \mathbf{W}_{\text{OFDM},m} \mathbf{R}_{\text{CP},m},
  \end{equation}
  where $\mathbf{W}_{\text{OFDM},m}$ is the $\bar{L}_{\text{OFDM},m}\times \bar{L}_{\text{OFDM},m}$ \ac{dft} matrix and the $\bar{L}_{\text{OFDM},m}\times(\bar{L}_{\text{OFDM},m}+\bar{L}_{\text{CP},m})$ \ac{cp} removal matrix is given by
  \begin{equation}
    \mathbf{R}_{\text{CP},m} = 
    \begin{bmatrix}
      \mathbf{0}_{\bar{L}_{\text{OFDM},m}\times \bar{L}_{\text{CP},m}} & \mathbf{I}_{\bar{L}_{\text{OFDM},m}}
    \end{bmatrix}.  
  \end{equation}
\end{subequations}
Now the processing response from the \ac{fc-fb} input to the $\ell$th \ac{ofdm} subcarrier can be expressed as
\begin{subequations}
  \begin{equation}
    \mathbf{G}^\text{(OFDM)}_{\ell,m}=
    \mathbf{C}_{m}^{(\ell)}
    \mathbf{T}^{\text{(diag)}}_{\text{RX},m}
    \mathbf{G}^\text{(sub)}_m,
  \end{equation}
  where the $Q_m\times P_m$ sub-matrix $\mathbf{G}_m^\text{(sub)}$ with $Q_m=\bar{L}_{\text{S},m}B_{\text{G},m}$ and  $B_{\text{G},m}=\lfloor(P_m+N)/N_{\text{S}}-(B_{\text{F},m})_2\rfloor$ is given by
  \begin{equation}
    [\mathbf{G}_m^\text{(sub)}]_{q,p} = [\mathbf{G}_m]_{q,p+S_{\text{G},m}} 
  \end{equation}
\end{subequations}
with $S_{\text{G},m} = \lfloor(N+[B_{\text{G},m}-1]N_\text{S}-P_m)/2\rfloor$ for $q=1,2,\dots,Q_m$ and for $p=1,2,\dots,P_m$. The block diagonal $Q_m\times Q_m$ matrix $\mathbf{T}^{\text{(diag)}}_{\text{RX},m}$ is constructed from block diagonal $B_{\text{OFDM},m}\bar{L}_{\text{OFDM},m}\times B_{\text{OFDM},m}(\bar{L}_{\text{OFDM},m}+\bar{L}_{\text{CP},m})$ matrix with $B_{\text{OFDM},m}=\lceil Q_m/(\bar{L}_{\text{OFDM},m}+\bar{L}_{\text{CP},m}) \rceil$ blocks
\begin{equation}
  \label{eq:tx_matrix2}
  \mathbf{T}^{\text{(diag)}}_{\text{RX},m} = \diag(\underbrace{\mathbf{T}_{\text{RX},m}, \mathbf{T}_{\text{RX},m}, \dots, \mathbf{T}_{\text{RX},m}}_{\text{$B_{\text{OFDM},m}$ blocks}}) 
\end{equation}
by selecting the first $Q_m$ rows and columns whereas $\mathbf{C}_{m}^{(\ell)}$ is the $Q_m\times Q_m$ down-sampling by $\bar{L}_{\text{OFDM},m}$ with $\ell$-sample offset matrix.

Stemming from the above fundamental modeling, the overall processing response from the \ac{ofdm} subcarrier $\ell$ of subband $m$ to subcarrier $\kay$ of subband $n$ can be expressed as
\begin{equation} 
  \mathbf{t}^{(m,n)}_{\ell,\kay} = \mathbf{G}_{\kay,n}^\text{(OFDM)}\mathbf{f}_{\ell,m}^\text{(OFDM)}.
\end{equation}
Then, the passband quality on an active subcarrier $\ell$ and on subband $m$ can be measured using the following normalized \ac{mse} measure:
\begin{equation}
  \label{eq:MSE}
  \text{MSE}^{(m)}_{\ell}=
  \norm{\mathbf{e}-\mathbf{t}^{(m,m)}_{\ell,\ell}}^2 + \sum\limits_{\kay=0,\kay\neq \ell} ^{\bar{L}_{\text{ACT},m}-1}\norm{\mathbf{t}^{(m,m)}_{\ell,\kay}}^2
\end{equation} 
where 
$
\mathbf{e}=
\begin{bmatrix}
  \mathbf{0}_{1\times U_m} &  
  1 &  
  \mathbf{0}_{1\times V_m} 
\end{bmatrix}^\transpose 
$ %  
with $U_m=\lceil S_{\text{G},m}\bar{L}_m/(N\bar{L}_{\text{OFDM},m})\rceil$ and $V_m=\lceil P_m/\bar{L}_{\text{OFDM},m}\rceil-U_m$ and $\bar{L}_{\text{ACT},m}$ is number of active subcarriers on the receiver side on subband $m$. Here, the first term measures the effect of time-domain dispersion at subcarrier index $\ell$, resulting in general into \ac{isi} between $U_m$ preceding and $V_m$ following OFDM symbols. The second term, in turn, contains the \ac{ici} induced by the desired symbol, $U_m$ preceding  symbols, and $V_m$ following \ac{ofdm} symbols.

The corresponding \acf{evm} in decibels is expressed using \eqref{eq:MSE} as 
\begin{equation}
\text{EVM}^{(m)}_\ell = 10\log_{10}\left(\text{MSE}^{(m)}_\ell\right).
\end{equation}
The worst-case \ac{evm} is defined as a maximum of the \ac{evm} values over the active subcarriers as given by
\begin{subequations}
  \begin{equation}
    \label{eq:EVMmax}
    \text{EVM}^{(m)}_\text{MAX}=\max_{\ell=0,1,\dots,\bar{L}_{\text{ACT},m}-1}\text{EVM}^{(m)}_\ell 
  \end{equation}
  whereas the average \ac{evm} is defined as the mean value of the normalized \ac{mse} values on active subcarriers, expressed in decibels as
  \begin{equation}
    \label{eq:EVMavg}
    \text{EVM}^{(m)}_\text{AVG}=
    10\log_{10}\left(\frac{1}{\bar{L}_{\text{ACT},m}}\sum_{\ell=0}^{\bar{L}_{\text{ACT},m}-1}\text{MSE}^{(m)}_\ell\right).
  \end{equation}
\end{subequations}

Next, to quantify the in-band unwanted emissions (inside one carrier), and in particular the interference leakage between different subbands as illustrated in Fig.~\ref{fig:SBLRdiagram}, we define the \acf{sblr} as the ratio of the power leaking from \ac{tx} subband $m$ to another \ac{rx} subband $n$, within the carrier, relative to the observable \ac{rx} power at subband $m$ as
\begin{subequations}
  \label{eq:sblr} 
  \begin{equation}
    \text{SBLR}^{(m,n)}=10\log_{10}\left(\frac{P^{(m,n)}_\text{i}}{P^{(m)}_\text{s}}\right),
  \end{equation}
  where the power leaking from active subcarriers on \ac{tx} subband $m$ to the active subcarriers of the unintended \ac{rx} on subband $n$ is given as
  \begin{equation}
    P_\text{i}^{(m,n)} = 
    \sum\limits_{\ell=0} ^{L_{\text{ACT},n}-1}
    \sum\limits_{\kay=0} ^{\bar{L}_{\text{ACT},m}-1}
    \norm{\mathbf{t}^{(m,n)}_{\ell,\kay}}^2
  \end{equation}
  and the observable reference power on the active subcarriers on target \ac{rx} subband $m$ is defined as
  \begin{equation}	
    P_\text{s}^{(m)}= 
    \sum\limits_{\ell=0} ^{L_{\text{ACT},m}-1}
    \sum\limits_{\kay=0} ^{\bar{L}_{\text{ACT},m}-1}
    \norm{\mathbf{t}^{(m,m)}_{\ell,\kay}}^2.
  \end{equation}
\end{subequations}

In the actual \ac{fc}-based subband filter optimization, we use the minimum stopband attenuation of the synthesis processing as a figure of merit for the subband band-limitation characteristics, since this measure gives an straightforward way to control the power leaking to adjacent subband independent of \ac{rx} processing and the number of used guard subcarriers. The magnitude squared response of the synthesis processing is given by
\begin{equation}
  \label{eq:MR}
  M(\omega)=
  \frac{1}{N\lambda}
  \sum_{n=0}^{N\lambda-1}
  \lvert F^{(n)}_m(\eu^{\iu\omega})\rvert^2,
\end{equation}
where the frequency responses $F^{(n)}_m(\eu^{\iu\omega})$ are evaluated using the impulse responses given by the $N\lambda$ time-variant columns of the matrix $\mathbf{F}_m$ as given by \eqref{eq:BDM}. 

The \ac{fc}-filtered \ac{F-ofdm} system design can now be stated as an optimization problem for finding the optimal values of the frequency-domain window ($L_{\text{TBW},m}$ non-trivial values of $\mathbf{D}_m$ in \eqref{eq:param_vect}) to
\begin{equation*}
  \begin{aligned}
    & \mspace{-35mu}\underset{d_{0,m},d_{1,m},\dots,d_{L_\text{TBW}-1,m}}{\text{minimize}}
    & & \text{EVM}_\text{MAX}^{(m)} \\
    & \text{subject to}
    & & M(\omega) \leq A_\text{s},\quad \text{for $\omega\in \Omega_\text{s}$},
  \end{aligned}
\end{equation*}
where $A_\text{s}$ is the desired minimum stopband attenuation and $\Omega_\text{s}$ is the stopband region.  This problem can be straightforwardly solved using non-linear optimization algorithms since the number of optimized parameters is typically low (no more than seven in the examples considered in this paper) when taking into account that only the transition band weights are used in the optimization. Notice that by varying the value of $A_\text{s}$ in the optimization, the \ac{sblr} and \ac{oob} emissions can be directly controlled. 

Due to the complexity of the system model, it is practically impossible to prove the convexity of the optimization problem through analytical means, and therefore, strictly speaking, the global optimality of the solution cannot be guaranteed. However, our studies using several different starting points for the optimization and using different numerical algorithms have shown that the optimization converges reliably to same solution and, therefore, we can assume optimum is also the global one.

It should be noted that we are not targeting to reach perfectly linear convolution through \ac{fc} processing, but our goal in the design is to keep the cyclic distortion effects at a level that does not significantly impact the link error rate performance, such that the non-implementation-related effects are dominating. The main cause for \ac{evm} degradation is the partial suppression of sidelobes of subcarriers close to the subband edges, which is unavoidable in \ac{F-ofdm}. In contrast to earlier schemes, this effect is directly minimized by our design approach.  

\begin{figure*}
  \centering
  \includegraphics[angle=0,trim=0 0 0 0,clip,width=1.0\columnwidth]{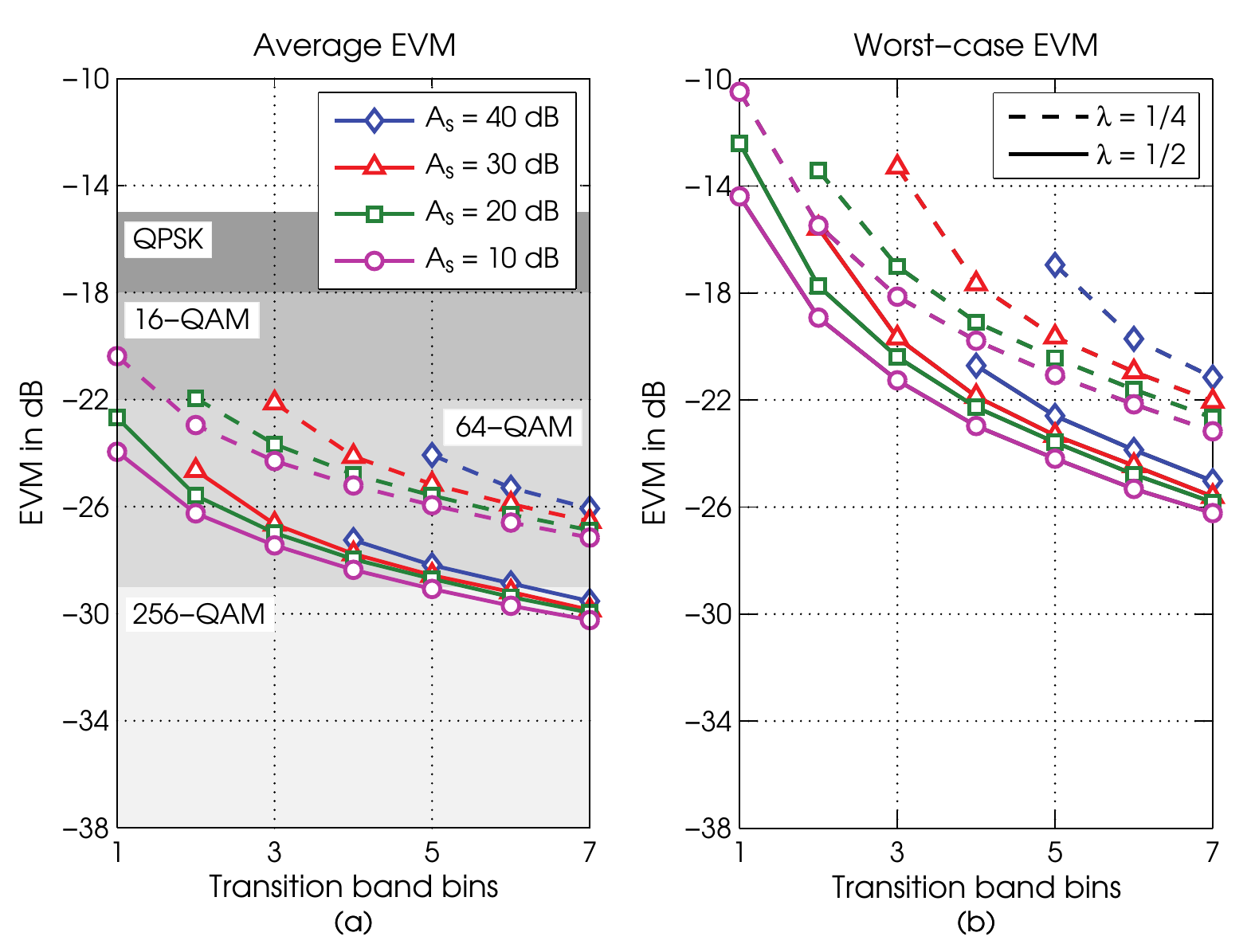}      
  \includegraphics[angle=0,trim=0 0 0 0,clip,width=1.0\columnwidth]{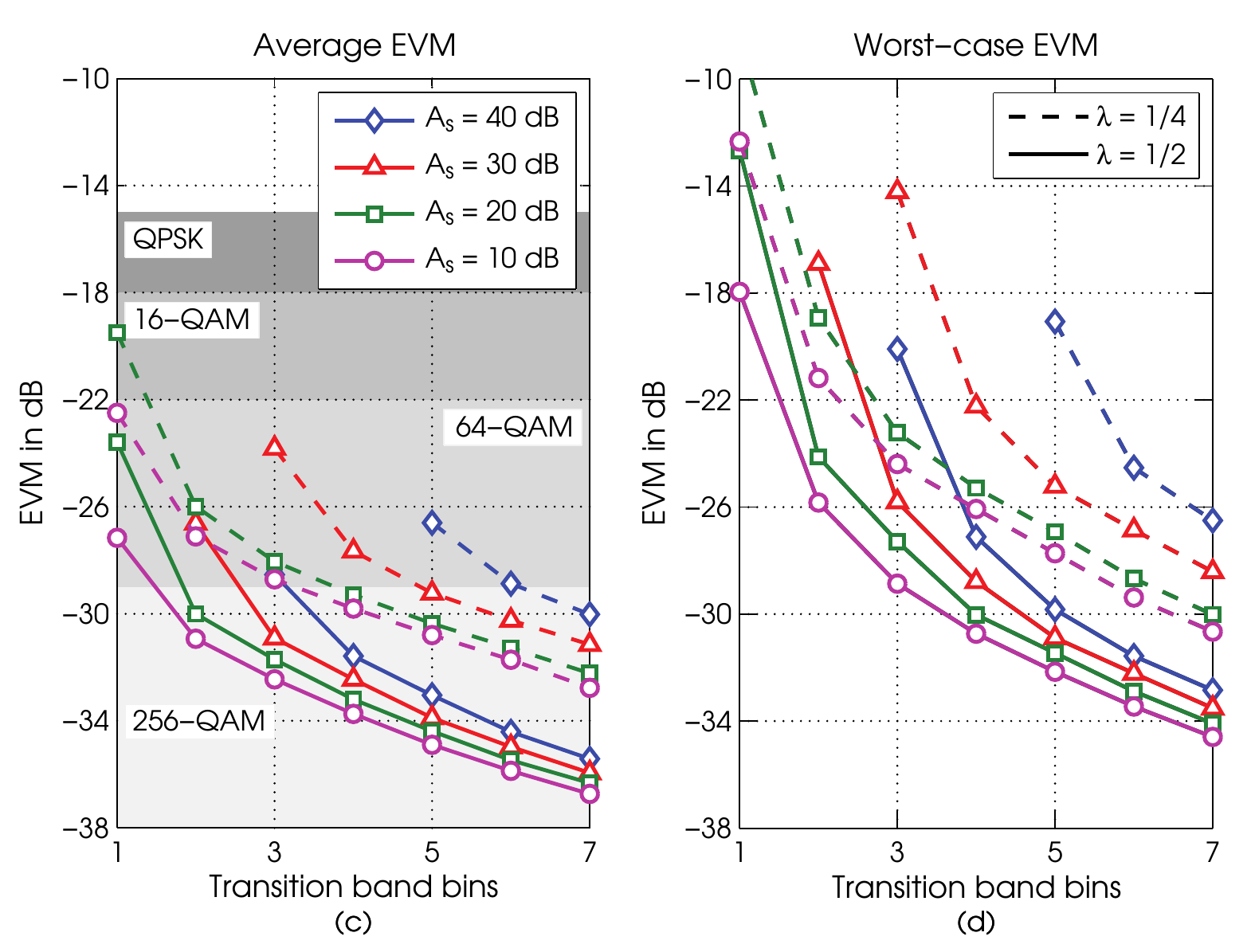}      
  \caption{Average and worst-case \acp{evm} as a function of transition bandwidth for a filtered group of 4\,\acp{prb} with different minimum stopband attenuation levels and \SI{25}{\%} and \SI{50}{\%} \ac{fc} overlaps. (a)-(b): Wideband \ac{tx}, \ac{rx} filtering only. (c)-(d): Isolated group of 4 \acp{prb} with both \ac{tx} and \ac{rx} filtering.}
  \label{fig:IBI_perf}
\end{figure*}

\subsection{\ac{5g-nr} Numerology and \ac{fc-fb} Parametrization}
%%%%%%%%%%%%%%%%%%%%%%%%%%%%%%%%%%%%%%%%%%%%%%%%%%%%%%%
Considering the \ac{fc-f-ofdm} transmitter in {Fig.~\ref{fig:FC-F-OFDM_TXblock}}, it was assumed in~\cite{C:Renfors2015:fc-f-ofdm} that the short \ac{fft}-length in \ac{fc} processing is the same as the \ac{ifft} length in \ac{ofdm} processing. In~\cite{C:Renfors16:adjustableCP}, this constraint is relaxed and it is assumed that the two transform lengths can be chosen independently. Considering the mixed numerology cases, we choose $N=\max\{N_{\text{OFDM},m}\}$, where $N_{\text{OFDM},m}$ is the \ac{ofdm} symbol duration in subband $m$ in high-rate samples, $f_\text{S}/N_{\text{OFDM},m}$ is the corresponding \ac{scs}, and $f_\text{S}$ is the high sampling rate. In the structure of {Fig.~\ref{fig:FC-F-OFDM_TXblock}}, the \ac{ofdm} processing module generates useful symbol duration of $L_{\text{OFDM},m}$ and the inserted \ac{cp}-length is denoted as $L_{\text{CP},m}$. The \ac{fc}-filtering process increases the sampling rate by the factor of $N/L_{m}$, resulting in overall symbol duration of $N_{\text{OVR}}=N(L_{\text{OFDM},m}+L_{\text{CP},m})/L_m= N_{\text{OFDM},m}+N_{\text{CP},m}$.  Here $L_{\text{OFDM},m}$ and $L_{\text{CP},m}$ need to take integer values. It is convenient, but not necessary, that $N_{\text{OFDM}}$ and $N_{\text{CP}}$ take integer values as well.  In the flexible numerology considered for \ac{3gpp} \ac{5g-nr}, the \ac{scs} is an integer power of two times \SI{15}{kHz}, say $2^\eta\times\SI{15}{kHz}$. Then one natural choice is $N=N_{\text{OFDM}}$ in the basic case with \SI{15}{kHz} \ac{scs} and $N_{\text{OFDM}}=N/2^\eta$ generally.

As a concrete example, we focus here on a \SI{10}{MHz} \ac{5g-nr} like multicarrier system utilizing \ac{cp-ofdm} baseline waveform.  The long transform length on the transmitter and receiver sides is fixed to $N=1024$. The sampling rate is $f_\text{S}=\SI{15.36}{MHz}$ with $N_{\text{OFDM}}=N/2^\eta=1024/2^\eta$ and $N_{\text{CP}}=72/2^\eta$.  Table~\ref{tab:params} shows example numerologies for \acp{scs} of \SI{15}{kHz} and \SI{30}{kHz}. The active subcarriers are always scheduled in \acp{prb} of 12 subcarriers. We notice that with narrow allocations, the short transform lengths are limited by the integer \ac{cp} length constraint.  While mixed numerology cases with wider \ac{scs} will be considered in Section V, in the numerical examples of this section we focus on the basic case with $\eta=0$.

It should be noted that in \ac{5g} numerology proposals, the length of the first \ac{cp} is different from the other ones. The symbol boundary alignment is done within \SI{0.5}{ms} intervals. With \SI{15}{kHz} \ac{scs}, the first \ac{cp} is extended in the 50 \ac{prb} case by 8 samples to 80 and in the 1 \ac{prb} and 4 \ac{prb} cases to 10 samples. Then a \SI{0.5}{ms} subframe corresponds exactly to 15 \ac{fc} processing blocks with \SI{50}{\%} overlap and to 10 \ac{fc} processing blocks with \SI{25}{\%} overlap. The same applies for higher \acp{scs} while the overall number of \ac{ofdm} symbols in a subframe is proportional to $2^\eta$.

\begin{table}[t]
  \caption{Example parametrizations for FC-F-OFDM based \ac{5g} physical layer with \SI{10}{MHz} carrier bandwidth}
  \label{tab:params}
  \centering
  \footnotesize{
   \begin{tabular}{cccccc}
     \toprule
     \multicolumn{1}{c}{SCS}  & 
     \multicolumn{1}{c}{No. act. subcarr.}  & 
     \multicolumn{1}{c}{$N$} &
     \multicolumn{1}{c}{${L}_{\text{OFDM},m}$} & 
     \multicolumn{1}{c}{${L}_{\text{CP},m}$} & 
     \multicolumn{1}{c}{${L}_m$} 
     \\
     \midrule
      \SI{15}{kHz} & 12 (1\,\ac{prb}) & 1024 &  128 &  9 &  128
     \\
      \SI{15}{kHz} & 48 (4\,\acp{prb}) & 1024 &  128 &  9 &  128
     \\
     \SI{15}{kHz} & 600 (50\,\acp{prb})  & 1024 & 1024 & 72 & 1024
     \\
      \SI{30}{kHz} & 12 (1\,\ac{prb}) & 1024 &  128 &  9 &  256
     \\
      \SI{30}{kHz} & 24 (2\,\acp{prb}) & 1024 &  128 &  9 &  256
     \\
     \SI{30}{kHz} & 300 (25\,\acp{prb})  & 1024 & 512 & 36 & 1024
     \\
	\bottomrule
    \end{tabular}}  
\end{table}

\subsection{Numerical Results for Passband \ac{evm} and Adjacent Subband \ac{sblr}}
%%%%%%%%%%%%%%%%%%%%%%%%%%%%%%
Here we evaluate first the passband \ac{evm} and adjacent subband \ac{sblr} characteristics of \ac{fc-f-ofdm} subband filtering for the 4-\ac{prb} configuration of ~Table~\ref{tab:params} with different filter transition bandwidths and two \ac{fc} overlap factors, $\lambda=1/2$ or $\lambda=1/4$. In these evaluations, the worst-case and average \acp{evm}, as given by \eqref{eq:EVMmax} and \eqref{eq:EVMavg}, are used as measures of the passband quality while the \ac{sblr} directly quantifies how much interference leakage there is between the neighboring subbands. The results are shown in Figs.~\ref{fig:IBI_perf} (a)--(b) for the case where a full-band \ac{cp-ofdm} signal is transmitted and optimized subband filtering is done only on the \ac{rx} side. The targeted group of 4\,\acp{prb} is in the central part of the carrier, such that adjacent \ac{prb} groups are present in the received signal. This can be seen as a basic unmatched filtering case, which is relevant for narrowband low-power \ac{mtc} receivers. The corresponding results can be seen in Figs.~\ref{fig:IBI_perf}(c)--(d) for the case where matched subband filtering is done on both \ac{tx} and \ac{rx} sides for an isolated group of 4\,\acp{prb}. {Figs.~\ref{fig:IBI_perf}(a),(c) show the average passband \acp{evm} and Figs.~\ref{fig:IBI_perf}(b),(d) show the worst-case \acp{evm}} as a function of the filter transition bandwidth expressed in \ac{fft} bin spacings and for cases where the lowest stopband attenuation levels (typically at the stopband edges) are at $A_\text{s}=\text{\{10, 20, 30, 40\}\,dB}$. 

We can see that the \ac{fc} overlap factor has quite significant effect on the performance. For narrow transition bandwidths, the tradeoff between \ac{evm} and minimum stopband attenuation is clear, whereas for wider transition bands, reduced stopband attenuation does not help to improve the passband performance significantly.  The worst-case \ac{evm} is considerably higher than the average. This is obviously due to the fact that on the edge subcarriers, the strict orthogonality is impaired.  This contribution to average \ac{evm} is more severe with narrowband allocations. 

These results can be evaluated in the context of the \ac{evm} requirements of \ac{lte}, stated as \{\SI{17.5}{\%}, \SI{12.5}{\%}, \SI{8}{\%}, \SI{3.5}{\%}\} or \{\SI{-15}{dB}, \SI{-18}{dB}, \SI{-22}{dB}, \SI{-29}{dB}\} for \{QPSK, 16-QAM, 64-QAM, 256-QAM\} modulations, respectively. We also respect the idea that the filtering effects should consume only a relatively small part of the stated \ac{evm} targets, to leave room for \ac{evm} degradation due to RF components. Then we can see that, with $\lambda=1/2$, even transition band of 1 \ac{fft} bin can be considered sufficient for QPSK from the average \ac{evm} point of view. For 64-QAM, transition band of 2 \ac{fft} bins is enough, also with one-sided filtering in the average \ac{evm} sense. For 256-QAM,  2 \ac{fft} bin transition band is sufficient in the two-sided filtering case to achieve average \ac{evm} below \SI{-29}{dB}, while 7 \ac{fft} bin transition band is required in the \ac{rx} filtering only case. 

Similar study on the 1 \ac{prb} allocation shows that the average \ac{evm} values are about \SI{5}{dB} higher than in  4\,\ac{prb} subband case and the worst-case values are \SIrange{1}{2}{dB} higher. Typically, 1\,\ac{prb} transmissions are used only on cell edge in coverage limited scenarios, in which the used modulation is most likely QPSK and not limited by the \ac{fc} processing induced passband \ac{evm}.  In the full-band case, with 50\,\ac{prb} filtering band, the average \ac{evm} is in the order of \SI{-40}{dB}, more than \SI{10}{dB} lower than in the 4\,\ac{prb} case, and also the worst-case \acp{evm} are improved by \SIrange{3}{4}{dB}. 

\begin{figure}[t]
  \centering 
  \includegraphics[angle=0,width=0.5\textwidth]{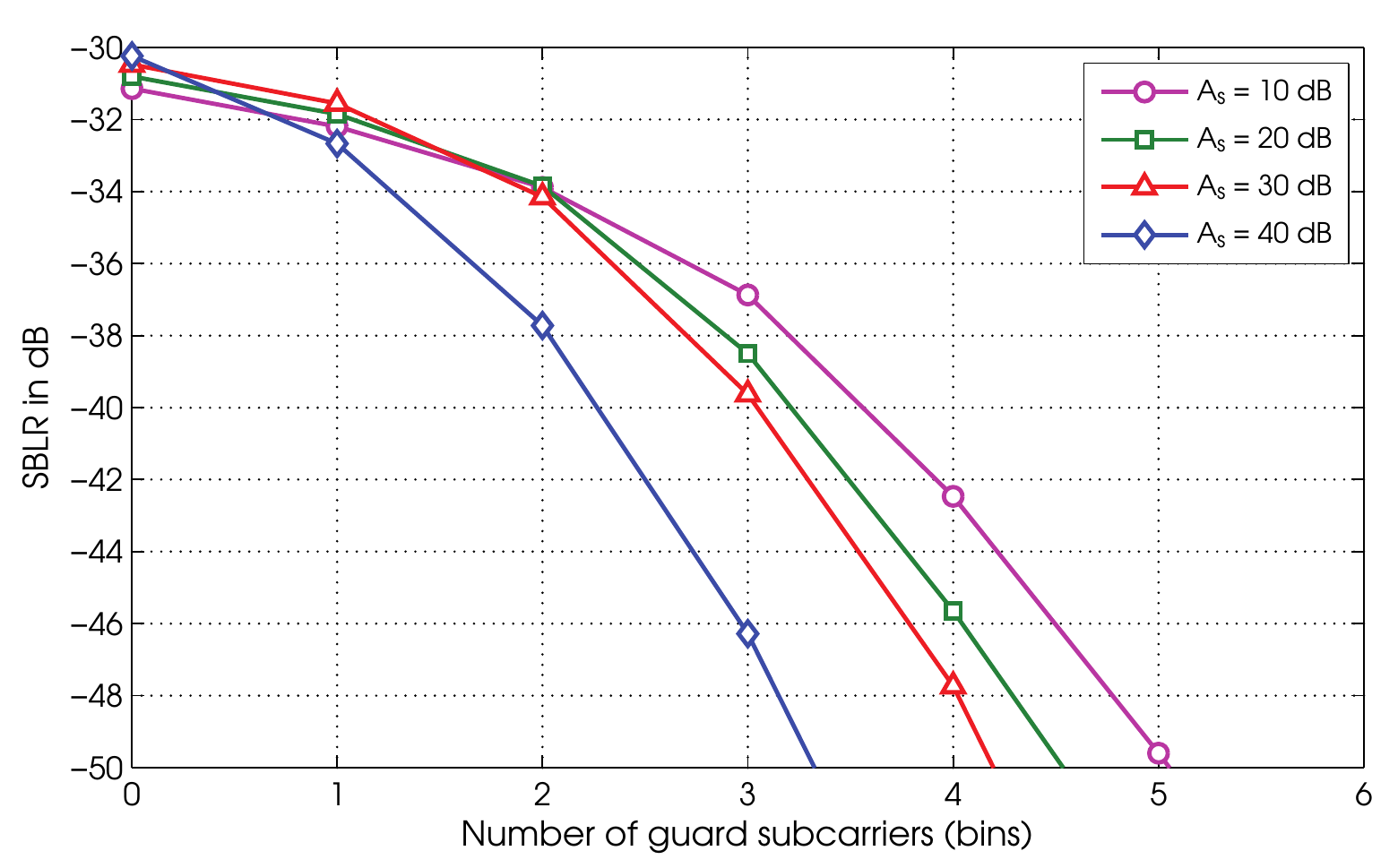}
  \caption{\ac{sblr} as a function of the number of guard subcarriers between two adjacent subbands of 4 \acp{prb}.} 
  \label{fig:SBLR}    
\end{figure}

Fig.~\ref{fig:SBLR} shows the corresponding \ac{sblr}, evaluated using \eqref{eq:sblr}, as function of the guard subcarriers between two adjacent 4-\ac{prb} subbands. As can be observed from this figure, the \ac{sblr} is well below \SI{-30}{dB} without any guard subcarriers between the subbands and reaches \SI{-45}{dB} with only \SIrange{3}{5}{} subcarrier guardband. In addition, it can be seen that the required minimum stopband attenuation in the optimization controls perfectly the subband band-limitation characteristics as desired. Now given the target \ac{evm} for the desired \ac{mcs} to be supported, one can select the parameter set that fulfills the passband \ac{evm} target and then evaluate the \ac{sblr} with different number of guard subcarriers to find out the required guard band to support the desired \ac{mcs}.

\begin{figure*} 
  \centering 
  \includegraphics[angle=0,width=0.99\columnwidth]{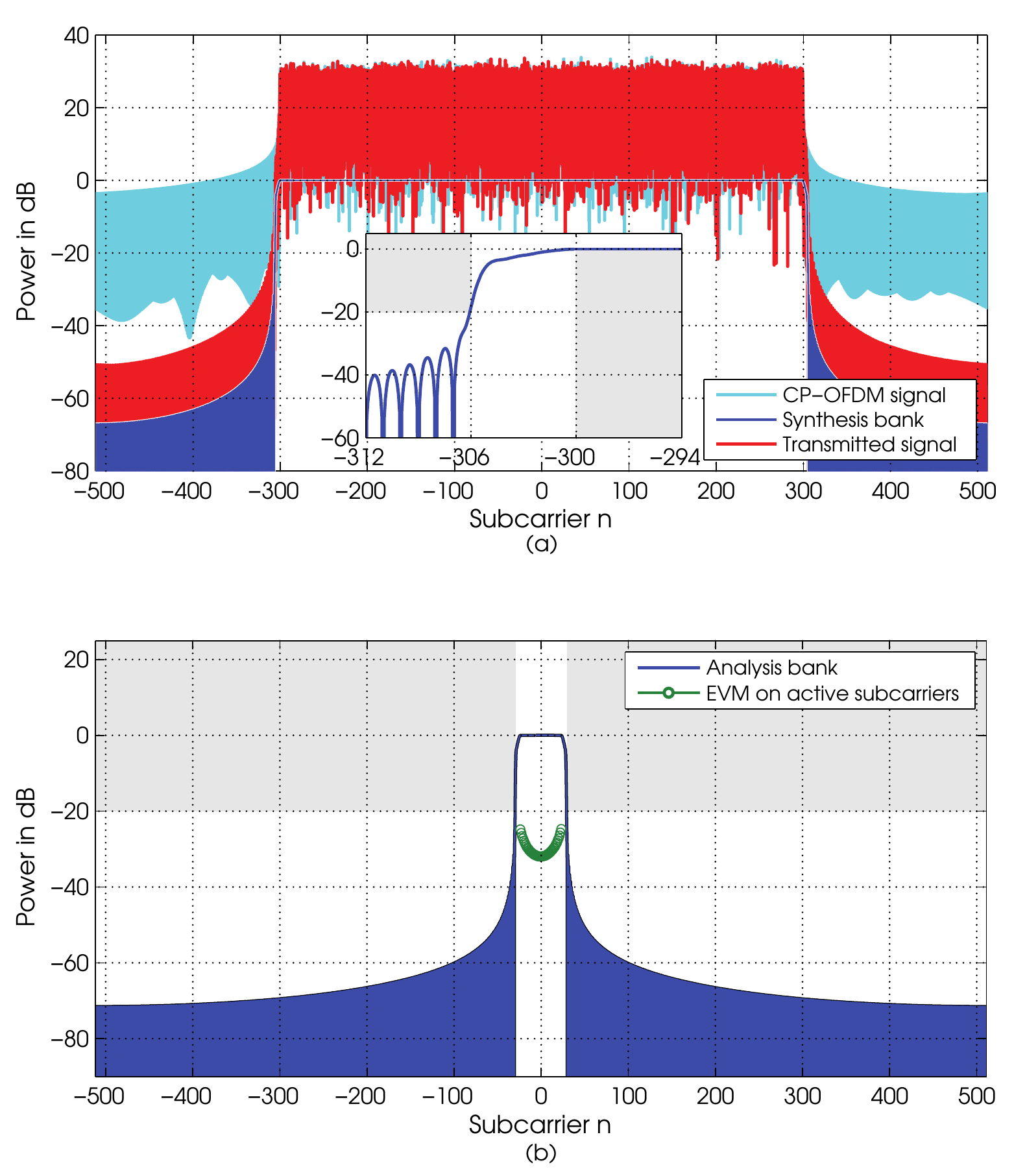}      
  \includegraphics[angle=0,width=0.99\columnwidth]{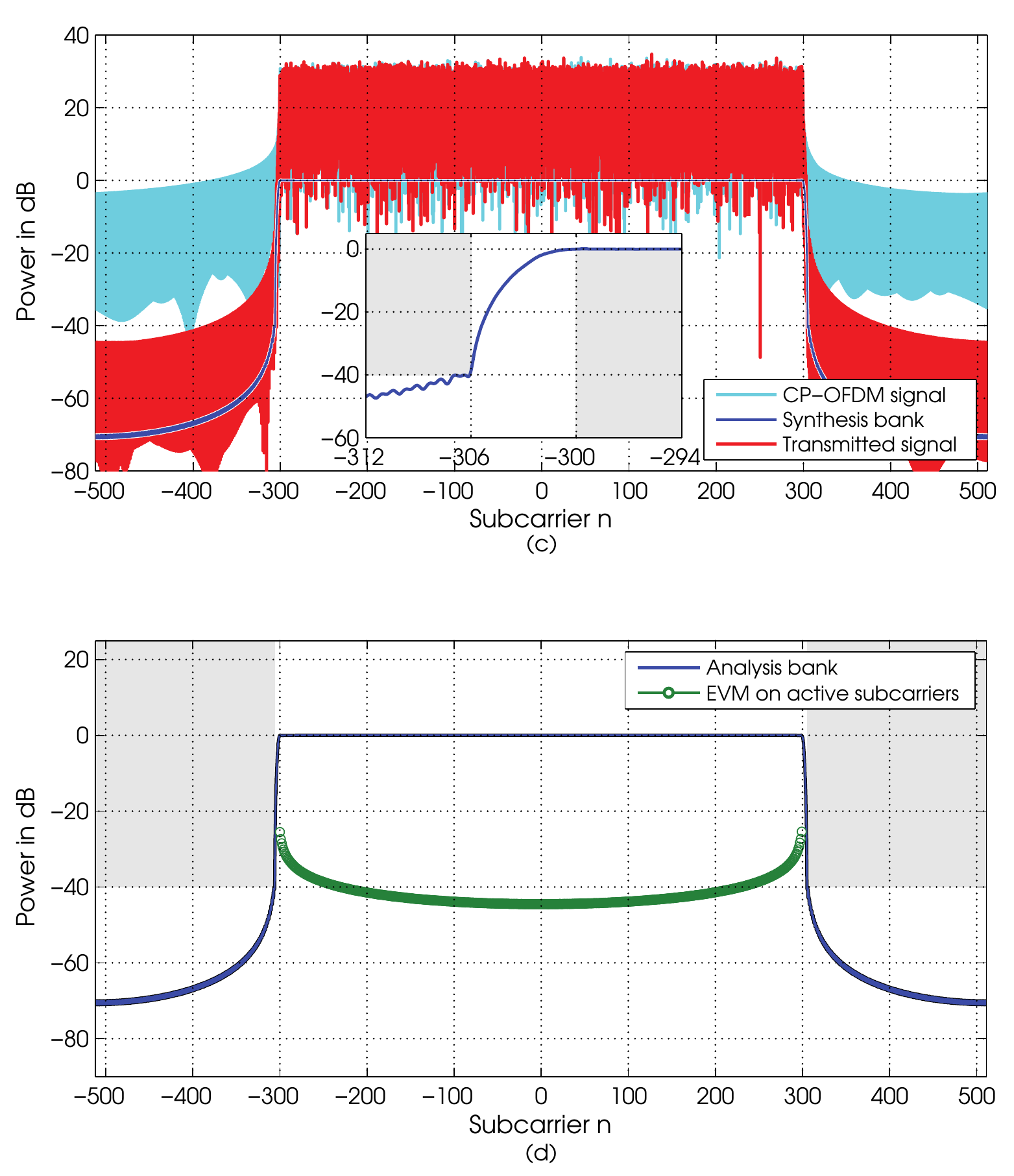}
  \caption{Upper figures: \ac{psd} of the generated \ac{fc}-filtered \ac{ofdm} signal in the case of 50 active \acp{prb}. Lower figures: Simulated active subcarrier \ac{evm} for two different \ac{rx} configurations. Also the analysis and synthesis filter bank responses are shown in these figures. In these cases, six transition band \ac{fft} bins are used. (a)-(b) The overlap in \ac{fc} processing is \SI{50}{\%} and the required minimum stopband attenuation is $A_\text{s}=\SI{20}{dB}$. The number of active \acp{prb} on the \ac{rx} side is 4. (c)-(d) The overlap in \ac{fc} processing is \SI{25}{\%} and the required minimum stopband attenuation is $A_\text{s}=\SI{40}{dB}$.  The number of active \acp{prb} on the \ac{rx} side is 50.}
  \label{fig:FC-F-OFDM_TX}    
\end{figure*}

Two examples of \ac{psd} plots for \ac{fc} filtered \ac{ofdm} with 50 active \acp{prb} are shown in Figs.~\ref{fig:FC-F-OFDM_TX}(a) and \ref{fig:FC-F-OFDM_TX}(c).  In Fig.~\ref{fig:FC-F-OFDM_TX}(a) the overlap factor in \ac{fc} processing is $\lambda=1/2$ and the minimum stopband attenuation is $A_\text{s}=\SI{20}{dB}$ whereas in Fig.~\ref{fig:FC-F-OFDM_TX}(c) the overlap factor is $\lambda=1/4$ and the minimum stopband attenuation is $A_\text{s}=\SI{40}{dB}$. Figs.~\ref{fig:FC-F-OFDM_TX}(b) and \ref{fig:FC-F-OFDM_TX}(d) show the simulated \acp{evm} on active subcarriers as well as the squared magnitude responses of the \acp{afb}. In Figs.~\ref{fig:FC-F-OFDM_TX}(b) and \ref{fig:FC-F-OFDM_TX}(d) the number of active \acp{prb} on the receiver side are 4 and 50, respectively.

\section{Link-Level Performance Evaluations at sub-\SI{6}{GHz} Frequency Bands}
%%%%%%%%%%%%%%%%%%%%%%%%%%%%%%%%%%%%%%%%%%%
In this section, performance results for \ac{fc} filtered \ac{cp-ofdm} and \ac{dft-s-ofdm} radio links are presented in terms of \ac{psd}, \ac{evm}, and \ac{bler} performance. In addition, the \ac{fc-f-ofdm} waveform is compared against several different \ac{5g} candidate waveforms, including traditional \ac{cp-ofdm} without any windowing or filtering which is used as a reference, \ac{wola} \cite{2015QualcommWola}, universally filtered \ac{ofdm} with \ac{cp} (\ac{cp}-\ac{uf-ofdm}) \cite{J:20145GNOW,2016RAN1UF-OFDM}, and \ac{f-ofdm} \cite{C:2015_Zhang_f-OFDM_for_5G}. 

\subsection{Reference Waveforms}
With \ac{cp}-\ac{uf-ofdm} \cite{J:20145GNOW,2016RAN1UF-OFDM}, the used filter is a Dolph-Chebyshev \ac{fir} filter of length $N_\text{FIR}$ and the stopband attenuation is a design parameter defining also the the \SI{3}{dB}-passband width of the filter. The \ac{uf-ofdm} processing was initially developed for zero-prefix \ac{ofdm},  but it can be equally well used with \ac{cp-ofdm} \cite{2016RAN1UF-OFDM}. The \ac{cp}-\ac{uf-ofdm} design ideology relies on a small number of different, predesigned filters typically optimized for 1, 2, or 4 \ac{prb} subbands.  In the later performance examples with \SI{15}{kHz} \ac{scs} and 4-\ac{prb} subbands, filter stopband attenuations of \SI{75}{dB} and \SI{37}{dB} are used with $N_\text{FIR}=73$ and $N_\text{FIR}=37$, respectively. With \SI{30}{kHz} \ac{scs} and 2-\ac{prb} subbands, the stopband attenuations of \SI{37}{dB} and \SI{30}{dB} are used with $N_\text{FIR}=73$ and $N_\text{FIR}=37$, respectively. \ac{tx} side pre-equalization is used to remove the \ac{tx} filter effect on the amplitude response, and on the \ac{rx} side, corresponding equalization is used to compensate the \ac{rx} filter passband attenuation.

The \ac{f-ofdm} was introduced in \cite{C:2015_Zhang_f-OFDM_for_5G}. The used filter is based on Hann-windowed sinc-function, where the sinc-function is defined based on the allocation bandwidth. The filter length is $N_\text{FIR}=512$. The filter is separately designed for different subband widths by tuning the sinc-function spectral width to match the allocation width. Because the main sinc-pulse length in time depends on the allocation width, the assumed filter causes minimal \ac{isi} with wide allocations but may cause significant \ac{isi} with narrow allocations, e.g., 1 \ac{prb} allocation. The subband wise filtering is performed in both, \ac{tx} and \ac{rx}, and the \ac{rx} filter is matched to the \ac{tx} filter. \ac{tx} side subband-wise pre-equalization and \ac{rx}-side compensation is used to alleviate the \ac{evm} increase caused by passband attenuation with \ac{f-ofdm}. Typically, a \ac{to} is defined for \ac{f-ofdm}. \ac{to} defines the passband width extension as an integer multiple of the used \ac{scs}. In the presented results it is assumed that \ac{to} is either 0 or 4 depending on the simulation case, as explained later.

The \ac{wola} processing with \ac{cp-ofdm} or \ac{dft-s-ofdm} is a widely known, computationally efficient method to improve the spectral containment of a \ac{cp-ofdm} signal \cite{3GPPTR25892,2015QualcommWola}. It has been introduced for \ac{5g-nr} as a low-complexity candidate method to allow improved \ac{sblr} to support mixed numerology and asynchronous traffic. In the presented results, only the simple single window scheme is assumed. Enhanced versions of \ac{wola} exist, and their potential impacts will be briefly discussed at the end of Section IV-I. 

In \ac{wola}, the \ac{cp-ofdm} symbol is extended by $N_\text{EXT}$ samples, and the number of extended samples equals to the window slope length $N_\text{WS}=N_\text{EXT}$. Window slope length defines in samples the rising and falling edge of the window. The window slope length used in the simulations is $N_\text{WS} = N_\text{CP} = 72$. This value is chosen to provide as good spectral containment as possible without significant degradation in the passband \ac{evm} after the \ac{pa}. The total window length in \ac{tx} is $N_\text{WIN,TX} = N_\text{FFT}+N_\text{CP}+N_\text{WS}$. After windowing, an overlap-and-add processing is used to partially overlap adjacent windowed \ac{cp-ofdm} symbols by $N_\text{EXT}$ samples to reduce the overhead caused by windowing and to retain the original symbol timing. The used window is a \ac{rc} window with roll-off of $N_\text{WS}/N_\text{WIN}$~\cite{C:Sahin2011_edgewin}.  In the \ac{rx} side, the \ac{wola} processing is performed within the \ac{cp-ofdm} symbol boundaries. The used window length is $N_\text{WIN,RX} = N_\text{FFT}+N_\text{WS}$. In other words, in \ac{rx} side the received \ac{cp-ofdm} symbol is not extended before \ac{wola} processing, as indicated also in \cite{2015QualcommWola}.

\subsection{Simulation Cases, Assumptions, and Performance Metrics}
In \ac{3gpp} \ac{tsg}-\ac{ran} WG1 way forward agreement it has been agreed that \ac{cp-ofdm} is the baseline waveform for \ac{5g-nr} physical layer \cite{2016RAN1WayForwardWaveforms}. Therefore, all \ac{dl} and \ac{ul} performance results are based on subband-filtered \ac{cp-ofdm} signals. In \cite{2016RAN1WayForwardOnULWaveforms}, also DFT-spread-OFDM was agreed to be supported in \ac{ul} in coverage limited scenarios. In Section \ref{subsec:FC-F-DFTs-OFDM}, examples of \ac{fc} filtered DFT-spread-OFDM signal are also given. 

The baseline physical layer definition and numerology follows the one defined for \ac{lte} operating in a \SI{10}{MHz} channel. The main evaluation parameters are given in Table \ref{tab:simulation_params}. As observed in Fig. \ref{fig:IBI_perf}, optimized \ac{fc} design with \ac{tbw} of 2 \ac{fft} bins and minimum stopband attenuation level $A_\text{s}=\SI{10}{dB}$ is sufficient to achieve average \acp{mse} below \SI{-29}{dB}, required for 256-QAM \cite{3GPPTS36104}. The values of Table \ref{tab:simulation_params} are used unless stated otherwise.  

All the presented results assume an ideal channel knowledge in the \ac{rx} and each simulated subframe contains only data symbols. A guard period is added to each subframe to allow rising and falling transients caused by filtering or windowing to take place. If filtering causes transients longer than the guard period, they are truncated with a \ac{rc} window to fit within the subframe. The link performance is evaluated in TDL-C channels \cite{3GPPTR38900} with \SI{300}{ns} and \SI{1000}{ns} \ac{rms} delay spread. In TDL channels the \ac{rms} delay spread is defined by a scaling factor indicated in the name.

The link performance results are provided for \ac{dl} and \ac{ul} following the simulation cases defined in \cite[Annex A]{3GPPTR38802}. \Case{1a} corresponds to interference free \ac{dl} scenario and \Case{1b} corresponds to interference free \ac{ul} scenarios, \Case{2} defines a mixed numerology \ac{dl} scenario, \Case{3} defines an asynchronous \ac{ul} scenario, and \Case{4} defines a mixed numerology \ac{ul} scenario. In all the simulations, the target signal has a \ac{scs} of \SI{15}{kHz}. In \Case{2} and \Case{4}, the interfering signal has \ac{scs} of \SI{30}{kHz}. In \Case{3}, the interfering signal has the same \ac{scs} as the target signal, but it is time shifted with 128 samples to model asynchronous interference. In all cases with interfering signal present, the interfering signal is assumed to be processed in similar manner as the evaluated waveform. In the mixed numerology scenarios (\Case{2} and \Case{4}) and asynchronous interference scenario (\Case{3}), \ac{gb} defines the increase in the distance of the edge-most SC centers with respect to the minimum \SI{15}{kHz} distance.

\begin{table}[t]
  \caption{Main evaluation parameters.}
  \label{tab:simulation_params}
  \centering
  \footnotesize{
   \begin{tabular}{lc}
      \toprule
      \multicolumn{1}{c}{Parameter} & 
      \multicolumn{1}{c}{Value} \\
      \midrule
      Carrier frequency & \SI{4}{GHz} \\
      UE mobility & \SI{3}{km/h} \\
      Channel bandwidth & \SI{10}{MHz} \\
      Sampling frequency & \SI{15.36}{MHz} \\
      Channel model & TDL-C \cite{3GPPTR38900} \\
      DL PA model & Modified Rapp \cite{2016RAN1RappPA} \\
      IBO in DL & \SI{11.6}{dB} \\
      UL PA model & Polynomial model \cite{2016RAN1PolynomialPA}\\
      Desired signal IBO in UL & \SI{8}{dB} \\
      Interfering signal IBO in UL & \SI{5.5}{dB} \\
      Channel codec & Turbo code \cite{3GPPTS36300}\\
      Guard period length at high rate & 72 samples \\
      Subband allocation granularity & \SI{720}{kHz} \\
      FC block length & $N=1024$ samples\\      
      Overlap factor in FC processing & 1/2 \\
      Transition bandwidth in FC processing & $L_{\text{TBW}}=2$\\
      Minimum stopband attenuation target & $A_\text{s}=\SI{10}{dB}$ \\
      Target OFDM symbol length & $N_{\text{OFDM}}=1024$ samples\\
      Target signal CP length & 72 samples \\
      Target signal subcarrier spacing & \SI{15}{kHz} \\
      Interfering OFDM symbol length & $N_{\text{OFDM}}=1024/512$ samples\\
      Interfering signal CP length & 72/36 samples \\
      Interfering signal subcarrier spacing & \SI{15/30}{kHz} \\
      OFDM symbols per subframe & 14/28 \\
      \bottomrule
    \end{tabular}}  
\end{table}

The presented \ac{evm} results are evaluated by inserting the \ac{pa} input or output signal to a waveform specific detector. No equalization is applied in the presented \ac{evm} results.  In the evaluation of the \ac{evm}, with \ac{f-ofdm} and \ac{cp}-\ac{uf-ofdm}, the \ac{fft} window is located at the center of the \ac{cp-ofdm} symbol in TDL-C \SI{300}{ns} channel and in the end of the \ac{cp-ofdm} symbol in TDL-C \SI{1000}{ns} channel. This shifting of the \ac{rx} \ac{fft} window reduces the filtering induced \ac{isi} in the TDL-C \SI{300}{ns} channel to provide a realistic performance comparison between waveform candidates. In practice, to support this the \ac{rx} has to have capability to estimate channel delay profile and adjust the \ac{rx} \ac{fft} window location per subband. With \ac{wola}, due to long window and the \ac{rx} \ac{wola} processing, the \ac{fft} window is always located in the middle of the \ac{cp-ofdm} symbol after \ac{rx} \ac{wola} processing. With \ac{fc-f-ofdm}, the \ac{rx} \ac{fft} window is always in the end of the \ac{cp-ofdm} symbol.

The \ac{psd} is evaluated per subframe. In the presented results, 100 independent realizations at the \ac{pa} output are averaged and filtered to model \SI{30}{kHz} measurement bandwidth used to define the \ac{lte} \ac{oobem} within \SI{1}{MHz} distance from the channel edge in \ac{dl} and \ac{ul}. The \ac{dl} \ac{oobem} is defined in \cite{3GPPTS36104} and \ac{ul} \ac{oobem} is defined in \cite{3GPPTS36101}. These are commonly agreed as a starting point for evaluating the \ac{oob} emissions for new waveform candidates in \ac{5g-nr}.

%%%%%%%%%%%%%% PA models %%%%%%%%%%%%%%%% 
\subsection{Power Amplifier Models}
\label{sec:PAModels}
The \ac{pa} models used in this paper have been introduced for performance evaluations for below \SI{6}{GHz} communications in \ac{3gpp} \ac{tsg}-\ac{ran} WG1. The \ac{dl} \ac{pa} model was introduced in \cite{2016RAN1RappPA} and \ac{ul} \ac{pa} model in \cite{2016RAN1PolynomialPA}. These models are used because they are openly available and commonly agreed to provide a good starting point for spectral containment evaluations related to \ac{5g-nr}.

The \ac{dl} \ac{pa} model is a modified Rapp model \cite{2016RAN1RappPA}. It mimics a \ac{bs} \ac{pa} including some crest factor reduction and digital predistortion schemes to linearize the \ac{bs} \ac{pa} to achieve the \ac{lte} \ac{oobem} and \ac{oob} \ac{aclr} of \SI{45}{dB} with a fully populated \SI{10}{MHz} \ac{lte} signal with 50 \acp{prb} and 64-QAM modulation. The modified Rapp model is defined by the amplitude-to-amplitude (AM-AM) distortion

\begin{subequations}
\begin{equation}
    F_\text{AM-AM}(x) = \frac{G}{\left( 1+ \left\lvert \frac{Gx}{V_\text{SAT}} \right\rvert^{2p} \right)^{1/(2p)}},
\end{equation}
and amplitude-to-phase (AM-PM) distortion
\begin{equation}
    F_\text{AM-PM}(x) = \frac{A \left\lvert \frac{Gx}{V_\text{SAT}} \right\rvert^q}
    {1+ \left\lvert\frac{Gx}{BV_\text{SAT}}\right\rvert^q},
\end{equation}
\end{subequations}
where $x$ is the instantaneous amplitude of the signal, gain is normalized to $G=1$, saturation voltage is $V_\text{SAT}=239.6$ at \SI{50}{\ohm} load, smoothness factors are $p=3$ and $q=5$, and tuning parameters are $A=-0.14$ and $B=1.2$. This model has a \SI{1}{dB} compression point at $P_\text{1\,dB}=\SI{57.6}{dBm}$. A backoff of \SI{11.6}{dB} is assumed for simulations, providing $P_\text{DL}=\SI{46}{dBm}$ total output power from the \ac{pa}. 

The \ac{ul} \ac{pa} model is a polynomial model of order nine obtained by fitting the polynomial to measurements from a commercial \ac{pa} \cite{2016RAN1PolynomialPA}. The polynomial coefficients are ordered from $p_9$ to $p_0$, given in logarithmic domain, and defined for the amplitude distortion as 
\begin{subequations}
\begin{equation}
\makeatletter%
\if@twocolumn%
\begin{array}{rl}
  p_\text{AM} = & [7.9726\times10^{-12},  1.2771\times10^{-9},  8.2526\times10^{-8}, \\
  & 2.6615\times10^{-6},  3.9727\times10^{-5},  2.7715\times10^{-5}, \\
  &-7.1100\times10^{-3}, -7.9183\times10^{-2},  8.2921\times10^{-1}, \\
  &  27.3535],
\end{array}
\else% \@twocolumnfalse
\begin{array}{rl}
  p_\text{AM} = &
  [7.9726\times10^{-12},  1.2771\times10^{-9},  8.2526\times10^{-8}, 2.6615\times10^{-6},  3.9727\times10^{-5},  \\
  & 2.7715\times10^{-5}, -7.1100\times10^{-3}, -7.9183\times10^{-2},  8.2921\times10^{-1}, 27.3535],
\end{array}
\fi
\makeatother
\end{equation}
and for phase distortion as
\begin{equation}
  \makeatletter%
  \if@twocolumn%
    \begin{array}{rl}
      p_\text{PM} = & [9.8591\times10^{-11},  1.3544\times10^{-8},  7.2970\times10^{-7}, \\
      & 1.8757\times10^{-5},  1.9730\times10^{-4},  -7.5352\times10^{-4}, \\
      & -3.6477\times10^{-2},  -2.7752\times10^{-1},  -1.6672\times10^{-2}, \\
      & 79.1553].
    \end{array}
  \else% \@twocolumnfalse
    \begin{array}{rl}
      p_\text{PM} = & [9.8591\times10^{-11},  1.3544\times10^{-8},  7.2970\times10^{-7}, 1.8757\times10^{-5},  1.9730\times10^{-4}, \\
      &  -7.5352\times10^{-4}, -3.6477\times10^{-2},  -2.7752\times10^{-1},  -1.6672\times10^{-2},  79.1553].
  \end{array}
\fi
\makeatother
\end{equation}
\end{subequations}

The polynomial model should be used only with input levels between \SI{-30}{dBm} and \SI{9}{dBm}. The input referred \SI{1}{dB} compression point is at $P_\text{1\,dB}=\SI{3.4}{dBm}$ and the model is parametrized to provide \SI{26}{dBm} \ac{pa} output power with \SI{20}{MHz} \ac{qpsk} modulated fully populated \ac{lte} uplink signal (100 \ac{prb} allocation), while meeting the minimum \ac{aclr} requirement of \SI{30}{dB} for \ac{e-utra} and \ac{ul} emission masks.

Below, all \ac{psd} and link performance results are obtained with the stated \ac{pa} models, while \ac{evm} results are given for both \ac{pa} input and output signals to illustrate the \ac{pa} induced error.  All \ac{evm} and link performance results are for matched \ac{tx} and \ac{rx} filtering cases. The numerical \ac{evm} values given below are for average \ac{evm} over the active subcarriers, unless otherwise noted.

% 55 PRB WF comparison
\begin{figure} 
  \centering
  \begin{subfigure}{\figwidth}
    \centering  
    \includegraphics[angle=0,width=\FIGURE_WIDTH]{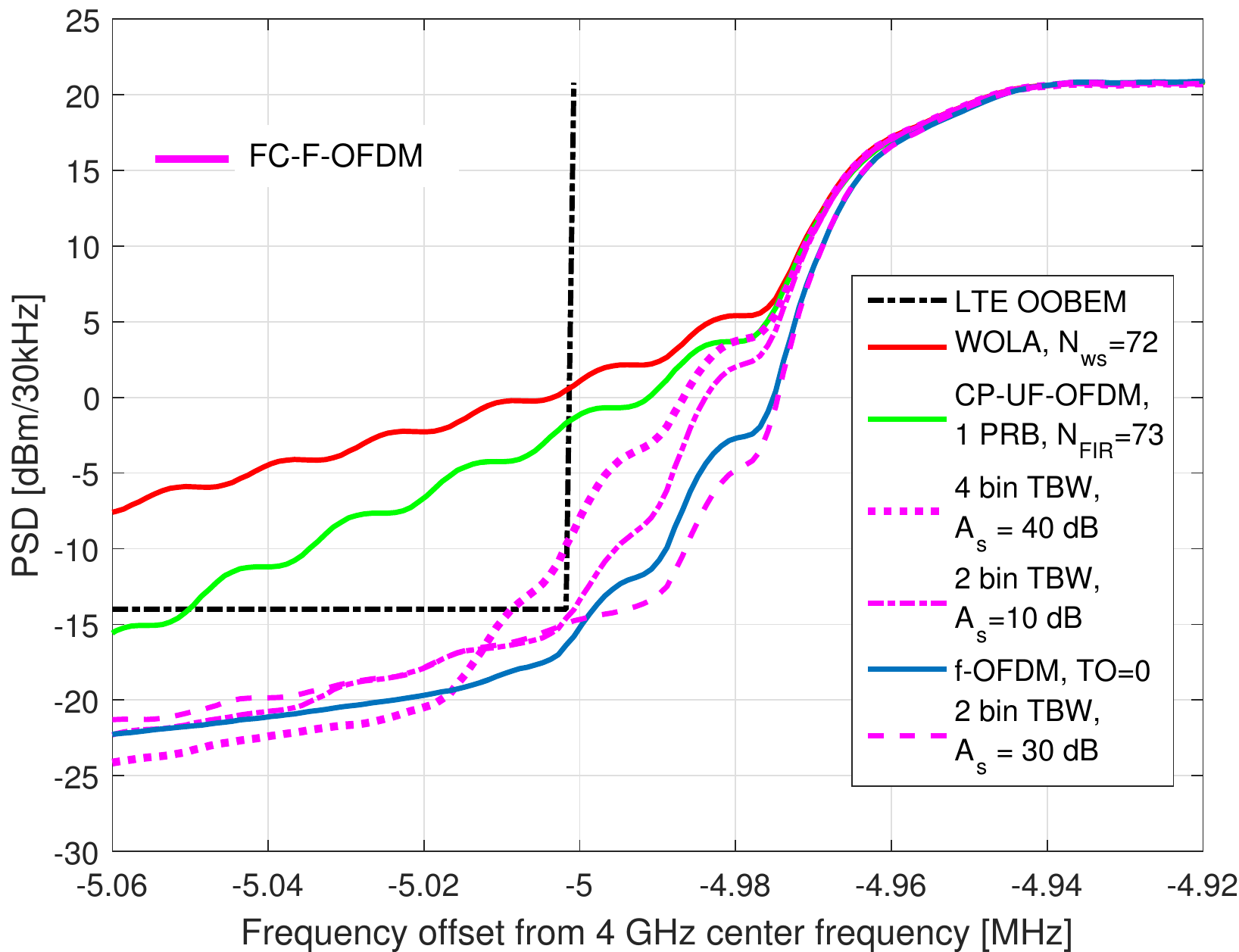}    
    \caption{}
  \end{subfigure}\hfill 
  \vspace{1mm}
  \begin{subfigure}{\figwidth}
    \centering  
    \includegraphics[angle=0,width=\FIGURE_WIDTH]{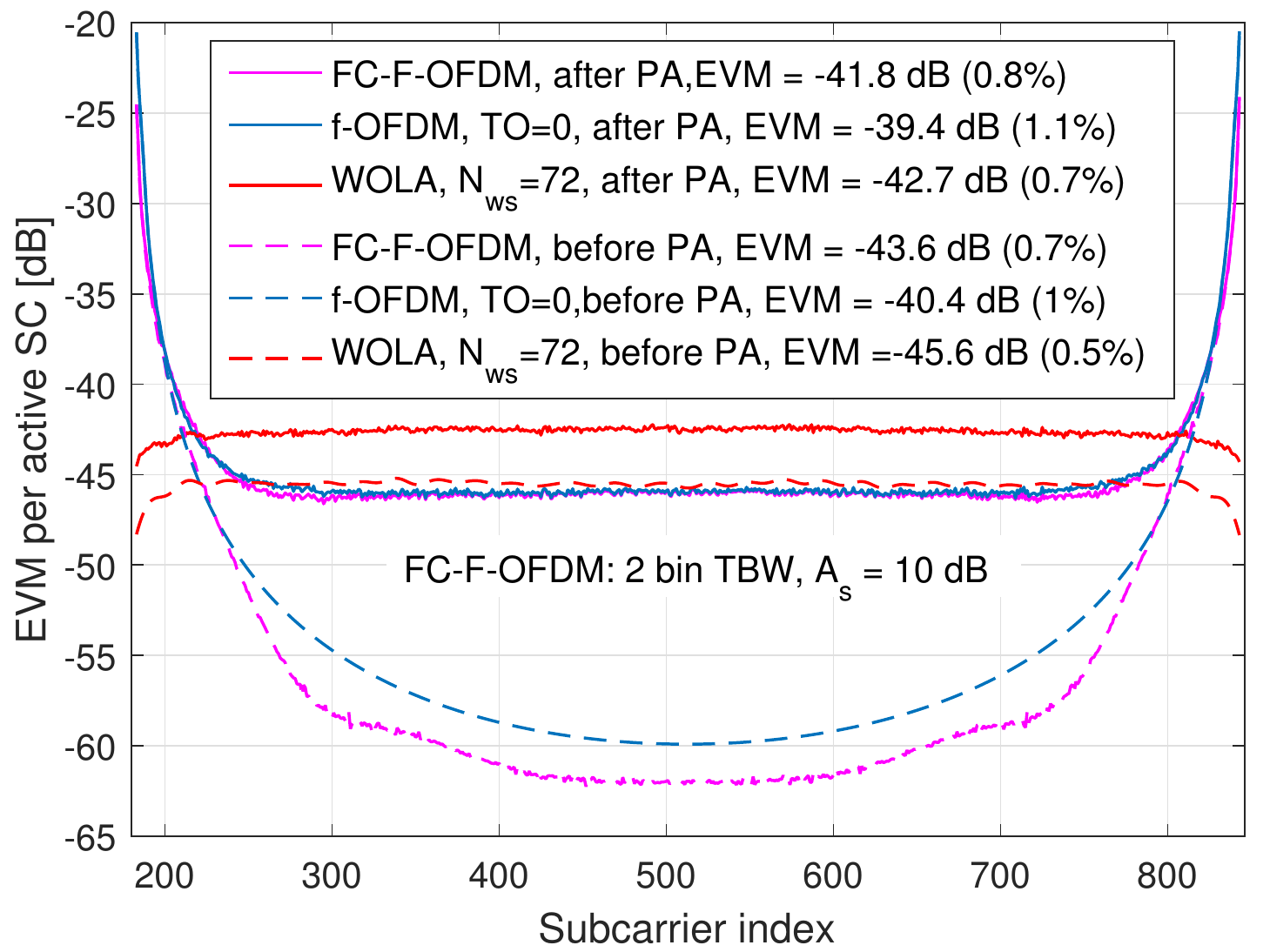}
    \caption{} 
  \end{subfigure}
  \caption{Results for 55 \ac{prb} allocation in a \SI{10}{MHz} channel with different enhanced \ac{cp-ofdm} waveforms. (a) Channel edge \ac{psd} with \ac{dl} \ac{pa} model. (b) Subcarrier wise \ac{evm} results with and without the \ac{dl} \ac{pa} model.}
  \label{fig:Case1a_55PRB_FC_WF}
\end{figure}

\subsection{Case 1: Interference free DL and UL link performance}
In \ac{3gpp} \ac{tsg}-\ac{ran} WG1 physical layer way forward agreement \cite{2016RAN1WayForwardWaveforms} it has been decided that \ac{5g-nr} should support higher bandwidth efficiency than the current \ac{lte} technology. Therefore, we start by evaluating the \ac{fc-f-ofdm} performance in the given \SI{10}{MHz} channel with 55 \ac{prb} allocation. The current \ac{lte} technology supports 50 \acp{prb} in a \SI{10}{MHz} channel. Increasing the maximum number of supported \acp{prb} per channel increases the filtering requirements and emphasizes the channel filter design required to achieve the current \ac{lte}-based \ac{oobem} requirements. At the same time, the \ac{evm} degradation of the edge \acp{prb} should be included in the evaluation because it would limit the \ac{mcs} range that can be applied in these \acp{prb}.

In Fig.~\ref{fig:Case1a_55PRB_FC_WF}, the performance of the \ac{fc-f-ofdm} is compared against \ac{f-ofdm} and \ac{wola} in terms of (a) \ac{psd} and (b) passband \ac{evm}. The results are shown assuming a 55 \ac{prb} allocation in a \SI{10}{MHz} \ac{lte} channel. In this example, the bandwidth utilization efficiency is \SI{99}{\%} and there is only \SI{100}{kHz} \ac{gb} between channel edge and first \ac{prb}. As expected, \ac{fc-f-ofdm} with 4 bin \ac{tbw} already violates the \ac{oobem}. Also, for 3 bin \ac{tbw} no optimized weights were found achieving the \ac{oobem}. With 2 bin \ac{tbw}, both attenuation targets $A_\text{s}=\SI{10}{dB}$ and $A_\text{s}=\SI{30}{dB}$ achieve the defined \ac{oobem}. The 2 bin \ac{tbw} with $A_\text{s}=\SI{10}{dB}$ design is the best choice when noting the clear improvement in the passband \ac{evm}. These results also give an example how the attenuation target in the optimization can be used to fine-tune the tradeoff between frequency selectivity and passband \ac{evm}. With 2 bin \ac{tbw} and $A_\text{s}=\SI{30}{dB}$, the filtered signal is well within the emission mask but the \ac{evm} is \SI{-37.1}{dB} (\SI{1.4}{\%}). By reducing design attenuation target to $A_\text{s}=\SI{10}{dB}$, the 2 bin \ac{tbw} \ac{evm} can be reduced to \SI{-41.8}{dB} (\SI{0.8}{\%}), while fulfilling the \ac{oobem}. 

When the bandwidth efficiency is increased from 50 \acp{prb}, \ac{wola} is unable to suppress the \ac{oob} emissions sufficiently and needs to be combined with some additional channel filtering to fulfill the \ac{oobem}. This indicates that simple \ac{wola} processing alone is not sufficient to support increased bandwidth utilization efficiency expected from \ac{5g-nr}. Similar observations can be made for \ac{cp}-\ac{uf-ofdm}, for which the example \ac{psd} is obtained with 1 \ac{prb} subband filter with \SI{20}{dB} stopband attenuation. The \ac{f-ofdm} performance is between the 2 bin \ac{tbw} cases presented in Fig.~\ref{fig:Case1a_55PRB_FC_WF} (a). \ac{wola} provides the best average passband \ac{evm} performance, \ac{fc-f-ofdm} achieves almost the same, and \ac{f-ofdm} has the largest \ac{evm} while being still sufficiently low to support fullband transmission with 256-QAM modulation. At the edges of the channel a clear degradation on the \ac{evm} is observed for \ac{fc-f-ofdm} and \ac{f-ofdm} due to the steep filtering. \ac{wola} has significantly better \ac{evm} at channel edges and it is in fact decreasing towards the channel edges. In the first \ac{prb}, the average \ac{evm} for \ac{fc-f-ofdm} is \SI{-30.2}{dB} and for \ac{f-ofdm} it is \SI{-26.6}{dB}. This indicates that \ac{fc-f-ofdm} allows to use 256-QAM modulation in all \acp{prb}, whereas with \ac{f-ofdm} the edge-most \acp{prb} are limited to 64-QAM modulation. 

% case1 TDLC1000 perf
\begin{figure} 
  \centering
  \begin{subfigure}{\figwidth}
    \centering    
    \includegraphics[angle=0,width=\FIGURE_WIDTH]{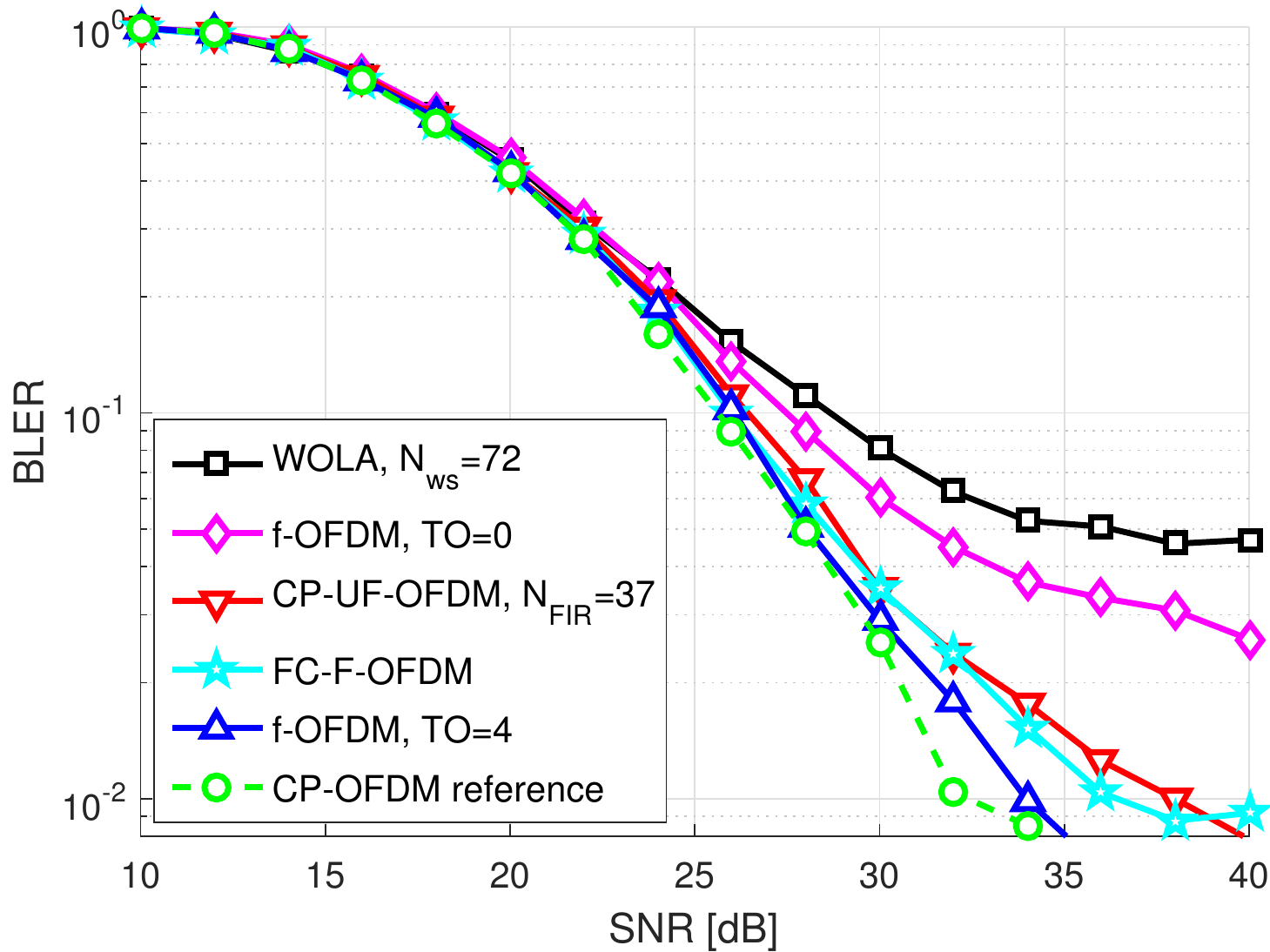}    
    \caption{} 
    \vspace{1mm}
  \end{subfigure}\hfill
  \begin{subfigure}{\figwidth}
    \centering 
    \includegraphics[angle=0,width=\FIGURE_WIDTH]{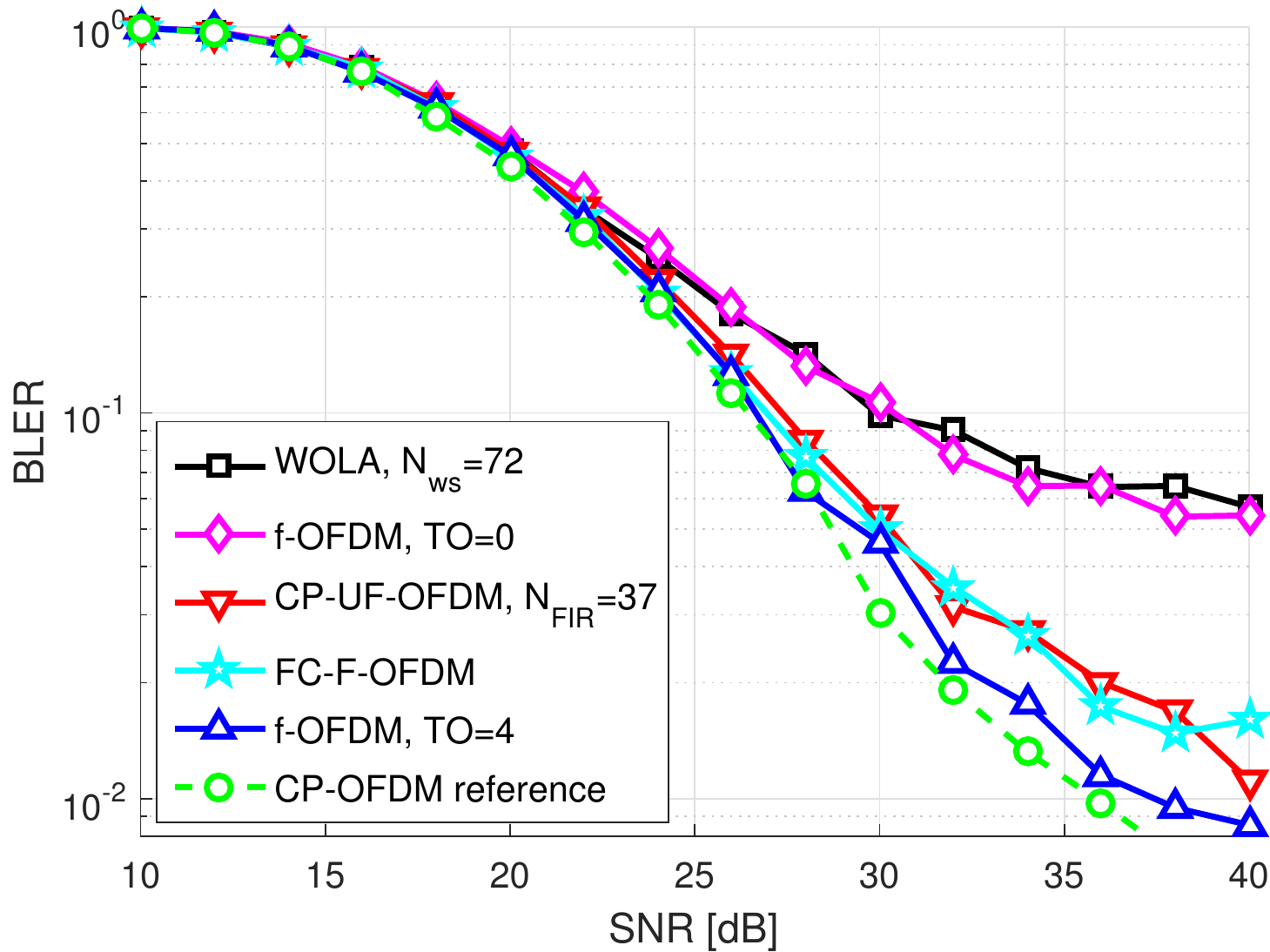}
    \caption{}  
  \end{subfigure}
  \caption{Performance comparison for (a) \Case{1a} (\ac{dl}) and (b) \Case{1b} (\ac{ul}) with 4 \ac{prb} allocation.}
  \label{fig:Case1_4PRB_TDLC1000_linkPerf}
\end{figure}

In the TDL-C \SI{300}{ns} channel, all waveform candidates perform well in \ac{dl} with \ac{mcs} 256-QAM, $R=4/5$ with allocations sizes from 4 \acp{prb} up to 55 \acp{prb}. In \ac{ul} all waveform candidates work up to \ac{mcs} 64-QAM, $R=3/4$. Higher \acp{mcs} do not work in \ac{ul} due to the passband distortion generated by the polynomial \ac{pa} model with the used \SI{8}{dB} input backoff. In the TDL-C \SI{1000}{ns} channel differences between waveform candidates become more clear as the channel induced \ac{isi} starts to have a significant role. In Fig.~\ref{fig:Case1_4PRB_TDLC1000_linkPerf} the link performance for different waveform candidates in (a) \ac{dl} and (b) \ac{ul} are shown for \ac{mcs} 64-QAM, $R=3/4$. In the presented results, \ac{f-ofdm} with \ac{to} of  4 provides performance closest to the reference \ac{cp-ofdm}. \ac{fc-f-ofdm} and \ac{cp}-\ac{uf-ofdm} with $N_\text{FIR}=37$ provide similar performance and are very close to the reference \ac{cp-ofdm} at \ac{bler} target of \SI{10}{\%} but diverge further at \ac{bler} target of \SI{1}{\%}. \ac{wola} and \ac{f-ofdm} with \ac{to} of 0 lose approximately \SI{2}{dB} to other waveform candidates at \ac{bler} target of \SI{10}{\%} and do not achieve \ac{bler} target of \SI{1}{\%}. With \ac{wola}, the \ac{tx} and \ac{rx} processing with long window slope length spreads the channel induced \ac{isi} which leads to degraded performance. This effect could be reduced with shorter \ac{rx} window lengths. With \ac{f-ofdm}, increasing the \ac{to} reduces the passband distortion caused by the subband filtering in this scenario. On the other hand, the mentioned methods for enhancing the passband performance of \ac{wola} and \ac{f-ofdm} reduce the spectral containment and degrade the performance in the following simulation cases where an interfering signal is introduced to the vicinity of the target signal.

\subsection{Case 2: Mixed numerology DL link performance}

% case2 TDLC300 perf
\begin{figure} 
  \centering
  \begin{subfigure}{\figwidth}
    \centering     
    \includegraphics[angle=0,width=\FIGURE_WIDTH]{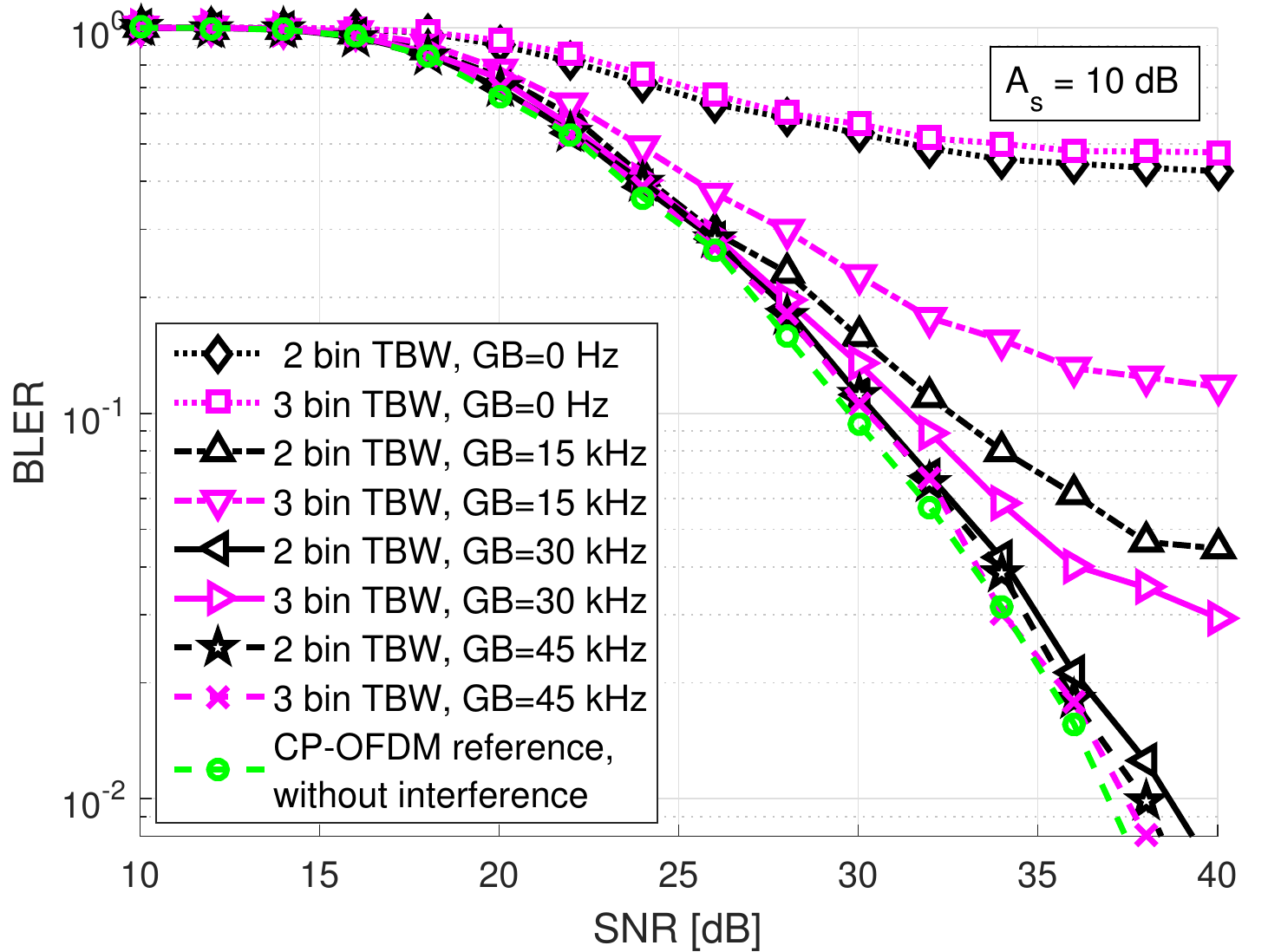}      
    \caption{}  
    \vspace{1mm}
  \end{subfigure}\hfill
  \begin{subfigure}{\figwidth}
    \centering 
    \includegraphics[angle=0,width=\FIGURE_WIDTH]{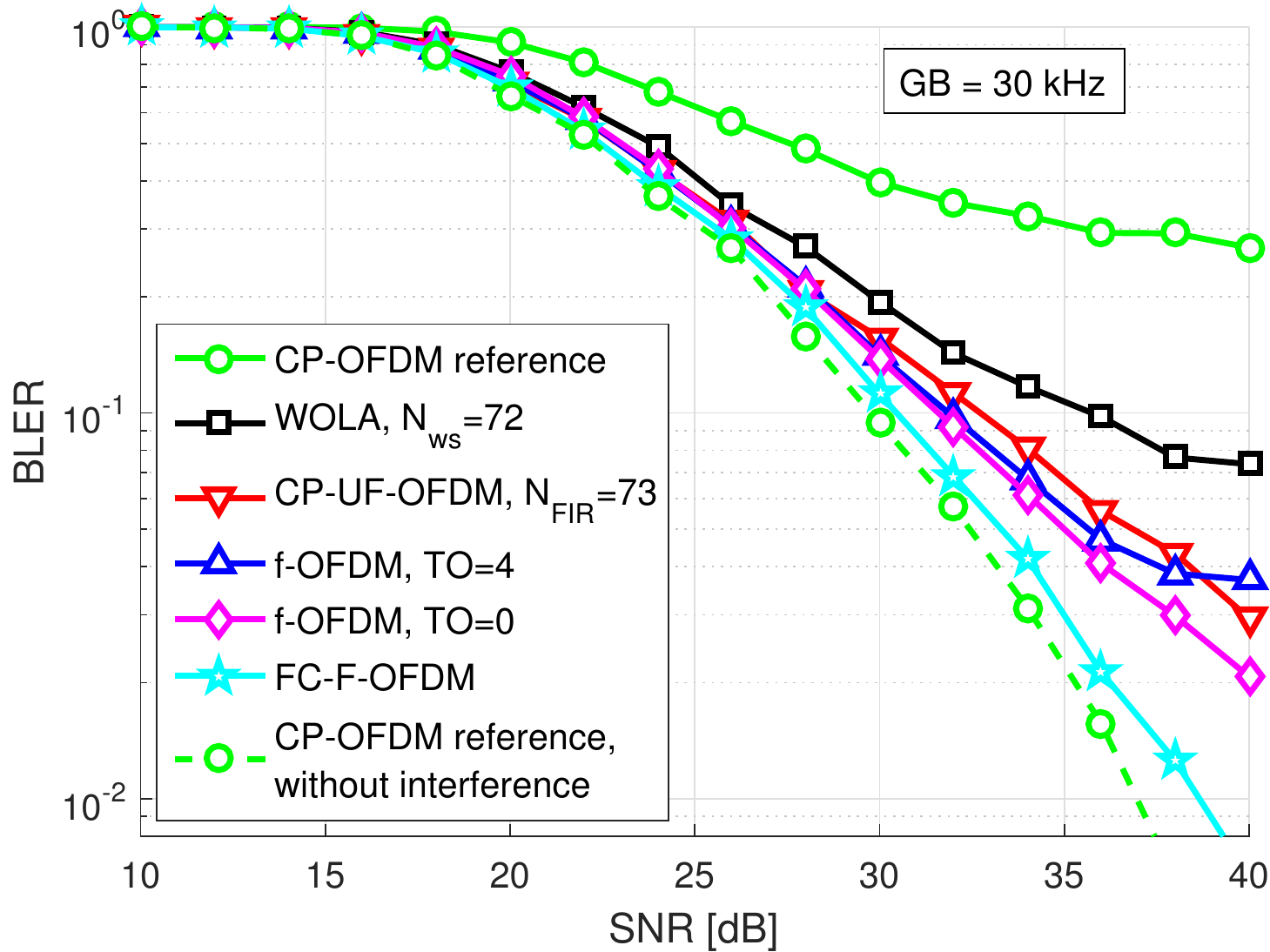}
    \caption{} 
  \end{subfigure}
  \caption{Performance comparison for \Case{2} mixed numerology \ac{dl} scenario in TDL-C \SI{300}{ns} channel. (a) \ac{fc-f-ofdm} performance with different GB values. (b) The relative performance with different waveforms with GB = \SI{30}{kHz}.
  }
  \label{fig:Case2_TDLC300_linkPerf}
\end{figure}

In Fig.~\ref{fig:Case2_TDLC300_linkPerf}, the \Case{2} \ac{dl} scenario is evaluated in the TDL-C \SI{300}{ns} channel with \ac{mcs} 256-QAM, $R=3/4$. The effect of \ac{gb} with two different \ac{fc} \acp{tbw} and $A_s=\SI{10}{dB}$ are shown in Fig.~\ref{fig:Case2_TDLC300_linkPerf} (a) and the link performance of all candidate waveforms assuming a GB=\SI{30}{kHz} are shown in Fig.~\ref{fig:Case2_TDLC300_linkPerf} (b).  In this \ac{dl} scenario, the required \ac{gb} is typically smaller than in the corresponding \ac{ul} scenario and it supports higher \ac{mcs} due to the more linear \ac{pa} model. It can be seen that a \ac{gb} of \SI{30}{kHz} is sufficient to achieve interference free performance with \ac{fc-f-ofdm}. Furthermore, with the same \ac{gb}, \ac{fc-f-ofdm} provides the best link performance among all waveform candidates.

% case2 TDLC1000 perf
\begin{figure}  
  \centering
  \begin{subfigure}{\figwidth}
    \centering    
    \def\FIGURE_WIDTH{1.02\columnwidth}
    \includegraphics[angle=0,width=\FIGURE_WIDTH]{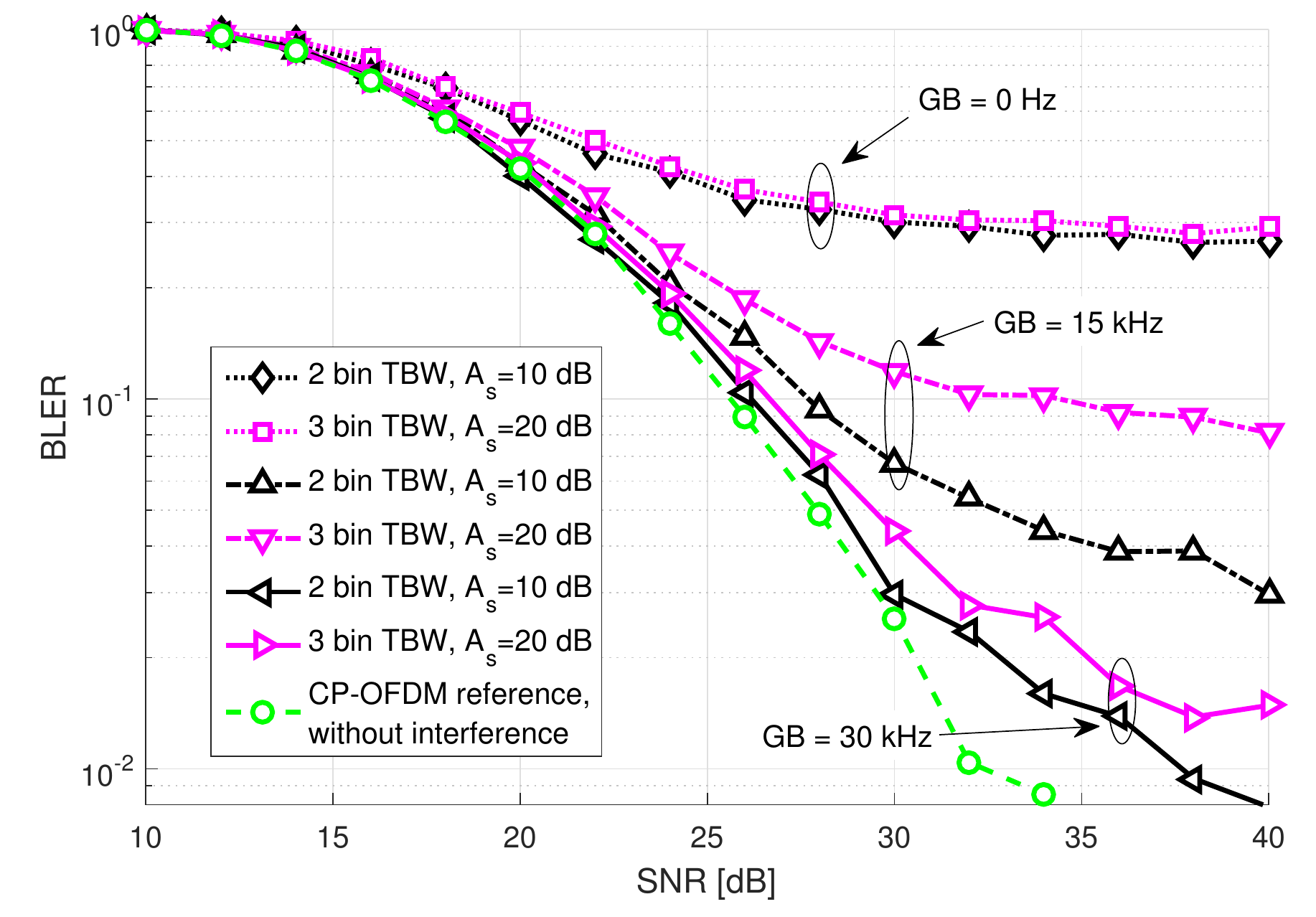}      
    \caption{}
    \vspace{1mm}
  \end{subfigure}\hfill
  \begin{subfigure}{\figwidth}
    \centering 
    \includegraphics[angle=0,width=\FIGURE_WIDTH]{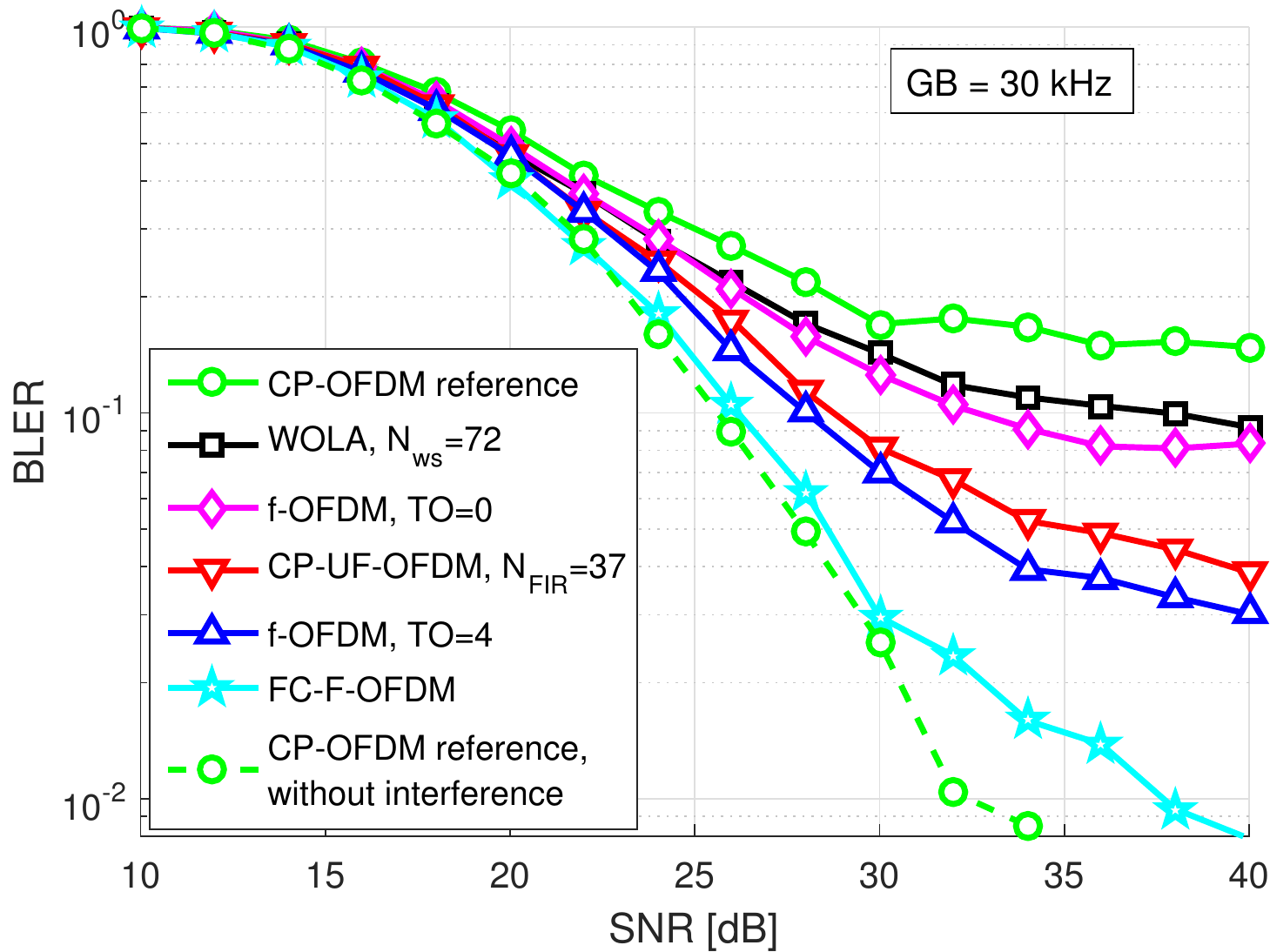}
    \caption{}    
  \end{subfigure}
  \caption{Performance comparison for \Case{2} mixed numerology \ac{dl} scenario in TDL-C \SI{1000}{ns} channel. (a) \ac{fc-f-ofdm} performance with different GB values and parameterizations. (b) Relative performance with different waveforms with GB = \SI{30}{kHz}.
  }
  \label{fig:Case2_TDLC1000_linkPerf}
\end{figure}

In Fig.~\ref{fig:Case2_TDLC1000_linkPerf}, the \Case{2} scenario is evaluated in the TDL-C \SI{1000}{ns} channel with \ac{mcs} 64-QAM, $R=3/4$. The outcomes of the evaluation are similar as for TDL-C \SI{300}{ns} channel: \ac{gb} of \SI{30}{kHz} is sufficient to isolate the interference between different numerologies with \ac{fc-f-ofdm} and that \ac{fc-f-ofdm} provides the best link performance with this \ac{gb}. In TDL-C \SI{1000}{ns} channel the difference between \ac{fc-f-ofdm} and other waveform candidates grows even bigger, showing a \SI{2}{dB} difference to closest candidate at \ac{bler} target of \SI{10}{\%}. 

\subsection{Case 3: Asynchronous UL link performance}

In \Case{3} and \Case{4} simulations, the target signal uses \ac{mcs} 64-QAM, $R=1/2$, with \ac{ibo} of \SI{8}{dB} which leads to \SI{22.5}{dBm} \ac{pa} output power, and the interfering signals use \ac{mcs} QPSK, $R=1/2$ with \ac{ibo} of \SI{5.5}{dB} which leads to \SI{24.7}{dBm} \ac{pa} output power. These \ac{ibo} values were chosen in such a manner that the in-band emission mask and \acp{oobem} are fulfilled for the desired and interfering signals, assuming a 50 \ac{prb} maximum allocation in the channel and evaluating separately either desired or interfering signal at the channel edge. The interfering signal was chosen to have lower \ac{mcs} and lower \ac{ibo} because this maximizes the interference leakage from the interfering signal and can be considered as the worst case scenario.

% TDLC1000 perf
\begin{figure} 
  \centering
  \begin{subfigure}[t]{\figwidth}
    \centering    
    \includegraphics[angle=0,width=\FIGURE_WIDTH]{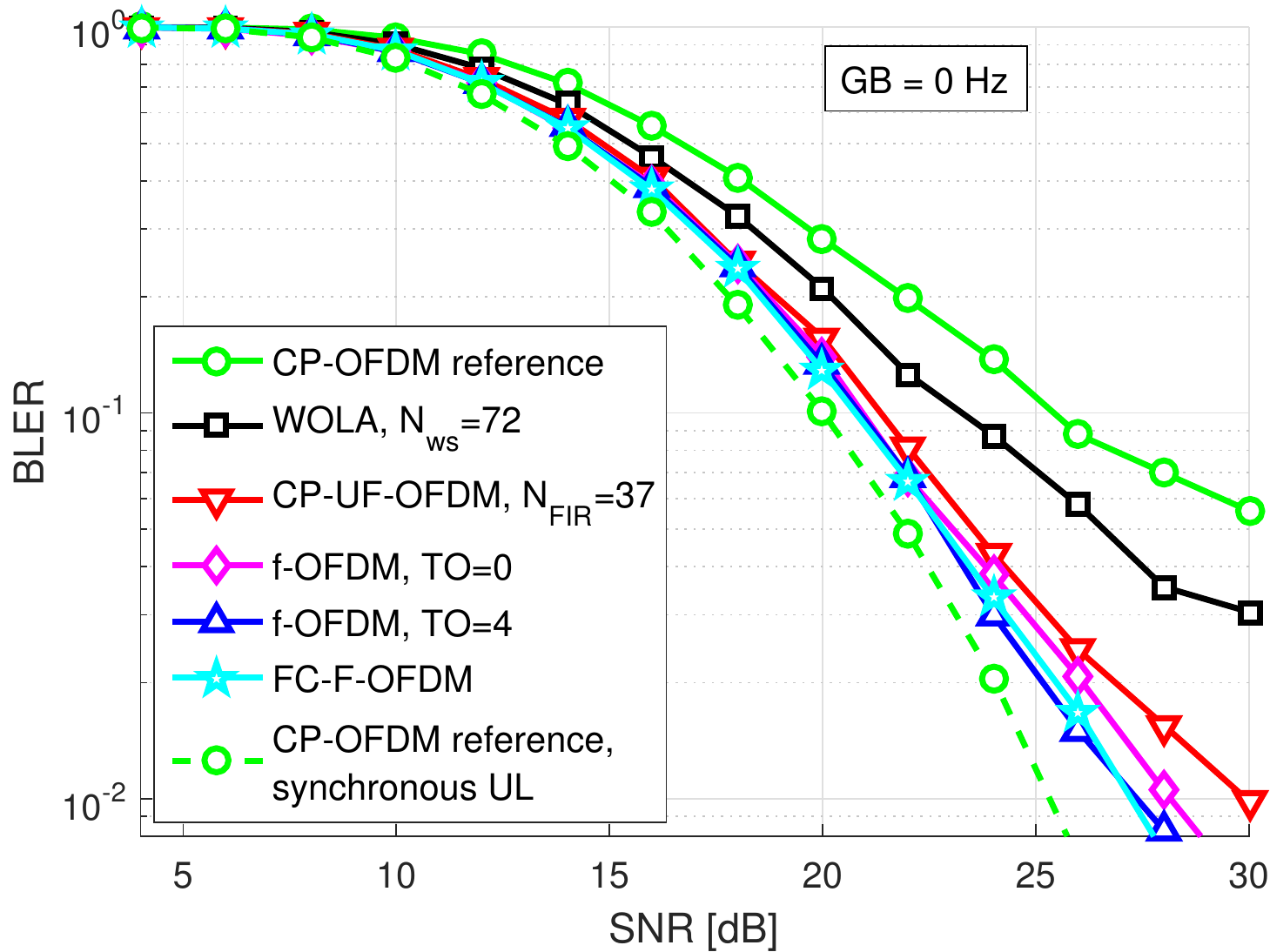}      
	\caption{}
    \vspace{1mm}
  \end{subfigure}\hfill 
  \begin{subfigure}[t]{\figwidth}
    \centering 
    \includegraphics[angle=0,width=\FIGURE_WIDTH]{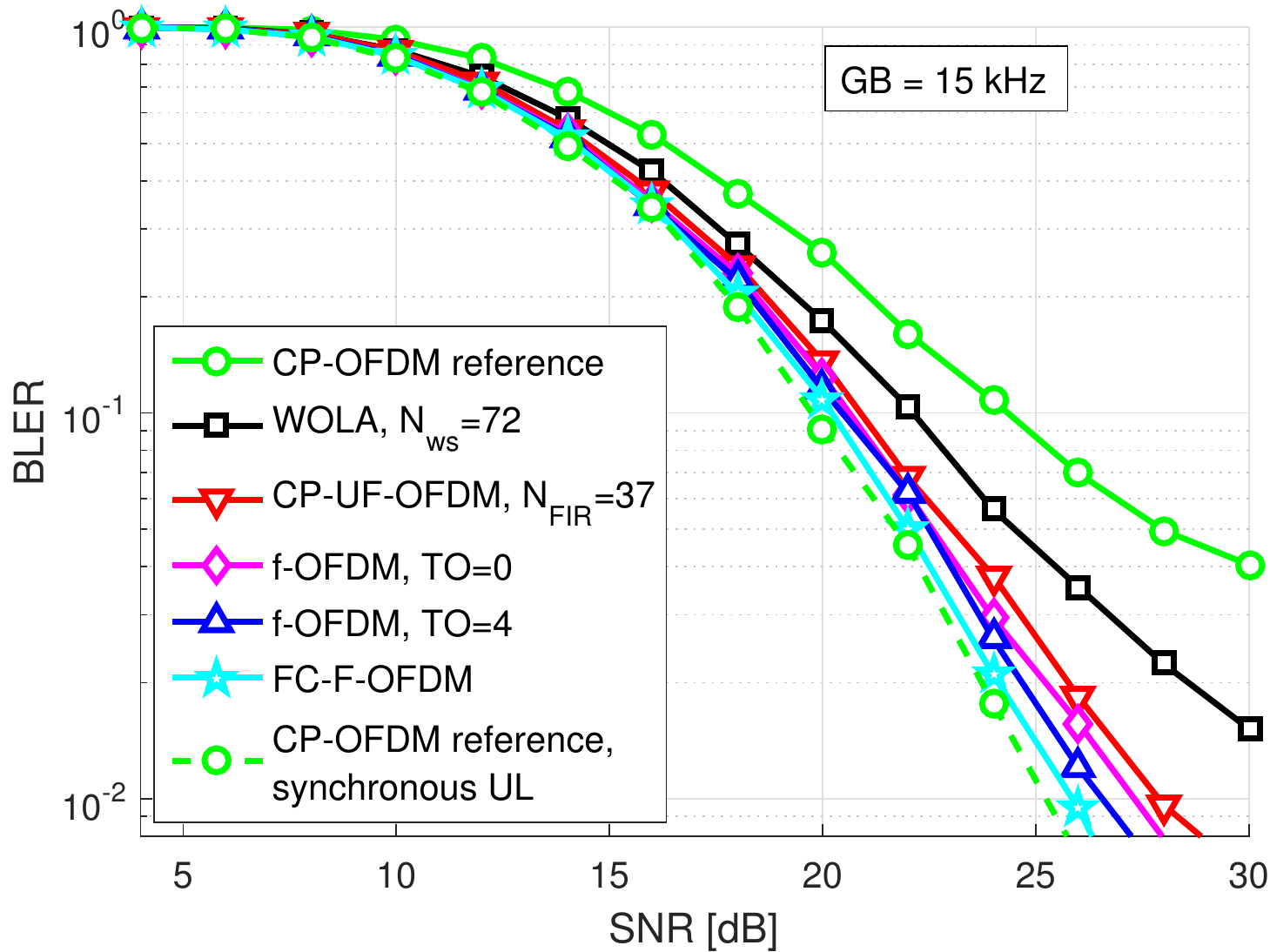}
    \caption{}  
  \end{subfigure}
  \caption{Performance comparison for \Case{3} asynchronous UL scenario in TDL-C \SI{1000}{ns} channel. The relative performance of different waveforms with (a) GB=\SI{0}{Hz} and (b) GB=\SI{15}{kHz} are shown.
  }
  \label{fig:Case3_TDLC1000_linkPerf}
\end{figure}

In Fig.~\ref{fig:Case3_TDLC1000_linkPerf}, the link performance in \Case{3} asynchronous \ac{ul} in TDL-C \SI{1000}{ns} channel with \ac{mcs} 64-QAM, $R=1/2$ is shown for (a) \acp{gb}=\SI{0}{Hz} and (b) \ac{gb}=\SI{15}{kHz} cases. In the case of \SI{15}{kHz} \ac{gb}, the \ac{fc-f-ofdm} achieves the synchronous \ac{ul} \ac{cp-ofdm} performance. Increasing the \ac{gb} from this does not considerably improve the performance of synchronous \ac{cp-ofdm} or asynchronous \ac{fc-f-ofdm}, as it only reduces the differences between the waveform candidates.

\subsection{Case 4: Mixed numerology UL link performance}

% case4 TDLC1000 perf
\begin{figure} 
  \centering
  \begin{subfigure}[b]{\figwidth}
    \centering      
    \includegraphics[angle=0,width=\FIGURE_WIDTH]{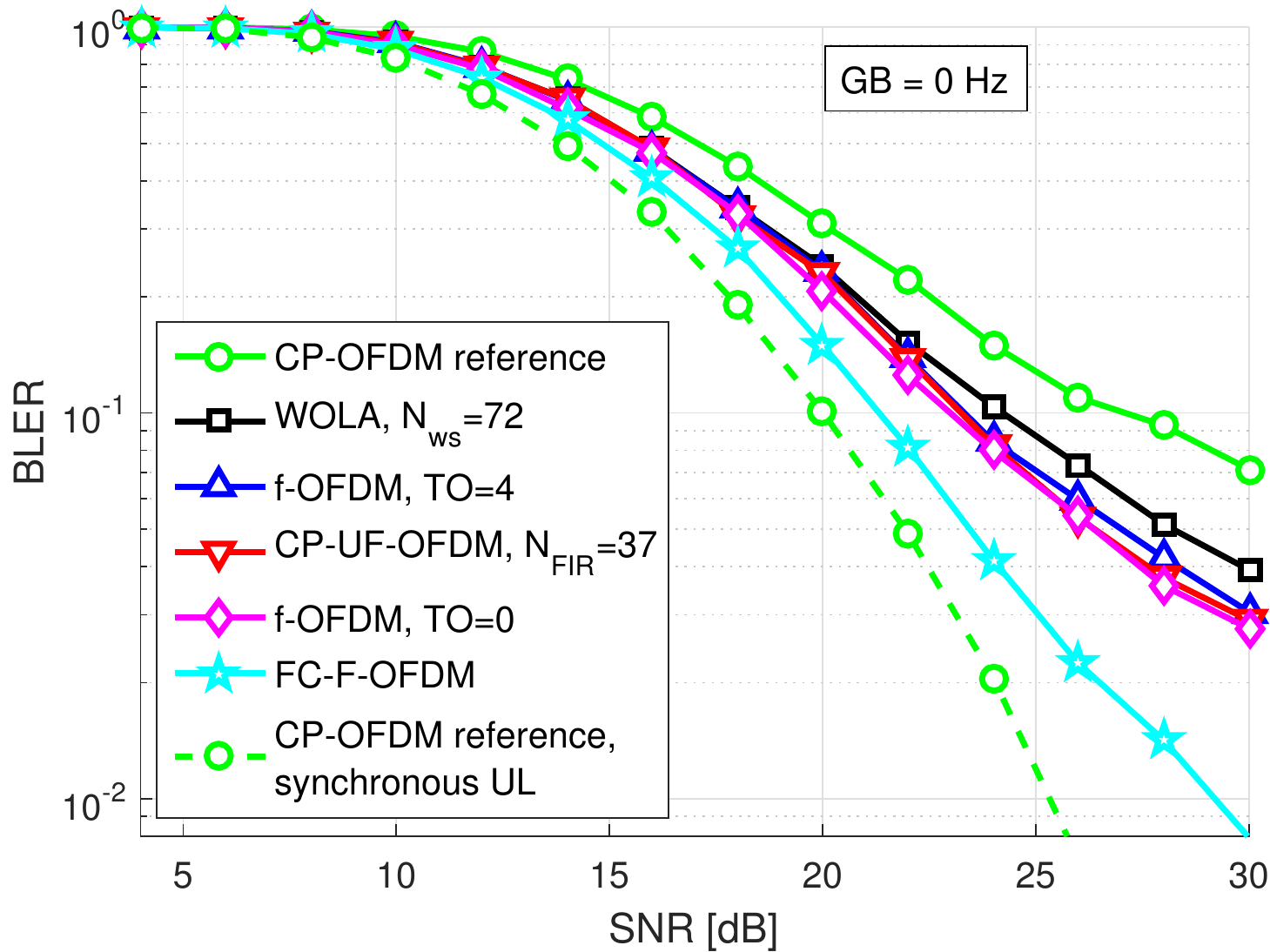}      
    \caption{} 
    \vspace{1mm}  
  \end{subfigure}\hfill
  \begin{subfigure}[b]{\figwidth}
    \centering 
    \includegraphics[angle=0,width=\FIGURE_WIDTH]{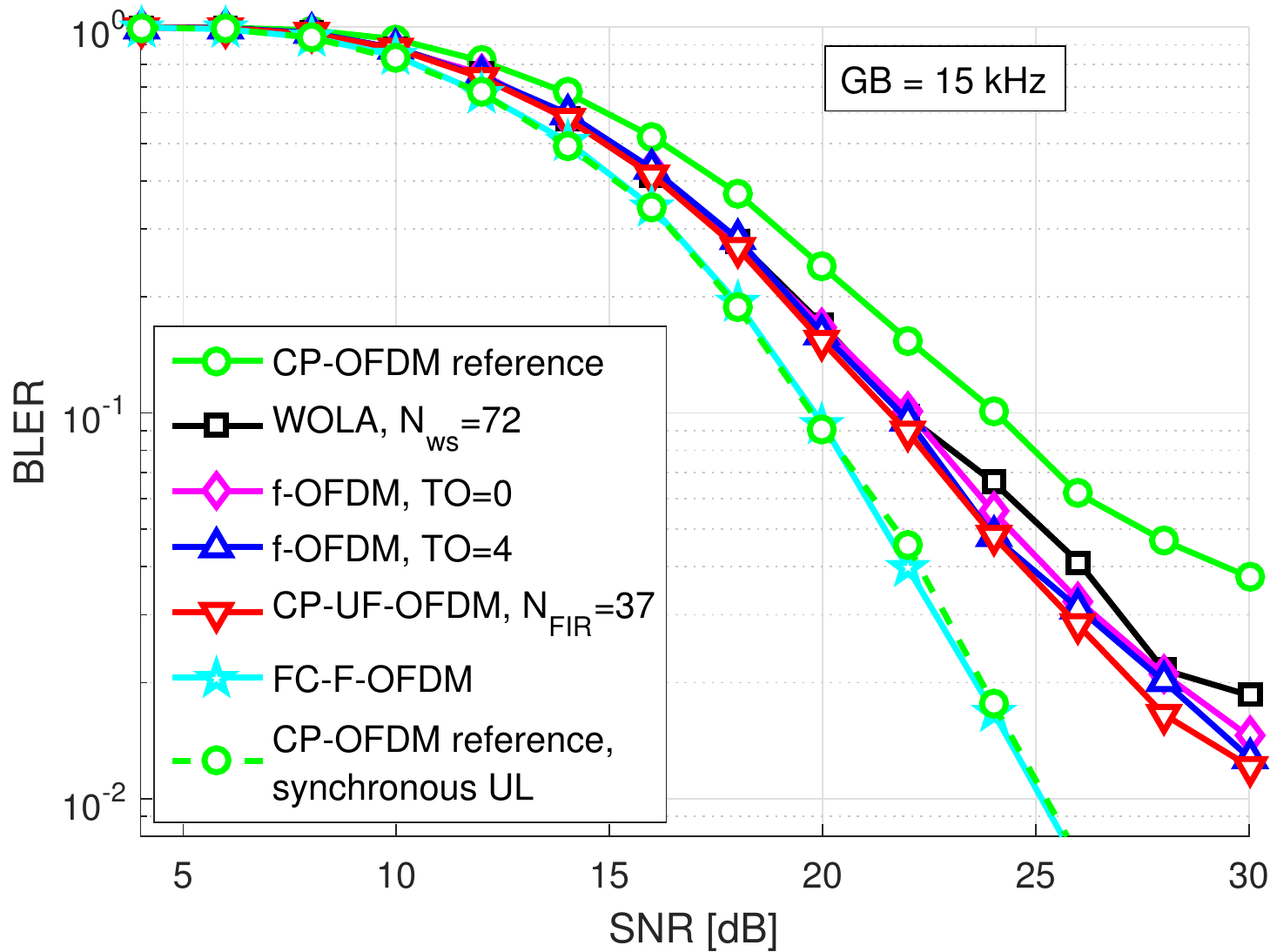}
    \caption{}  
  \end{subfigure}
  \caption{Performance comparison for \Case{4} mixed numerology \ac{ul} scenario in TDL-C \SI{1000}{ns} channel. The relative performance of different waveforms with (a) GB=\SI{0}{Hz} and (b) GB=\SI{15}{kHz} are shown.
  }
  \label{fig:Case4_TDLC1000_linkPerf}
\end{figure}

In Fig.~\ref{fig:Case4_TDLC1000_linkPerf}, the performance of different waveform candidates is shown for \Case{4} mixed numerology \ac{ul} scenario in TDL-C \SI{1000}{ns} channel with \ac{mcs} 64-QAM, $R=1/2$ for the desired signal. The results are very similar to \Case{3} shown in Fig.~\ref{fig:Case3_TDLC1000_linkPerf}. The \ac{cp-ofdm} reference in synchronous \ac{ul} is the one that was used in \Case{3} results to provide a realistic lower bound for the \ac{bler} performance. The link performance is given (a) without a \ac{gb} and (b) with \ac{gb}=\SI{15}{kHz}. In both cases, the given \ac{fc} parameterization provides the best link performance. With \acp{gb}=\SI{15}{kHz} the \ac{fc-f-ofdm} is able to achieve the synchronous \ac{ul} \ac{cp-ofdm} reference while the other waveform candidates lose approximately \SI{2}{dB} at \ac{bler} target of \SI{10}{\%}.

\subsection{FC Filtered DFT-spread-OFDM}
\label{subsec:FC-F-DFTs-OFDM}

\begin{figure} 
  \centering
  \begin{subfigure}{\figwidth}
    \centering 
    \includegraphics[angle=0,width=\FIGURE_WIDTH]{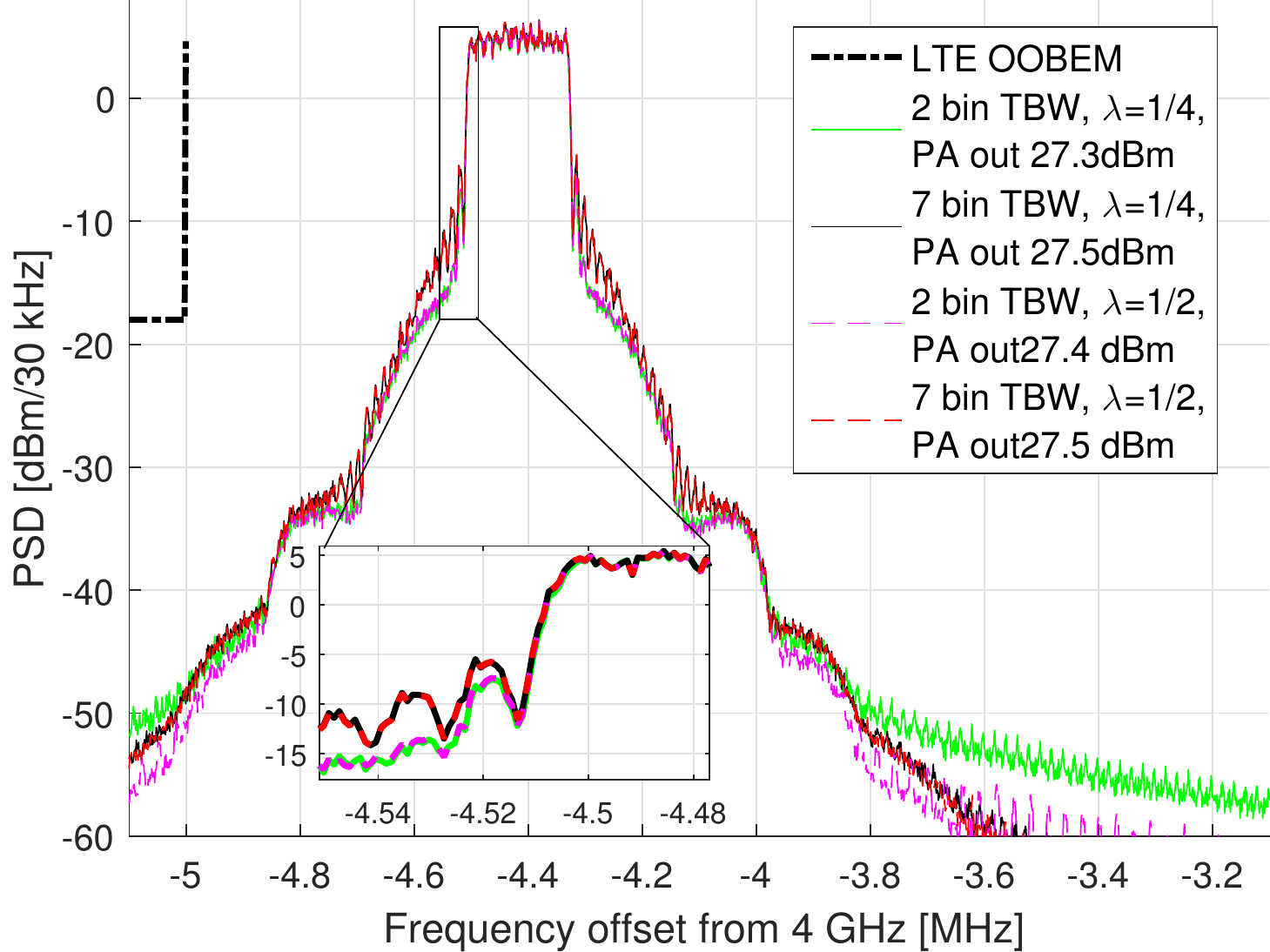}      
    \caption{}
    \vspace{3mm}
  \end{subfigure}\hfill
  \begin{subfigure}{\figwidth}  
    \centering 
    \includegraphics[angle=0,width=\FIGURE_WIDTH]{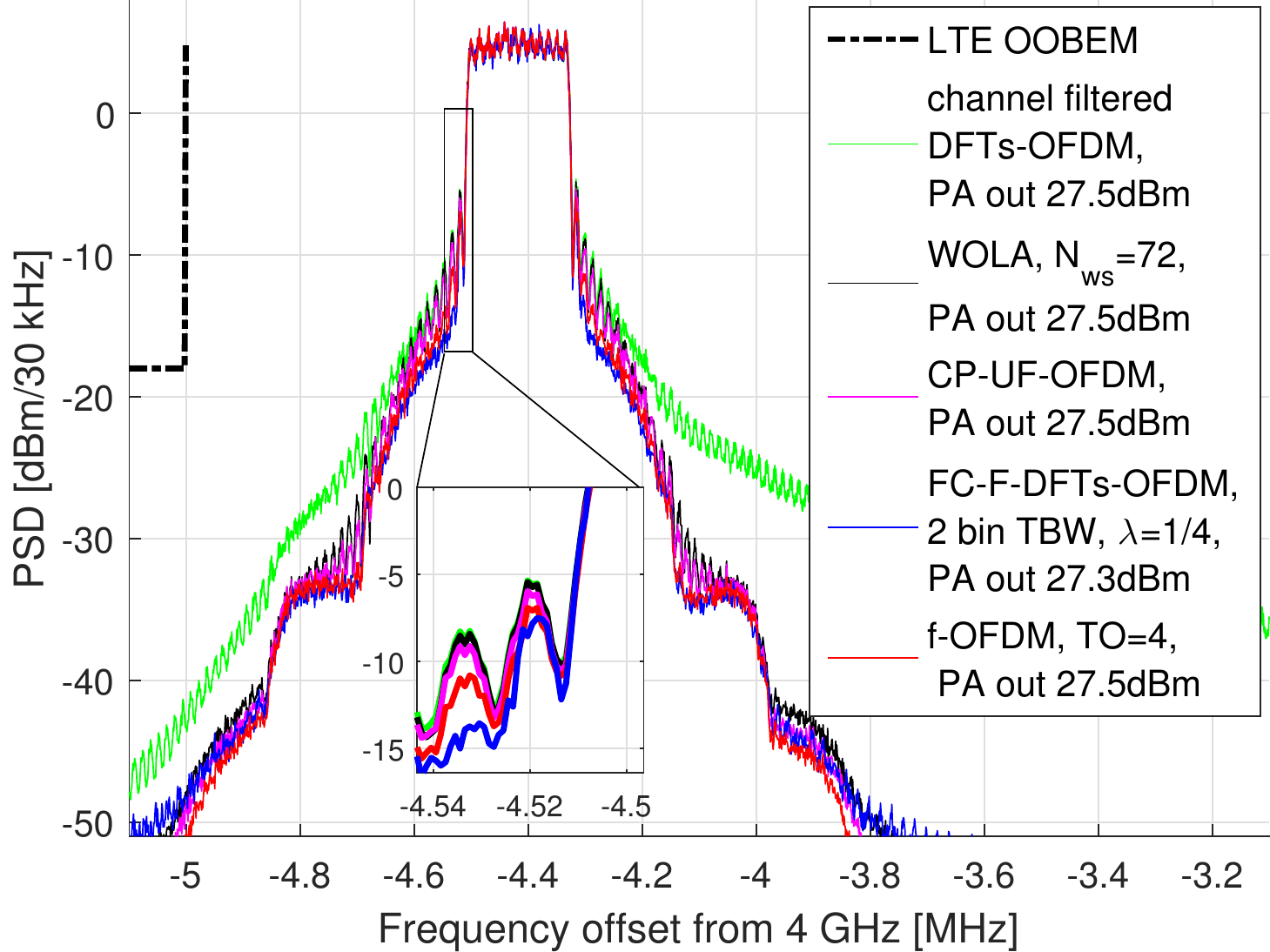}
    \caption{}   
  \end{subfigure}
  \caption{\acp{psd} and maximum \ac{pa} output powers for 1 \ac{prb} transmission with (a) \ac{fc} filtered \ac{dft}-spread-\ac{ofdm} signal and (b) channel filtered and subband-filtered or windowed \ac{dft-s-ofdm} signals.}
  \label{fig:1PRB_DFTs}
\end{figure}

In \cite{2016RAN1WayForwardOnULWaveforms}, it was agreed that \ac{dft}-spread-\ac{ofdm} is supported in \ac{ul} in coverage limited scenarios. Therefore, in addition to spectral confinement, the maximum achievable \ac{pa} output power is of great importance. In Fig. \ref{fig:1PRB_DFTs}, \acp{psd} and maximum \ac{pa} output powers are shown for \ac{fc} filtered \ac{dft}-spread-\ac{ofdm} signal, assuming a 1 \ac{prb} allocation within a carrier of 50 \acp{prb}. The transmitted signal is using \ac{mcs} QPSK, $R=1/2$, and the \ac{pa} maximum output power is searched by brute force simulations where the minimum input backoff is searched with \SI{0.1}{dB} step while fulfilling the in-band emission mask, \ac{oobem}, and \ac{evm} requirements. In \cite{3GPPTS36101}, the given \ac{evm} requirement for QPSK modulation is \SI{17.5}{\%}. Here, \ac{evm} target of \SI{12}{\%} was used for \ac{pa} induced \ac{evm}. This threshold was selected to leave sufficient headroom for other \ac{tx} imperfections, e.g., I/Q imbalance, phase noise, etc.

In Fig. \ref{fig:1PRB_DFTs} (a) different \acp{tbw} are compared with overlap factor $\lambda=1/2$ and $\lambda=1/4$. The first observation is that the actual parameterization of the \ac{fc} filtered \ac{dft-s-ofdm} has relatively small effect on the maximum \ac{pa} output power and on the overall frequency response after the polynomial \ac{pa} model. In the magnified subfigure the detailed differences between different \acp{tbw} are visible. The selection of larger \ac{tbw} will affect \ac{sblr} over the neighboring \ac{prb}, but otherwise the performance is dictated by the \ac{pa} induced spectral spreading. The observation that the lower overlap factor provides similar \ac{pa} output powers and spectral containment allows to consider the overlap ratio $\lambda=1/4$ for \ac{ue} to reduce the \ac{ue} implementation complexity at least in coverage limited scenarios. 

Furthermore, from Fig. \ref{fig:1PRB_DFTs}, it is clear that as we push to increase the bandwidth efficiency by increasing the number of \acp{prb} allocated in a certain channel, it is necessary to limit the maximum allowed \ac{pa} output power in the edge-most \acp{prb}. This is due to the significant spreading of the signal with small power backoffs, which would cause the presented examples to violate the \ac{lte} \ac{oobem} if maximum allocation size would be 54 or 55 \acp{prb}. Therefore, it is most likely that the higher bandwidth efficiency is first applied to \ac{dl}, where more linear \acp{pa} with efficient linearization and crest factor reduction algorithms are applied. As the \ac{ue} \ac{pa} technology evolves to support higher \ac{mcs} for \ac{ul}, the improved linearity will also allow larger bandwidth efficiency with the given coverage targets. Alternatively, only \acp{ue} using a high \ac{mcs} could be scheduled to channel edges to fulfill the \ac{oobem}. In this case, the passband \ac{mse} of the used subband filtering scheme is critical not to limit the set of usable \acp{mcs}.

In Fig. \ref{fig:1PRB_DFTs} (b) the performance of \ac{fc} filtered \ac{dft}-spread-\ac{ofdm} with 2 bin \ac{tbw} is compared against a reference channel filtered \ac{dft-s-ofdm} and other subband constrained \ac{dft-s-ofdm} candidates. In current \ac{lte} \acp{ue}, channel filtering or windowing is required to achieve the \ac{oobem}. From this example, a clear benefit of subband filtering on the in-band spectral containment is observed with respect to the channel filtered case. All of the waveform candidates provide very similar performance, but the magnified subfigure shows that the used \ac{fc} filtered \ac{dft-s-ofdm} has the lowest leakage power nearby the allocation edge. In general, the maximum \ac{pa} output power is approximately equal with all different subband processed signals allowing similar coverage as with channel filtering.

\subsection{Complexity vs. Performance Comparison}
%%%%%%%%%%%%%%%%%%%%%%%%%%%%%%%%%%%%%%%%%%%%%%%%%%
A basic computational complexity metric, the number of multiplications per QAM symbol, is shown in Table~\ref{tab:results} for different subband configurations with \ac{fc-fb} and time-domain filtering alternatives. The \ac{fc-fb} complexity evaluation is based on the principles explained in~\cite{C:Renfors2015:fc-f-ofdm}.  With \SI{50}{\%} overlap, the multiplication rates are \SIrange{2.4}{5.4}{} times the multiplication rates of a basic \ac{cp-ofdm} transmitter or receiver, depending on the \ac{prb} configuration. Reducing the overlap factor from \SI{50}{\%} to \SI{25}{\%} reduces the multiplication rate by about \SI{30}{\%}.
 
Effective time-domain implementation of an \ac{F-ofdm} transmitter includes the steps listed below. 
\begin{enumerate}
\item Length $L$ \ac{ifft} taking $(L(\log_2(L)-3)+4)/N_\text{SYMB}$ multiplications per symbol using the split-radix algorithm.  $N_\text{SYMB}$ is the number of QAM symbols per \ac{ofdm} symbol in the subband.
\item Inserting \ac{cp}
\item Interpolating lowpass filter with filter length of approximately $N_\text{FIR}L/N$, where $N_\text{FIR}$ is the required length at the output sampling rate of \SI{15.36}{MHz}. This is because \ac{fir} filter order is inversely proportional to the relative transition bandwidth. Making use of coefficient symmetry and noting that two real filters are needed (for in-phase and quadrature components), the multiplication rate becomes: $N_\text{FIR}L (L+L_\text{CP})/(N N_\text{SYMB})$.
\item Mixing at the output sampling rate taking $4(N+L_\text{CP}N/L)/N_\text{SYMB}$ multiplications per symbol.
\end{enumerate}

\begin{table*}[t]
  \caption{FC based \ac{rx} or \ac{tx} complexity for different subband configurations with overlap factors of $\lambda=1/4$ and $\lambda=1/2$ in comparison with time-domain subband filtering approaches. }
  \label{tab:results}
  \centering
  \footnotesize{
   \begin{tabular}{ccS[table-format=2.1]ccc}
      \toprule
      \multicolumn{1}{c}{} & 
      \multicolumn{3}{c}{FC-F-OFDM} &
      \multicolumn{1}{c}{CP-UF-OFDM, $N_\text{FIR}=73$} &
      \multicolumn{1}{c}{f-OFDM, $N_\text{FIR}=512$} \\       
      \cmidrule(lr){2-4}\cmidrule(lr){5-5}\cmidrule(lr){6-6}
      \multicolumn{1}{c}{No. active} & 
      \multicolumn{1}{c}{Overlap} & 
      \multicolumn{1}{c}{Complexity} & 
      \multicolumn{1}{c}{Complexity} &
      \multicolumn{1}{c}{Complexity} & 
      \multicolumn{1}{c}{Complexity} \\ 
      %\multicolumn{2}{c}{Time-domain FIR} \\
      \multicolumn{1}{c}{subcarriers} & 
      \multicolumn{1}{c}{factor $\lambda$} &  
      \multicolumn{1}{c}{(muls/symbol)} & 
      \multicolumn{1}{c}{relative to \ac{ofdm}} &
      \multicolumn{1}{c}{(muls/symbol)} & 
      \multicolumn{1}{c}{(muls/symbol)} 
      \\
      \midrule
      \multirow{2}[2]{*}{1\,\ac{prb}}
      & $1/2$ & 1441.83 & $\times$2.41 &
      \multirow{2}[2]{*}{512} &
      \multirow{2}[2]{*}{1139} \\
      & $1/4$ &  979.83 & $\times$1.64 \\
      \midrule
      \multirow{2}[2]{*}{4\,\acp{prb}}
      & $1/2$ &  360.46 & $\times$2.41 &
      \multirow{2}[2]{*}{128} &
      \multirow{2}[2]{*}{285} \\
      & $1/4$ &  244.96 & $\times$1.64 \\
      \midrule
      \multirow{2}[2]{*}{50\,\acp{prb}}
      & $1/2$ &   64.11 & $\times$5.36 &
      \multirow{2}[2]{*}{133} &
      \multirow{2}[2]{*}{935} \\
      & $1/4$ &   46.89 & $\times$3.92 \\
      \midrule
      \multirow{2}[2]{*}{12$\times$4\,\acp{prb}}
      & $1/2$ &   61.51 & $\times$4.94 &
      \multirow{2}[2]{*}{128} &
      \multirow{2}[2]{*}{285} \\
      & $1/4$ &   44.75 & $\times$3.59 \\
      \bottomrule
    \end{tabular}}  
\end{table*}

When the transmitted signal is composed of multiple subbands, the subbands are processed independently of each other, and in case of equal subband widths, the overall multiplication rate (per symbol) is equal to that of the single subband case. 

It can be concluded that time-domain implementation is effective in case of single or few narrow subbands, but for high number of subbands, or wide subbands, the \ac{fc-f-ofdm} scheme is clearly more effective in terms of multiplication rate. Notably, \ac{fc-f-ofdm} can provide very good spectrum localization with low complexity compared with corresponding time-domain realizations of \ac{F-ofdm}.  Furthermore, the proposed approach facilitates direct optimization of the frequency-domain window coefficients. Also high flexibility is achieved by a very small number of filter coefficients, determined by the transition bandwidth. In practice, it may be useful to have a few different sets of coefficients for different cases, e.g., wider transition band for the outer edges of the channel to achieve high \ac{oob} attenuation. Different transition band weight masks can also be combined in asymmetric manner. We have also seen that \ac{fc-f-ofdm} gives in many cases the best, and practically always at least equally good \ac{evm} and in-band emission performance among the \ac{F-ofdm} schemes. 

On the other hand, the \ac{wola} scheme needs only minor increase in complexity compared to the basic \ac{cp-ofdm} and it has very good \ac{evm} performance. However, \ac{wola} parametrization compatible with tentative \ac{5g-nr} numerology provides only limited improvement in the spectrum localization, and it is probably not sufficient for all intended \ac{5g} scenarios.  Also enhanced time-domain windowing schemes are available in the literature. The edge windowing idea was originally proposed in \cite{C:Sahin2011_edgewin} and later in \cite{2016RAN1_multiwin} for \ac{5g}. The main benefit of this approach is that subcarriers in the center of a subband enjoy from almost full effective \ac{cp} length, while long window transitions are applied only for subcarriers close to the subband edges.  In our simulation examples, the overall \ac{cp-ofdm} symbol duration is not extended in the \ac{wola} case, which leads to degraded performance with high delay spread channels, like TDL-C \SI{1000}{ns}. By using edge windowing, the multipath delay-spread tolerance could be improved for the center subcarriers in case of wide subcarrier allocations, while the spectral containment results would experience relatively small degradation compared to the presented results. On the other hand, implementation complexity would be significantly increased, and edge windowing would also add some complexity to the resource allocation function~\cite{C:Sahin2011_edgewin}.

\section{Conclusions}
%%%%%%%%%%%%%%%%%%%%%
We have proposed a straightforward and efficient technique for designing and optimizing the \ac{fc-f-ofdm} based physical layer waveform processing in \ac{5g} mobile cellular radio networks, building on the concept of subband filtering on top of the baseline \ac{cp-ofdm} waveform. For given transition bandwidth, different tradeoffs between the passband \ac{evm} performance and in-band emissions, in terms of the subband leakage ratio, can be obtained in a flexible manner. The current optimization framework uses the minimum stopband attenuation value as an optimization constraint, allowing to control the band-limitation characteristics of the subband signals in an efficient manner. In general, good match was demonstrated between the simplified system model used in the optimization and actual simulated radio link performance in tentative \ac{5g-nr} test scenarios, incorporating different levels of asynchronism between \ac{ul} users, different numerologies for different subbands, as well as different radio channel propagation characteristics at sub-\SI{6}{GHz} bands.  

The obtained results show that the optimization based design of \ac{fc-f-ofdm} physical layer achieves the \ac{evm} requirements and often significantly exceeds the \ac{sblr} and the overall radio link performance of the existing time-domain filtering based realizations of \ac{F-ofdm} or windowing based \ac{cp-ofdm}. In general, time-domain filtering based schemes are computationally effective for cases with single narrow filtering subband (e.g. 1--4 \acp{prb}), so they could be particularly useful for low-rate user devices. In full-band cases (e.g. 50 \acp{prb} in a \SI{10}{MHz} \ac{lte} channel), however, both with single or multiple subbands, the proposed \ac{fc-f-ofdm} has significantly lower complexity. It is able to provide effective filtering for the whole carrier with narrow transition bands, allowing, as a concrete example, to increase the number of \acp{prb} from 50 to 55 in case of a \SI{10}{MHz} carrier. Furthermore, \ac{fc-f-ofdm} has the flexibility to construct arbitrary subband configurations, as groups of \acp{prb}, with minimal coefficient storage requirements, which provides substantial implementation benefits. Our future work will focus on revisiting the formulation of the optimization framework for \ac{fc-f-ofdm}, using an explicit constraint on the adjacent subband \ac{sblr}, instead of the currently used minimum stopband attenuation requirement.

\bibliographystyle{IEEEtran} 
\bibliography{jour_short,conf_short,References} 
 
\begin{IEEEbiography}[{\includegraphics[width=1in,height=1.25in,clip,keepaspectratio]{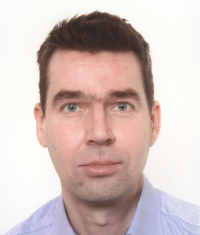}}]{\href{https://fi.linkedin.com/in/ylikaakinen}{Juha Yli-Kaakinen}}%
received the degree of Diploma Engineer in electrical engineering and Doctor of Technology (with honors) from the Tampere University of Technology (TUT), Tampere, Finland, in 1998 and 2002, respectively.

Since 1995, he has held various research positions at TUT. His research interests are in digital signal processing, especially in digital filter and filter bank optimization for communication systems and VLSI implementations.
 
% His main research areas include signal processing algorithms 
\end{IEEEbiography}

\begin{IEEEbiography}[{\includegraphics[width=1in,height=1.25in,clip,keepaspectratio]{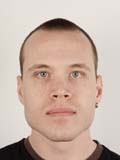}}]{Toni Levanen}%
received the M.Sc. (with honours) and D.Sc. degrees in Digital communications from Tampere University of Technology (TUT), Finland, in 2007 and 2014, respectively. He is currently a researcher with the Laboratory of Electronics and Communications Engineering, TUT. In addition to his contributions in academic research, he has worked in industry on wide variety of development and research projects for communications systems. His current research interests include physical layer design for 5G NR, traffic and interference modeling in 5G small cells and millimeter-wave communications, and high-mobility support in ultra-dense small cell networks.
\end{IEEEbiography}

\begin{IEEEbiography}[{\includegraphics[width=1in,height=1.25in,clip,keepaspectratio]{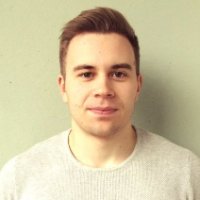}}]{Sami Valkonen}%
received his B.Sc. in 2015 from Tampere University of Technology. He is currently pursuing his M.Sc. degree on wireless mobile communications with research interest in time-domain windowed and subband-filtered CP-OFDM waveforms for 5G new radio.
\end{IEEEbiography}

\begin{IEEEbiography}[{\includegraphics[width=1in,height=1.25in,clip,keepaspectratio]{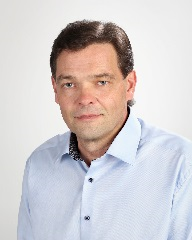}}]{Kari Pajukoski}%
received his B.S.E.E. degree from the Oulu University of Applied Sciences in 1992. He is a Fellow with the Nokia Bell Labs. He has a broad experience from cellular standardization, link and system simulation, and algorithm development for products. He has more than 100 issued US patents, from which more than 50 have been declared “standards essential patents”. He is author or co-author of more than 300 standards contributions and 30 publications, including conference proceedings, journal contributions, and book chapters. 
\end{IEEEbiography}

\begin{IEEEbiography}[{\includegraphics[width=1in,height=1.25in,clip,keepaspectratio]{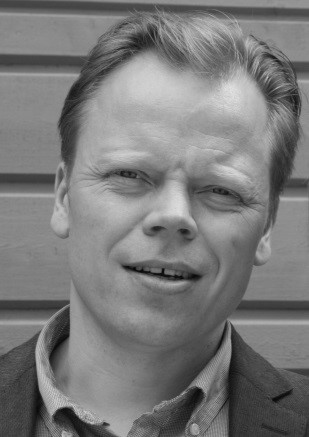}}]{Juho Pirskanen}%
received his M.Sc. in Electrical Engineering from Tampere University of Technology in 2000. Since then, he has worked in several different positions on wireless radio research, technology development, and standardization at Nokia Bell Labs, Nokia Networks, Nokia, Renesas Mobile and Broadcom Corporation. He has participated actively standardization of 3G, HSPA, LTE, 5G and WLAN technologies by leading the standardization delegation in 3GPP and in IEEE802.11. His latest research activities include 3GPP 5G NR physical layer, control layer, and network design.
\end{IEEEbiography}

\begin{IEEEbiography}[{\includegraphics[width=1in,height=1.25in,clip,keepaspectratio]{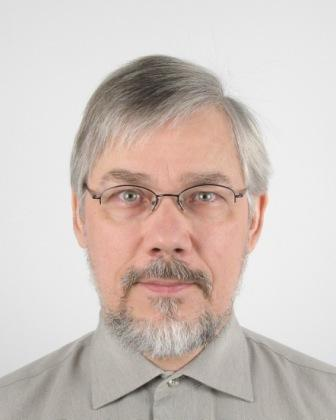}}]{\href{http://www.cs.tut.fi/~mr/}{Markku Renfors}}%
(S'77--M'82--SM'90--F'08) received the D.Tech.~degree from Tampere University of Technology (TUT), Tampere, Finland, in 1982. Since 1992, he has been a Professor with the Department of Electronics and Communications Engineering, TUT, where he was the Head from 1992 to 2010. His research interests include filter bank based multicarrier systems and signal processing algorithms for flexible communications receivers and transmitters. Dr. Renfors was a corecipient of the Guillemin Cauer Award (together with T. Saram\"aki) from the IEEE Circuits and Systems Society in 1987.
\end{IEEEbiography}

\begin{IEEEbiography}[{\includegraphics[width=1in,height=1.25in,clip,keepaspectratio]{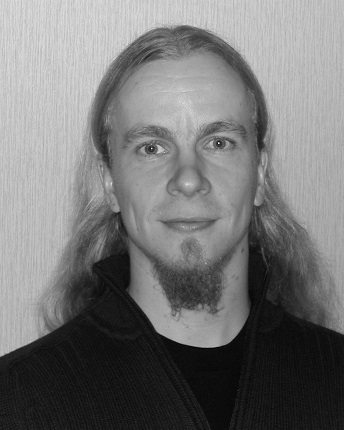}}]{\href{http://www.cs.tut.fi/~valkama/}{Mikko Valkama}}%
was born in Pirkkala, Finland, on November 27, 1975. He received the M.Sc. and Ph.D. Degrees (both with honours) in electrical engineering (EE) from Tampere University of Technology (TUT), Finland, in 2000 and 2001, respectively. In 2002, he received the Best Ph.D. Thesis -award by the Finnish Academy of Science and Letters for his dissertation entitled "Advanced I/Q signal processing for wideband receivers: Models and algorithms". In 2003, he was working as a visiting researcher with the Communications Systems and Signal Processing Institute at SDSU, San Diego, CA. Currently, he is a Full Professor and Department Vice-Head at the Department of Electronics and Communications Engineering at TUT, Finland. He has been involved in organizing conferences, like the IEEE SPAWC'07 (Publications Chair) held in Helsinki, Finland. His general research interests include communications signal processing, estimation and detection techniques, signal processing algorithms for software defined flexible radios, cognitive radio, full-duplex radio, radio localization, 5G mobile cellular radio, digital transmission techniques such as different variants of multicarrier modulation methods and OFDM, and radio resource management for ad-hoc and mobile networks.
\end{IEEEbiography}

% Acronyms that won't appear in the printed list
%
% Acronyms for the printed list
% Rewritten using lowercase
\begin{acronym}[FBMC/OQAM2]
\acro{3gpp}[3GPP]{3rd Generation Partnership Project}
\acro{5g}[5G]{5th generation}
\acro{5g-nr}[5G-NR]{5th generation new radio}
\acro{aclr}[ACLR]{adjacent channel leakage ratio}
\acro{afb}[AFB]{analysis filter bank} 
\acro{bler}[BLER]{block error rate}
\acro{bs}[BS]{base station}
\acro{cb-fmt}[CB-FMT]{cyclic block-filtered multitone}
\acro{cp-ofdm}[CP-OFDM]{cyclic prefix orthogonal frequency-division multiplexing}
\acro{cp}[CP]{cyclic prefix} 
\acro{dft}[DFT]{discrete Fourier transform} 
\acro{dft-s-ofdm}[DFT-s-OFDM]{DFT-spread-OFDM}
\acro{dl}[DL]{downlink} 
\acro{evm}[EVM]{error vector magnitude}
\acro{e-utra}[E-UTRA]{evolved UMTS terrestrial radio access}
\acro{F-ofdm}[F-OFDM]{subband filtered CP-OFDM}
\acro{f-ofdm}[f-OFDM]{filtered-OFDM}
\acro{fbmc-coqam}[FBMC-COQAM]{filterbank multicarrier with circular offset-QAM}
\acro{fbmc/oqam}[FBMC/OQAM]{filter bank multicarrier with offset-QAM subcarrier modulation} 
\acro{fbmc}[FBMC]{filter bank multicarrier}
\acro{fb}[FB]{filter bank}
\acro{fc-fb}[FC-FB]{fast-convolution filter bank}
\acro{fc}[FC]{fast-convolution}
\acro{fc-f-ofdm}[FC-F-OFDM]{FC-based F-OFDM}
\acro{fft}[FFT]{fast Fourier transform} 
\acro{fir}[FIR]{finite impulse response}
\acro{fmt}[FMT]{filtered multitone} 
\acro{gb}[GB]{guard band}
\acro{gfdm}[GFDM]{generalized frequency-division multiplexing}
\acro{ibe}[IBE]{in-band emission}
\acro{ibi}[IBI]{in-band interference}
\acro{oob}[OOB]{out-of-band}
\acro{oobem}[OOBEM]{out-of-band emission mask}
\acro{ici}[ICI]{inter-carrier interference}
\acro{idft}[IDFT]{inverse discrete Fourier transform}
\acro{ifft}[IFFT]{inverse fast Fourier transform}
\acro{isi}[ISI]{inter-symbol interference}
\acro{lpsv}[LPSV]{linear periodically shift variant} 
\acro{lptv}[LPTV]{linear periodically time-varying} 
\acro{lte}[LTE]{long-term evolution}
\acro{mcm}[MCM]{multicarrier modulation}
\acro{mc}[MC]{multicarrier}
\acro{mcs}[MCS]{modulation and coding scheme}
\acro{mse}[MSE]{mean-squared error} 
\acro{npr}[NPR]{near perfect reconstruction}
\acro{ofdm}[OFDM]{orthogonal frequency-division multiplexing} \acro{ofdma}[OFDMA]{orthogonal frequency-division multiple access} 
\acro{oqam}[OQAM]{offset quadrature amplitude modulation}
\acro{prb}[PRB]{physical resource block}
\acro{rbg}[RBG]{resource block group}
\acro{pa}[PA]{power amplifier} 
\acro{pr}[PR]{perfect reconstruction}
\acro{psd}[PSD]{power spectral density} 
\acro{qpsk}[QPSK]{quadrature phase-shift keying}
\acro{rb}[RB]{resource block}
\acro{rc}[RC]{raised cosine} 
\acro{rms}[RMS]{root mean squared \acroextra{[error]}}
\acro{rrc}[RRC]{square root raised cosine}
\acro{rx}[RX]{receiver}
\acro{sc}[SC]{single-carrier}
\acro{sc-fdma}[SC-FDMA]{single-carrier frequency-division multiple access} 
\acro{scs}[SCS]{subcarrier spacing}
\acro{sdr}[SDR]{software defined radio}
\acro{sfb}[SFB]{synthesis filter bank}
\acro{tbw}[TBW]{transition-band width}
\acro{to}[TO]{tone offset}
\acro{tmux}[TMUX]{transmultiplexer}
\acro{tsg}[TSG]{technical specification group}
\acro{ran}[RAN]{radio access network}
\acro{tx}[TX]{transmitter} 
\acro{ue}[UE]{user equipment} 
\acro{wola}[WOLA]{weighted overlap-add} 
\acro{uf-ofdm}[UF-OFDM]{universal filtered OFDM}
\acro{ul}[UL]{uplink} 
\acro{zp}[ZP]{zero prefix} 
\acro{wola}[WOLA]{windowed overlap-and-add} 
\acro{sblr}[SBLR]{subband leakage ratio}
 \acro{ibo}[IBO]{input back-off}
\acro{3gpp}[3GPP]{3rd Generation Partnership Project}
\acro{4g}[4G]{4th Generation} 
\acro{adsl}[ADSL]{Asymmetric Digital Subscriber Line}
\acro{af}[AF]{Amplify-and-Forward} 
\acro{am/am}[AM/AM]{Amplitude Modulation/Amplitude
Modulation \acroextra{[NL PA models]}}
\acro{am/pm}[AM/PM]{Amplitude Modulation/Phase Modulation
\acroextra{[NL PA models]}} \acro{amr}[AMR]{Adaptive Multi-Rate}
\acro{ap}[AP]{Access Point} \acro{app}[APP]{A Posteriori
Probability} \acro{awgn}[AWGN]{Additive White Gaussian Noise}
\acro{bcjr}[BCJR]{Bahl-Cocke-Jelinek-Raviv \acroextra{algorithm}}
\acro{ber}[BER]{Bit Error Rate} \acro{bicm}[BICM]{Bit-Interleaved
Coded Modulation} \acro{blast}[BLAST]{Bell Labs Layered Space Time
\acroextra{[code]}} 
\acro{b-pmr}[B-PMR]{Broadband PMR} \acro{bpsk}[BPSK]{Binary
Phase-Shift Keying} 
\acro{cazac}[CAZAC]{Constant Amplitude Zero Auto-Correlation}
\acro{ccc}[CCC]{Common Control Channel}
\acro{ccdf}[CCDF]{Complementary Cumulative Distribution Function}
\acro{cdf}[CDF]{Cumulative Distribution Function}
\acro{cdma}[CDMA]{Code-Division Multiple Access}
\acro{cfo}[CFO]{Carrier Frequency Offset} \acro{cfr}[CFR]{Channel
Frequency Response} \acro{ch}[CH]{Cluster Head}
\acro{cir}[CIR]{Channel Impulse Response} \acro{cma}[CMA]{Constant
Modulus Algorithm} \acro{cna}[CNA]{Constant Norm Algorithm}
\acro{cqi}[CQI]{Channel Quality Indicator} \acro{cr}[CR]{Cognitive
Radio} \acro{crlb}[CRLB]{Cram\'er-Rao Lower Bound}
\acro{crn}[CRN]{Cognitive Radio Network}
\acro{crs}[CRS]{Cell-specific Reference Signal}
\acro{csi}[CSI]{Channel State Information}
\acro{csir}[CSIR]{Channel State Information at the Receiver}
\acro{csit}[CSIT]{Channel State Information at the Transmitter}
\acro{d2d}[D2D]{Device-to-Device} \acro{dc}[DC]{Direct Current}
\acro{df}[DF]{Decode-and-Forward} \acro{dfe}[DFE]{Decision
Feedback Equalizer} 
\acro{dmo}[DMO]{Direct Mode Operation}
\acro{dmrs}[DMRS]{DeModulation Reference Signals}
\acro{dsa}[DSA]{Dynamic Spectrum Access} \acro{dzt}[DZT]{Discrete
Zak Transform} \acro{ed}[ED]{Energy Detector}
\acro{egf}[EGF]{Extended Gaussian Function}
\acro{em}[EM]{Expectation Maximization} \acro{emse}[EMSE]{Excess
Mean Square Error}
%\acro{emphatic}[EMPhAtiC]{FP7-ICT project}
\acro{epa}[EPA]{Extended Pedestrian-A \acroextra{[channel model]}}
\acro{etsi}[ETSI]{European Telecommunications Standards Institute}
\acro{eva}[EVA]{Extended Vehicular-A \acroextra{[channel model]}}
\acro{fb-sc}[FB-SC]{FilterBank Single-Carrier} 
\acro{fdma}[FDMA]{Frequency-Division Multiple Access}
\acro{fec}[FEC]{Forward Error Correction} 
\acro{flo}[FLO]{Frequency-Limited Orthogonal}
\acro{fpga}[FPGA]{Field Programmable Gate Array} 
\acro{fs-fbmc}[FS-FBMC]{Frequency Sampled FBMC-OQAM} 
\acro{ft}[FT]{Fourier Transform}
\acro{glrt}[GLRT]{Generalized Likelihood Ratio Test}
\acro{gmsk}[GMSK]{Gaussian Minimum-Shift Keying}
\acro{hpa}[HPA]{High Power Ampliﬁer}
\acro{iid}[i.i.d.]{independent and identically distributed}
\acro{i/q}[I/Q]{In-phase/Quadrature \acroextra{[complex data
signal components]}} \acro{iam}[IAM]{Interference Approximation
Method}
\acro{iota}[IOTA]{Isotropic Orthogonal Transform Algorithm}
\acro{itu}[ITU]{International Telecommunication Union}
\acro{itu-r}[ITU-R]{International Telecommunication Union
Radiocommunication \acroextra{sector}}
\acro{kkt}[KKT]{Karush-K\"uhn-Tucker} \acro{kpi}[KPI]{Key
Performance Indicator} \acro{le}[LE]{Linear Equalizer}
\acro{llr}[LLR]{Log-Likelihood Ratio} \acro{lmmse}[LMMSE]{Linear
Minimum Mean Squared Error} \acro{lms}[LMS]{Least Mean Squares}
\acro{ls}[LS]{Least Squares} 
\acro{lte-a}[LTE-A]{Long-Term Evolution-Advanced}
\acro{lut}[LUT]{Look Up Table} \acro{ma}[MA]{Multiple Access Relay
Channel} \acro{mac}[MAC]{Medium Access Control}
\acro{mai}[MAI]{Multiple Access Interference}
\acro{map}[MAP]{Maximum A Posteriori} 
\acro{mer}[MER]{Message Error Rate} \acro{mf}[MF]{Matched Filter}
\acro{mgf}[MGF]{Moment Generating Function}
\acro{mimo}[MIMO]{Multiple-Input Multiple-Output}
\acro{miso}[MISO]{Multiple-Input Single-Output}
\acro{ml}[ML]{Maximum Likelihood} \acro{mlse}[MLSE]{Maximum
Likelihood Sequence Estimation} \acro{mma}[MMA]{Multi-Modulus
Algorithm} \acro{mmse}[MMSE]{Minimum Mean-Squared Error}
\acro{mos}[MOS]{Mean Opinion Score} \acro{mrc}[MRC]{Maximum Ratio
Combining} \acro{ms}[MS]{Mobile Station}
\acro{msk}[MSK]{Minimum-Shift
Keying} 
\acro{mtc}[MTC]{machine type communications}
\acro{mu}[MU]{Multi-User} 
\acro{mud}[MUD]{MultiUser
Detection} \acro{mui}[MUI]{MultiUser Interference}
\acro{music}[MUSIC]{MUltiple SIgnal Classification}
\acro{nbi}[NBI]{NarrowBand Interference}
\acro{nc}[NC]{Non-Contiguous}
\acro{nc-ofdm}[NC-OFDM]{Non-Contiguous OFDM}
\acro{nl}[NL]{NonLinear} \acro{nmse}[NMSE]{Normalized Mean-Squared
Error} 
\acro{obo}[OBO]{Output Back-Off} 
\acro{ofdp}[OFDP]{Orthogonal
Finite Duration Pulse}\acro{omp}[OMP]{Orthogonal Matching Pursuit}
\acro{oqpsk}[OQPSK]{Offset Quadrature Phase-Shift Keying}
\acro{osa}[OSA]{Opportunistic Spectrum Access} \acro{pam}[PAM]{Pulse Amplitude Modulation}
\acro{papr}[PAPR]{Peak-to-Average Power Ratio}
\acro{pci}[PCI]{Perfect Channel Information}
\acro{per}[PER]{Packet Error Rate} \acro{pf}[PF]{Proportional
Fair} \acro{phy}[PHY]{Physical layer}  \acro{plc}[PLC]{Power Line
Communications}
\acro{pmr}[PMR]{Professional (or Private) Mobile
Radio} \acro{ppdr}[PPDR]{Public Protection and Disaster Relief}
% \acrodefplular{rb}[RB]{resource blocks}
\acro{prose}[ProSe]{Proximity Services} 
\acro{psk}[PSK]{Phase-Shift Keying}
 \acro{pswf}[PSWF]{Prolate Spheroidal Wave Function}
\acro{pts}[PTS]{Partial Transmit Sequence}
\acro{ptt}[PTT]{Push-To-Talk} \acro{pu}[PU]{Primary User}
\acro{pucch}[PUCCH]{Physical Uplink Control Channel}
\acro{pusch}[PUSCH]{Physical Uplink Shared Channel}
\acro{qam}[QAM]{Quadrature Amplitude Modulation}
\acro{qoe}[QoE]{Quality of Experience} \acro{qos}[QoS]{Quality of
Service}
\acro{ram}[RAM]{Random Access Memory} \acro{rat}[RAT]{Radio Access
Technology} 
\acro{rf}[RF]{Radio Frequency} \acro{rls}[RLS]{Recursive
Least Squares}  \acro{roc}[ROC]{Receiver Operating
Characteristics}
\acro{rrm}[RRM]{Radio Resource Management}
\acro{rssi}[RSSI]{Received Signal Strength Indicator}
\acro{sc-fde}[SC-FDE]{Single-Carrier Frequency-Domain
Equalization} 
\acro{sdm}[SDM]{Space-Division
Multiplexing} \acro{sdma}[SDMA]{Space-Division  Multiple Access}
% \acro{sdr}[SDR]{Signal-to-Distortion power Ratio}
\acro{sel}[SEL]{Soft Envelope Limiter} \acro{ser}[SER]{Symbol
Error Rate} 
\acro{sfbc}[SFBC]{Space Frequency Block Code}
\acro{sic}[SIC]{Successive Interference Cancellation}
\acro{simo}[SIMO]{Single-Input Multiple-Output}
\acro{sinr}[SINR]{Signal-to-Interference-plus-Noise Ratio}
\acro{sir}[SIR]{Signal-to-Interference Ratio}
\acro{siso}[SISO]{Single-Input Single-Output}
\acro{softio}[SI-SO]{Soft-Input Soft-Output}
\acro{slm}[SLM]{Selected Mapping}
\acro{slnr}[SLNR]{Signal-to-Leakage-plus-Noise  Ratio}
\acro{sm}[SM]{Spatial Multiplexing}
\acro{sndr}[SNDR]{Signal-to-Noise-plus-Distortion Ratio}
\acro{snr}[SNR]{Signal-to-Noise Ratio} \acro{ss}[SS]{Spectrum
Sensing} \acro{sspa}[SSPA]{Solid-State Power Amplifiers}
\acro{stbc}[STBC]{Space Time Block Code}
\acro{stbicm}[STBICM]{Space-Time Bit-Interleaved Coded Modulation}
\acro{stc}[STC]{Space-Time Coding} \acro{sthp}[STHP]{Spatial
Tomlinson Harashima Precoder} \acro{su}[SU]{Secondary Users}
\acro{svd}[SVD]{Singular Value Decomposition} \acro{td}[TD]{Time
Domain} \acro{tdd}[TDD]{Time-Division Duplex}
\acro{tdma}[TDMA]{Time-Division Multiple Access}
\acro{teds}[TEDS]{TETRA Enhanced Data Service}
\acro{tetra}[TETRA]{Terrestrial Trunked Radio}
\acro{tfl}[TFL]{Time Frequency Localization} \acro{tgf}[TGF]{Tight
Gabor Frame} \acro{tlo}[TLO]{Time-Limited Orthogonal}
\acro{tmo}[TMO]{Trunked Mode Operation} 
\acro{tr}[TR]{Tone Reservation} 
\acro{ula}[ULA]{Uniform Linear
Array} \acro{v-blast}[V-BLAST]{Vertical Bell Laboratories Layered
Space-Time \acroextra{[code]}} \acro{veh-a}[Veh-A]{Vehicular-A
\acroextra{[channel model]}} \acro{veh-b}[Veh-B]{Vehicular-B
\acroextra{[channel model]}} \acro{wimax}[WiMAX]{Worldwide
Interoperability for Microwave Access} \acro{wlan}[WLAN]{Wireless
Local Area Network} \acro{wlf}[WLF]{Widely Linear Filter}
\acro{zf}[ZF]{Zero Forcing} \acro{zt}[ZT]{Zak Transform}
\end{acronym}

%%% Local Variables: 
%%% fill-column: 80
%%% mode: latex
%%% TeX-master: "ICC2016_JYK"
%%% End:  
 
 % file with acronym definitions

\end{document}